# Sparse Affine Sampling: Ambiguity-Free and Efficient Sparse Phase Retrieval

Ming-Hsun Yang, Y.-W. Peter Hong, and Jwo-Yuh Wu

**Abstract**

Conventional sparse phase retrieval schemes can recover sparse signals from the magnitude of linear measurements only up to a global phase ambiguity. This work proposes a novel approach that instead utilizes the magnitude of affine measurements to achieve ambiguity-free signal reconstruction. The proposed method relies on two-stage approach that consists of support identification followed by the exact recovery of nonzero signal entries. In the noise-free case, perfect support identification using a simple counting rule is guaranteed subject to a mild condition on the signal sparsity, and subsequent exact recovery of the nonzero signal entries can be obtained in closed-form. The proposed approach is then extended to two noisy scenarios, namely, sparse noise (or outliers) and non-sparse bounded noise. For both cases, perfect support identification is still ensured under mild conditions on the noise model, namely, the support size for sparse outliers and the power of the bounded noise. Under perfect support identification, exact signal recovery can be achieved using a simple majority rule for the sparse noise scenario, and reconstruction up to a bounded error can be achieved using linear least-squares (LS) estimation for the non-sparse bounded noise scenario. The obtained analytic performance guarantee for the latter case also sheds light on the construction of the sensing matrix and bias vector. In fact, we show that a near optimal performance can be achieved with high probability by the random generation of the nonzero entries of the sparse sensing matrix and bias vector according to the uniform distribution over a circle. Computer simulations using both synthetic and real-world data sets are provided to demonstrate the effectiveness of the proposed scheme.

This work was supported in part by the Ministry of Science and Technology of Taiwan (MOST) under grants MOST 107-2634-F-009-002, 108-2634-F-009-002, 109-2634-F-007-015 and 110-2634-F-007-021. The work of Y.-W. Peter Hong and J.-Y. Wu is also supported by MOST Joint Research Center for AI Technology and AII Vista Healthcare.

M.-H. Yang and J.-Y. Wu are with the Department of Electrical and Computer Engineering, the Institute of Communications Engineering, and the College of Electrical Engineering, National Chiao Tung University, Hsinchu 30010, Taiwan (e-mail: archenemy.cm00g@nctu.edu.tw; jywu@cc.nctu.edu.tw). Y.-W. Peter Hong is with the Institute of Communications Engineering, National Tsing Hua University, Hsinchu 30013, Taiwan (e-mail: ywhong@ee.nthu.edu.tw).

# I. INTRODUCTION

## A. Overview

Conventional phase retrieval is concerned with the reconstruction of a signal $\mathbf{s} \in \mathbb{C}^N$ from magnitude-squared linear measurements

$$y_m = \left|\underline{\boldsymbol{\phi}}_m \mathbf{s}\right|^2 + v_m, \quad m = 1, \ldots, M, \tag{1}$$

where $M$ is the total number of measurements, $\underline{\boldsymbol{\phi}}_m = [\phi_{m,1} \cdots \phi_{m,N}] \in \mathbb{C}^{1 \times N}$ is the known sensing vector, and $v_m \in \mathbb{R}$ is the noise. This problem arises in many engineering and scientific applications, such as optical imaging, X-ray crystallography [1], [2], astronomy [3], ptychography [4], etc., mainly because sensing devices in optical systems can only capture the intensity of light waves. It also arises in wireless communications, e.g., channel estimation for millimeter-wave systems, for the purpose of removing unreliable phase information caused by large carrier frequency offset [5], [6]. In the noise-free case, sufficient conditions guaranteeing unique signal recovery (up to a global phase ambiguity) were investigated in [7]–[9]. The development of signal reconstruction algorithms alongside the related performance guarantees have been addressed in [10]–[14]. In the aforementioned works, a complete set of measurements, i.e., $M > N$, is needed, which would entail high data gathering and storage costs, especially when the ambient dimension $N$ is large. However, high-dimensional signals typically lie in low-dimensional subsets, or subspaces, of the ambient domain, thereby admitting sparse representations with respect to a certain basis. Motivated by this fact, sparse phase retrieval, which aims to recover sparse signals from sub-Nyquist amplitude samples (i.e., $M < N$ in (1)), has received considerable attention in recent years [15]–[18]. When noise is absent, it was shown in [15] that every $K$-sparse signal is uniquely determined by $4K - 1$ random Gaussian intensity measurements of the form (1) with a high probability. In the presence of noise, [16] has shown that $\mathcal{O}(K \log(N/K))$ measurements are sufficient for stable signal recovery. Many sparsity-promoting algorithms for efficient signal reconstruction have also been developed in [19]–[24].

It is worth noting that all the aforementioned works ensure signal recovery only up to a global phase ambiguity, which, if unresolved, can significantly degrade the signal reconstruction performance. Many existing works either resort to heuristic numerical search, or restrict their attention to certain signal types (e.g., nonnegativity) when resolving the phase ambiguity. However, additional numerical search incurs higher computational complexity, while extra requirements on the signal limit the applicability of the algorithms.

*B. Paper Contribution*

This paper proposes a novel sparse phase retrieval scheme that is both *exact* (i.e., free from the phase ambiguity) and *computationally efficient*. Instead of adopting magnitude-squared linear measurements, as done in (1), we consider the use of $M < N$ noisy magnitude-squared *affine* measurements, i.e.,

$$y_m = \left|\underline{\boldsymbol{\phi}}_m \mathbf{s} + b_m\right|^2 + v_m, \quad m = 1, \ldots, M, \tag{2}$$

for the reconstruction of the $K$-sparse signal $\mathbf{s} \in \mathbb{C}^N$ with unknown support $\mathcal{T} \subset \{1, \ldots, N\}$, where $|\mathcal{T}| = K \ll N$, and $b_m \in \mathbb{C}$ is a known scalar bias that facilitates the proposed ambiguity-free signal recovery process. Conceptually, the bias $b_m$ in (2) acts as a "reference" signal whose distance to the observation $\underline{\boldsymbol{\phi}}_m \mathbf{s}$ varies with the phase rotation on $\mathbf{s}$. Such variety in the affine measurements offers a simple and systematic approach to ambiguity-free signal recovery, as will be shown later. This is in sharp contrast with the conventional linear measurement considered in (1), where different phase rotation is indistinguishable from the squared distance $|\underline{\boldsymbol{\phi}}_m \mathbf{s}|^2$ measured with respect to the common reference point, namely, the origin. The affine measurement model given in (2) naturally arises in holography [29], in which the bias $b_m$ models the intensity of the background referenced light wave. It has also found applications in the probing of crystal structures via image processing techniques [30], where $b_m$ reflects the prior structural information being encoded into the measurement process.

In addition to the use of affine measurements, we further employ sparse sensing [5], [34], [35] (i.e., sparse sensing vectors $\underline{\boldsymbol{\phi}}_m$, $m = 1, \ldots, M$) to reduce the data gathering cost and to facilitate low-complexity signal recovery. Sparse sensing has often been adopted in modern sensing and data acquisition systems. For instance, in single-pixel cameras [36], the inbuilt digital micromirror devices can be mathematically modeled as a set of binary "zero-one" sensing vectors, which are sparse provided that most mirrors are rotated to off (or zero) states. In millimeter-wave channel estimation [5], the product of the channel and the hybrid beamforming weight matching the channel eigen-space is a sparse vector, leading to a sparse sensing measurement model. In our paper, we specifically consider the so-called $(N, M, d, K, \alpha)$ union-free family (UFF) of sparse sensing matrices, which has been adopted in [37]–[40] for efficient compressive sensing (CS) signal reconstruction. Based on the *sparse affine* measurements, we propose a two-stage sparse signal reconstruction scheme which first identifies the signal support and subsequently leverages the presence of the bias to conduct ambiguity-free signal recovery. The main contributions of this paper can be summarized as follows.

1) We first examine the ideal noise-free case and propose a simple counting rule that is shown to achieve perfect support identification subject to a mild condition on the signal sparsity. Once the signal support is correctly identified, ambiguity-free recovery of the nonzero signal entries can be done perfectly by exploiting the geometry underpinning the magnitude of the affine measurements. The analysis on the noise-free case provides key insights on the advantages of using affine measurements for sparse phase retrieval compared to the use of linear measurements in most existing works.

2) Then we extend the results in the noise-free case to the following two noisy scenarios:

   a) *Sparse noise:* Under sparse noise (or outlier) corruption, perfect support identification can also be achieved using a simple counting rule subject to a stronger requirement on the signal sparsity. In this case, exact entry-wise ambiguity-free signal recovery can be done by means of a simple majority rule over a finite set of candidate solutions.

   b) *Non-sparse bounded noise:* Under non-sparse bounded noise, we first show that perfect support identification via a similar counting rule is also attainable under mild conditions on the sensing matrix, the bias vector, and the noise level. In this case, ambiguity-free signal recovery can be achieved with low-complexity by utilizing a standard least-squares (LS) approach. An analytic performance guarantee (characterized in terms of the squared error bound) is derived accordingly to justify the stable reconstruction of the proposed LS estimator. The results also offer interesting insights into the geometry of the sensing matrices and bias vectors against the bounded noise corruption.

3) Finally, we provide guidelines for the construction of sensing matrices and bias vectors under non-sparse bounded noise corruption. We first derive the optimal sensing matrices and bias vectors that minimize the analytic error bound. The optimal solutions are of equal-magnitude and depend on the true signal support, which is unknown beforehand. To facilitate our designs, we instead consider a probabilistic construction using, e.g., random samples uniformly distributed on the circle. We can show that, with high probability, such randomly generated solutions are near-optimal in the sense that the resultant squared error bound is at most a constant multiple of the theoretical minimum.

*C. Connection to Previous Works*

In the current literature, the greedy sparse phase retrieval (GESPAR) [25], the sparse truncated amplitude flow (SPARTA) [26], and the compressive reweighted amplitude flow (CRAF) [27] rely on a similar protocol as our proposed approach, namely, support identification followed by signal retrieval over the estimated support. Built on the conventional linear measurement model given in (1), signal reconstruction using GESPAR, SPARTA and CRAF is subject to a phase ambiguity. GESPAR iteratively updates the support estimate using the 2-opt algorithm [28], and then conducts signal reconstruction using the damped Gauss-Newton algorithm. SPARTA performs iterative support identification using the power method initialized with a properly selected support estimate; then, signal recovery is done by a series of truncated gradient iterations in conjunction with hard thresholding. CRAF adopts a sparse spectral procedure for support identification and a series of reweighted gradient iterations in conjunction with hard thresholding for signal recovery. In contrast to GESPAR, SPARTA and CRAF, we employ only a simple counting rule for support identification and either the majority rule (under sparse noise) or linear LS estimation (under bounded noise) for signal reconstruction. The proposed scheme is not only computationally efficient, but also resolves the phase ambiguity existing in all prior methods.

It is worthwhile to note that sparse phase retrieval using affine measurements has also been addressed in [33]. However, the authors focused exclusively on the noise-free case and derived necessary and sufficient conditions on $\underline{\phi}_m$'s and $b_m$'s for exact signal identification, without addressing the issue of algorithm design. To the best of our knowledge, our work is the first to develop efficient algorithms, along with their theoretical performance guarantees, for ambiguity-free sparse phase retrieval under the noisy affine measurement model.

*D. Paper Organization and Notation List*

The rest of this paper is organized as follows. Section II introduces the problem statement. Section III presents the proposed signal reconstruction algorithm in the noise-free case. Section IV then shows the proposed algorithm and its performance guarantee under noise corruption. Section VI presents the simulation results. Finally, Section VII concludes this paper.

We use bold capital letters for matrices (e.g., $\boldsymbol{\Phi}, \mathbf{A}$), bold small case letters for vectors (e.g., $\mathbf{y}, \mathbf{s}$) and non-bold letters for scalars (e.g., $b, \alpha$). $\mathbf{A}^T$ and $\mathbf{A}^H$ denote the transpose and Hermitian of the matrix $\mathbf{A}$, respectively. $a_{m,n}$ and $y_m$ are used to denote the $(m,n)$th entry of the matrix $\mathbf{A}$ and the $m$th entry of the vector $\mathbf{y}$, respectively. To conserve notation and without causing confusion, $|\mathcal{S}|$ denotes the cardinality of the finite set $\mathcal{S}$. For any matrix $\mathbf{A} = [a_{m,n}]_{M \times N} \in \mathbb{C}^{M \times N}$, $|\mathbf{A}|^2 \in \mathbb{R}^{M \times N}$ is the matrix with $|a_{m,n}|^2$ as the $(m, n)$th entry. For a complex scalar $b \in \mathbb{C}$, the variables $b^{(\text{Re})} \in \mathbb{R}$, $b^{(\text{Im})} \in \mathbb{R}$, and $b^* \in \mathbb{C}$ denote its real part, imaginary part, and complex conjugate, respectively; clearly, $b^* = b^{(\text{Re})} - ib^{(\text{Im})}$. For any $t_1 < t_2 \in \mathbb{R}$, $\theta \sim \mathcal{U}(t_1, t_2)$ implies that $\theta$ is a continuous uniform random variable in the interval $(t_1, t_2)$.

## II. PROBLEM STATEMENT AND BASIC ASSUMPTIONS

We consider the problem of recovering a $K$-sparse signal $\mathbf{s} = [s_1 \cdots s_N]^T \in \mathbb{C}^N$ (i.e., a signal vector with at most $K$ nonzero entries) from $M < N$ magnitude-squared affine measurements as described in (2). The measurements can be written in a vector-matrix form given as follows:

$$\mathbf{y} = [y_1 \cdots y_M]^T = |\mathbf{\Phi s} + \mathbf{b}|^2 + \mathbf{v}, \qquad (3)$$

where $\mathbf{\Phi} = [\underline{\phi}_1^T \cdots \underline{\phi}_M^T]^T \in \mathbb{C}^{M \times N}$ is the sensing matrix (with $\underline{\phi}_m \in \mathbb{C}^{1 \times N}$ being its $m$th row), $\mathbf{b} = [b_1 \cdots b_M]^T \in \mathbb{C}^M$ is the bias vector, and $\mathbf{v} = [v_1 \cdots v_M]^T \in \mathbb{R}^M$ is the noise vector. We assume that the sensing matrix $\mathbf{\Phi}$ is sparse in the sense that the $n$th column of $\mathbf{\Phi}$ (denoted by $\phi_n$) has support $\mathcal{C}_n \subset \{1, \ldots, M\}$ with cardinality $|\mathcal{C}_n| = d$ for all $n \in \{1, \ldots, N\}$. In this way, each nonzero signal element is captured by exactly $d$ measurements in (3). Moreover, we assume that the overlapping support between any two columns is at most $r$, i.e., $|\mathcal{C}_n \cap \mathcal{C}_{n'}| \leq r$, for all $n \neq n'$. Sparse sensing matrices of this kind are formally referred to as the UFF family with parameter $(N, M, d, 1, \frac{r+1}{d})$ [40], [41] that have been widely adopted in the CS [37]–[39] and group testing [42]–[44] literature, and can be constructed using deterministic approaches such as DeVore's recipe [45].

*Assumption 1:* The sensing matrix $\mathbf{\Phi}$ belongs to the $(N, M, d, 1, \frac{r+1}{d})$-UFF family, i.e., $|\mathcal{C}_n| = d$, for all $n$, and $|\mathcal{C}_n \cap \mathcal{C}_{n'}| \leq r$, for all $n \neq n'$.

An example of the abovementioned sensing matrix is shown in Fig. 1 for the case where $N = 8$, $M = 5$, $d = 2$ and $r = 1$. The proposed signal recovery scheme exploits the sparse and affine nature of the sampling process in (3). Sparse sensing, while economizing the storage cost, ought to be properly selected so as to ensure efficient data acquisition. In fact, to yield good data gathering efficiency, the sparse sensing matrix $\mathbf{\Phi}$ should probe the signal $\mathbf{s}$ well enough so that $\mathbf{\Phi s}$ ends up with sufficiently many nonzero terms. This can be accomplished if the column supports are more dispersed over the measurement domain index set $\{1, \ldots, M\}$. That is, the support between different columns do not overlap significantly, resulting in a small value of $r$.

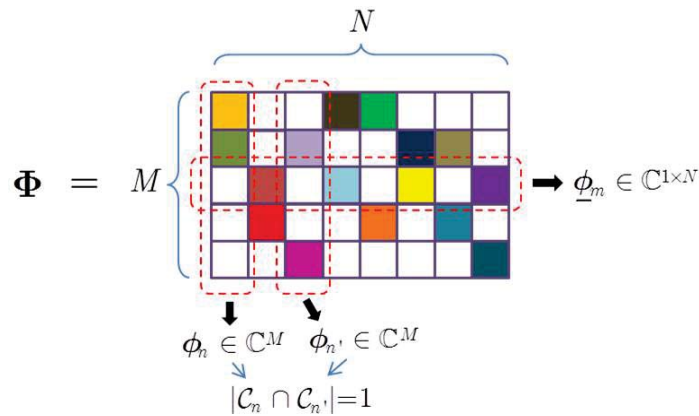

Fig. 1. Example of a sparse sensing matrix $\boldsymbol{\Phi}$ in UFF family with $N=8$, $M=5$, $d=2$, and $r=1$. The columns and rows of $\boldsymbol{\Phi}$ are denoted by $\{\boldsymbol{\phi}_n\}_{n=1}^N$ and $\{\underline{\boldsymbol{\phi}}_m\}_{m=1}^M$, respectively.

Furthermore, to enable element-wise signal recovery, as will be shown in Section III, the following assumption on the nonzero entries of $\boldsymbol{\Phi}$ and the bias values $\{b_m\}_{m=1}^M$ are also necessary throughout the paper.

*Assumption 2:* For any $n \in \{1, \ldots, N\}$ and any distinct indices $m_1, m_2, m_3 \in \mathcal{C}_n$, the three points $\frac{b_{m_1}}{\phi_{m_1,n}}$, $\frac{b_{m_2}}{\phi_{m_2,n}}$ and $\frac{b_{m_3}}{\phi_{m_3,n}}$ on the complex plane must not be collinear, i.e., $\frac{b_{m_1}}{\phi_{m_1,n}} - \frac{b_{m_2}}{\phi_{m_2,n}} \neq a\left(\frac{b_{m_1}}{\phi_{m_1,n}} - \frac{b_{m_3}}{\phi_{m_3,n}}\right)$ for any $a \in \mathbb{R}$.

Assumption 2 can be easily satisfied when the nonzero entries of $\boldsymbol{\Phi}$ and the bias values $\{b_m\}_{m=1}^M$ are independently drawn from continuous probability distributions, such as Gaussian or uniform distributions.

## III. NOISE-FREE SPARSE AFFINE PHASE RETRIEVAL

In this section, we first address the issue of signal retrieval from (3) in the noise-free case, where $\mathbf{v} = \mathbf{0}$. Generalization of the results to the noisy case will be discussed in Section IV. The proposed signal retrieval method consists of two stages: *support identification* and *recovery of nonzero signal entries*, which will be introduced in Sections III-A and III-B, respectively.

### A. Support Identification

In the noise-free case, we write (3) as

$$\mathbf{y} = \left|\sum_{n \in \mathcal{T}} s_n \boldsymbol{\phi}_n + \mathbf{b}\right|^2. \tag{4}$$

Since **s** is $K$-sparse and each column support of $\boldsymbol{\Phi}$ satisfies $|\mathcal{C}_n| = d$, at most $|\bigcup_{n \in \mathcal{T}} \mathcal{C}_n| \leq Kd$ of the entries in **y** will capture nonzero parts of the sparse signal **s**. The rest of the entries in **y** that do not capture any nonzero signal entry yield only the square of the the corresponding bias term, that is, $y_m = |b_m|^2$ whenever $m \notin \bigcup_{n \in \mathcal{T}} \mathcal{C}_n$. In most practical cases, we have $y_m = |\underline{\boldsymbol{\phi}}_m \mathbf{s} + b_m|^2 \neq |b_m|^2$, for $m \in \mathcal{C}_n$ and $n \in \mathcal{T}$, since $\underline{\boldsymbol{\phi}}_m$ is nonzero in its $n$th entry, and so is **s**. Motivated by this observation, we propose the following support identification scheme that is built on a simple counting rule:

$$\hat{\mathcal{T}} = \left\{ n \in \{1, \ldots, N\} \,\bigg|\, \sum_{m \in \mathcal{C}_n} 1\left\{ y_m \neq |b_m|^2 \right\} > \eta \right\}, \tag{5}$$

where $1\{\cdot\}$ is the indicator function, and $0 < \eta < d$ is a decision threshold. We show below that, under quite mild conditions on the signal sparsity, the proposed support estimate in (5) is guaranteed to be exact, i.e., $\hat{\mathcal{T}} = \mathcal{T}$.

*Theorem 3.1:* Let us consider the noise-free model in (4). Then, under Assumptions 1, 2 and with $\frac{d+r-2}{2r} > K$, the support estimate $\hat{\mathcal{T}}$ (defined in (5)) with $\eta \in [Kr, d - Kr + r - 2)$ exactly identifies the signal support $\mathcal{T}$.

*Proof:* See Appendix A. ∎

The above theorem shows that perfect support identification can be achieved by means of a simple counting rule in (5) if the signal support is small enough and the column support of the sensing matrix is large enough to successfully capture the nonzero signal entries. In fact, for $K \geq 2$, the sparsity condition is satisfied with an $(N, M, d, 1, \frac{1}{2K})$-UFF family of matrices since, by setting $\frac{r+1}{d} = \frac{1}{2K} < \frac{1}{2K-1} < \frac{d+2K-3}{(2K-1)d}$, it follows that $\frac{d+r-2}{2r} > K$. Notably, as shown in [40], an $(N, M, d, 1, \frac{1}{2K})$-UFF family of matrices can be generated using error-correcting codes with $M = \mathcal{O}(K^2 \log N)$ and $d = \mathcal{O}(K \log N)$.

*B. Signal Recovery*

With perfect support estimation (i.e., $\hat{\mathcal{T}} = \mathcal{T}$), we are then ready to address the issue of signal recovery for all signal entries associated with the support $\mathcal{T}$, i.e., $\{s_n\}_{n \in \mathcal{T}}$. We can see from (4) that the $m$th entry of **y** (i.e., $y_m$) captures the $n$th nonzero entry $s_n$ if and only if $m \in \mathcal{C}_n$. Let

$$\tilde{\mathcal{C}}_n \triangleq \mathcal{C}_n \setminus \bigcup_{n' \in \mathcal{T} \setminus \{n\}} \mathcal{C}_{n'} \tag{6}$$

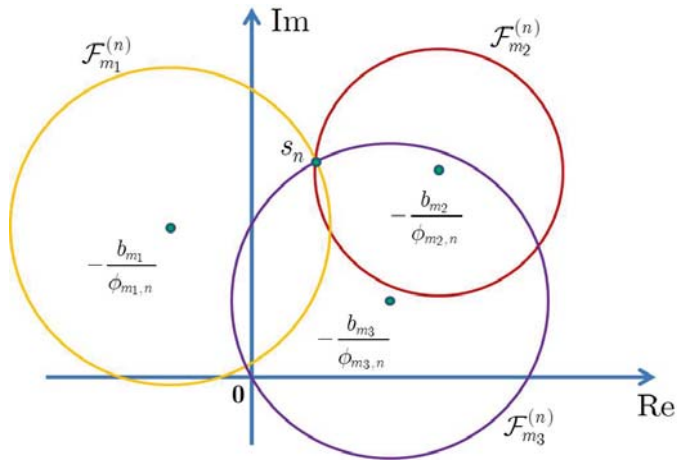

Fig. 2. Depiction of the intersection of three solution sets $\mathcal{F}_{m_1}^{(n)}$, $\mathcal{F}_{m_2}^{(n)}$ and $\mathcal{F}_{m_3}^{(n)}$.

be the subset of indices in $\mathcal{C}_n$ but not in the supports of all other columns indexed by $\mathcal{T}$. Clearly, if $m \in \tilde{\mathcal{C}}_n$, the $m$th measurement captures only the signal component $s_n$ and can be written as

$$y_m = |\phi_{m,n} s_n + b_m|^2. \qquad (7)$$

Notably, $\tilde{\mathcal{C}}_n$ is nonempty if the signal sparsity level $|\mathcal{T}| = K$ is small enough. Indeed, since

$$\left|\tilde{\mathcal{C}}_n\right| = \left|\mathcal{C}_n \setminus \bigcup_{i \in \mathcal{T} \setminus \{n\}} (\mathcal{C}_i \cap \mathcal{C}_n)\right| \geq |\mathcal{C}_n| - \sum_{i \in \mathcal{T} \setminus \{n\}} |\mathcal{C}_i \cap \mathcal{C}_n| \stackrel{(a)}{\geq} d - (K-1)r, \qquad (8)$$

where (a) holds by Assumption 1, it follows that $|\tilde{\mathcal{C}}_n| > Kr + 2 \geq 3$ provided that $\frac{d+r-2}{2r} > K$ and $K, r \geq 1$. The measurements in the form of (7) allow the use of a simple method to recover the nonzero entries of $\mathbf{s}$.

In particular, notice that, for $n \in \mathcal{T}$, the signal component $s_n$ must satisfy (7) and, thus, must belong to the set

$$\mathcal{F}_m^{(n)} = \left\{ s \in \mathbb{C} \,\middle|\, s = -\frac{b_m}{\phi_{m,n}} + \frac{\sqrt{y_m}}{|\phi_{m,n}|} e^{i\theta_m}, \text{ for any } \theta_m \in (0, 2\pi] \right\}, \qquad (9)$$

for all $m \in \tilde{\mathcal{C}}_n$, i.e., $s_n \in \cap_{m \in \tilde{\mathcal{C}}_n} \mathcal{F}_m^{(n)}$. The set $\mathcal{F}_m^{(n)}$ can be depicted as a circle centered at $-\frac{b_m}{\phi_{m,n}}$ with radius $\frac{\sqrt{y_m}}{|\phi_{m,n}|}$. Hence, as long as the center points $-\frac{b_{m_1}}{\phi_{m_1,n}}$, $-\frac{b_{m_2}}{\phi_{m_2,n}}$, and $-\frac{b_{m_3}}{\phi_{m_3,n}}$ are not collinear for at least 3 distinct indices $m_1$, $m_2$, and $m_3$ in $\tilde{\mathcal{C}}_n$ (which is satisfied by Assumption 2), $s_n$ can be easily obtained by taking the intersection of their respective solution sets $\mathcal{F}_{m_1}^{(n)}$, $\mathcal{F}_{m_2}^{(n)}$, and $\mathcal{F}_{m_3}^{(n)}$, as illustrated in Fig. 2. We have the following theorem.

**Algorithm 1** Fast Signal Recovery Algorithm

**Input:** $\mathbf{y}, \mathbf{\Phi}, \mathbf{b}, \eta$.

**Output:** Estimated signal $\hat{\mathbf{s}}$.

1: Initialize index set $\hat{\mathcal{T}} = \emptyset$.
2: **for** $n = 1, \ldots, N$ **do**  ▷ Stage I: Support identification
3:     Compute $\sum_{m \in \mathcal{C}_n} \mathbf{1}\{y_m \neq |b_m|^2\}$.
4:     **if** $\sum_{m \in \mathcal{C}_n} \mathbf{1}\{y_m \neq |b_m|^2\} > \eta$ **then**
5:         $\hat{\mathcal{T}} \leftarrow \hat{\mathcal{T}} \cup \{n\}$.
6:     **else**
7:         $\hat{s}_n = 0$.
8:     **end if**
9: **end for**
10: **for** $n \in \hat{\mathcal{T}}$ **do**  ▷ Stage II: Nonzero signal entry retrieval
11:     Select three elements $m_1, m_2, m_3 \in \mathcal{C}_n \setminus \bigcup_{j \in \hat{\mathcal{T}} \setminus \{n\}} \mathcal{C}_j$, and then compute $\hat{s}_n$ via (10).
12: **end for**

*Theorem 3.2:* Under Assumption 2, it follows that, for any $n \in \mathcal{T}$ and distinct indices $m_1$, $m_2$, and $m_3$ in $\tilde{\mathcal{C}}_n$, the intersection of sets $\mathcal{F}_{m_1}^{(n)}$, $\mathcal{F}_{m_2}^{(n)}$ and $\mathcal{F}_{m_3}^{(n)}$ yields a singleton set, i.e., the solution of $s_n$ to the equation in (7), for $m = m_1$, $m_2$, and $m_3$ is unique. The solution is given by

$$s_n = \frac{-i}{2\mathrm{Im}\left(\left(\frac{b_{m_1}}{\phi_{m_1,n}} - \frac{b_{m_2}}{\phi_{m_2,n}}\right)\left(\frac{b_{m_1}}{\phi_{m_1,n}} - \frac{b_{m_3}}{\phi_{m_3,n}}\right)^*\right)} \left[\left(\frac{b_{m_1}}{\phi_{m_1,n}} - \frac{b_{m_2}}{\phi_{m_2,n}}\right)\left(\frac{y_{m_1} - |b_{m_1}|^2}{|\phi_{m_1,n}|^2} - \frac{y_{m_3} - |b_{m_3}|^2}{|\phi_{m_3,n}|^2}\right) \right.$$
$$\left. - \left(\frac{b_{m_1}}{\phi_{m_1,n}} - \frac{b_{m_3}}{\phi_{m_3,n}}\right)\left(\frac{y_{m_1} - |b_{m_1}|^2}{|\phi_{m_1,n}|^2} - \frac{y_{m_2} - |b_{m_2}|^2}{|\phi_{m_2,n}|^2}\right)\right], \quad (10)$$

where $\mathrm{Im}(x)$ presents the imaginary part of $x$.

*Proof:* See Appendix B ∎

Hence, as long as $\hat{\mathcal{T}} = \mathcal{T}$, Theorem 3.2 suggests the following simple signal recovery protocol: (i) for $n \in \mathcal{T}$, select three elements $m_1, m_2, m_3 \in \tilde{\mathcal{C}}_n$ and find the associated solution sets $\mathcal{F}_{m_1}^{(n)}$, $\mathcal{F}_{m_2}^{(n)}$ and $\mathcal{F}_{m_3}^{(n)}$ using (9); (ii) compute $s_n$ according to the formula (10). The proposed signal recovery scheme is summarized in Algorithm 1.

## IV. SIGNAL RECONSTRUCTION FROM NOISY MEASUREMENTS

In this section, we extend the proposed two-step signal reconstruction scheme in Section III to the noisy case, where the signal model in (3) can be written as

$$\mathbf{y} = \left| \sum_{n \in \mathcal{T}} s_n \boldsymbol{\phi}_n + \mathbf{b} \right|^2 + \mathbf{v}. \tag{11}$$

In particular, we consider two noisy scenarios, namely, the sparse noise (or outlier) case in Section IV-A and the non-sparse bounded noise case in Section IV-B. We show that perfect signal recovery is possible under mild conditions in the former case and provide theoretical performance guarantees on the reconstruction error in the latter case (c.f., Section IV-C).

### A. Case I: Sparse Noise (or Outlier)

In the case with sparse noise, we assume that the noise vector $\mathbf{v}$ in (11) is $K_v$-sparse with unknown support $\mathcal{V} \subset \{1, \ldots, M\}$. Conceptually, for small $K_v$, few measurements in $\mathbf{y}$ are contaminated by noise. As a result, we can still resort to the counting-based rule in (5) for support identification, but with a larger threshold $\eta$ to tolerate outlier corruption. This is formalized in the following theorem.

*Theorem 4.1:* Let us consider the signal model in (11) with the noise $\mathbf{v}$ being $K_v$-sparse. Under Assumptions 1, 2 and with $\frac{d+r-2}{2} > Kr + K_v$, the support estimate $\hat{\mathcal{T}}$ in (5) with $\eta \in [Kr + K_v, d - Kr - K_v + r - 2)$ exactly identifies the signal support $\mathcal{T}$.

*Proof:* See Appendix C. ∎

Suppose that the conditions in Theorem 4.1 hold so that $\hat{\mathcal{T}} = \mathcal{T}$. To reconstruct the nonzero entries in $\mathbf{s}$, namely, $s_n$ for all $n \in \mathcal{T}$, we again utilize the measurements in (7) but with the noise taken into consideration, i.e.,

$$y_m = |\phi_{m,n} s_n + b_m|^2 + v_m, \ m \in \tilde{\mathcal{C}}_n. \tag{12}$$

In the presence of noise, $s_n$ may no longer belong to the set $\mathcal{F}_m^{(n)}$, defined in (9). As a result, the recovery of $s_n$ may no longer be performed by simply taking the intersection of solution sets $\mathcal{F}_{m_1}^{(n)}$, $\mathcal{F}_{m_2}^{(n)}$ and $\mathcal{F}_{m_3}^{(n)}$, for distinct indices $m_1, m_2, m_3 \in \tilde{\mathcal{C}}_n$, as done in the noise-free case. Nevertheless, thanks to the sparse nature of $\mathbf{v}$, many $v_m$'s in (12) are zero, meaning that the nonzero $s_n$ may still belong to the majority of sets $\mathcal{F}_m^{(n)}$ such that $m \in \tilde{\mathcal{C}}_n$. Motivated by this observation, we propose to approximate $s_n$ by the value that lies in the most number of the sets $\mathcal{F}_m^{(n)}$, for $m \in \tilde{\mathcal{C}}_n$, i.e.,

$$\hat{s}_n = \arg \max_{s \in \bigcup_{m \in \tilde{\mathcal{C}}_n} \mathcal{F}_m^{(n)}} \sum_{m \in \tilde{\mathcal{C}}_n} 1\left\{s \in \mathcal{F}_m^{(n)}\right\}. \tag{13}$$

Note that the above solution requires the search over all possible solutions in $\bigcup_{m \in \tilde{\mathcal{C}}_n} \mathcal{F}_m^{(n)}$, which is prohibitive in practice. However, by the sparsity of $\mathbf{v}$ and the geometry of the sets $\mathcal{F}_m^{(n)}$, for all $m \in \tilde{\mathcal{C}}_n \setminus \mathcal{V}$, it is possible to reduce the search over a finite set of candidate solutions. In particular, note that $s_n \in \mathcal{F}_m^{(n)}$, for all $m \in \tilde{\mathcal{C}}_n \setminus \mathcal{V}$. Under the assumption that $\frac{d+r-2}{2} > Kr + K_v$ (i.e., for $K_v$ sufficiently small), we have

$$|\tilde{\mathcal{C}}_n \setminus \mathcal{V}| \geq |\tilde{\mathcal{C}}_n| - |\mathcal{V}| = |\tilde{\mathcal{C}}_n| - K_v \stackrel{(a)}{\geq} d - (K-1)r - K_v$$
$$\stackrel{(b)}{>} Kr + K_v + 2 = Kr + |\mathcal{V}| + 2 > |\mathcal{V}| + 2, \tag{14}$$

where (a) follows from (8) and (b) follows from the assumption that $\frac{d+r-2}{2} > Kr + K_v$. Consequently, it follows that the cardinality $|\tilde{\mathcal{C}}_n \setminus \mathcal{V}|$ can be even larger than half of $|\tilde{\mathcal{C}}_n|$ since

$$|\tilde{\mathcal{C}}_n \setminus \mathcal{V}| = \frac{|\tilde{\mathcal{C}}_n \setminus \mathcal{V}| + |\tilde{\mathcal{C}}_n \setminus \mathcal{V}|}{2} > \frac{|\tilde{\mathcal{C}}_n \setminus \mathcal{V}| + |\mathcal{V}| + 2}{2} = \frac{|\tilde{\mathcal{C}}_n|}{2} + 1. \tag{15}$$

Let us partition the sets $\{\mathcal{F}_m^{(n)}\}_{m \in \tilde{\mathcal{C}}_n}$ into $\lfloor \frac{|\tilde{\mathcal{C}}_n|}{2} \rfloor$ disjoint pair-wise groups, say, $\{\mathcal{F}_{m_1}^{(n)}, \mathcal{F}_{m_2}^{(n)}\}$, $\{\mathcal{F}_{m_3}^{(n)}, \mathcal{F}_{m_4}^{(n)}\}, \ldots, \{\mathcal{F}_{m_{2\lfloor \frac{|\tilde{\mathcal{C}}_n|}{2} \rfloor - 1}}^{(n)}, \mathcal{F}_{m_{2\lfloor \frac{|\tilde{\mathcal{C}}_n|}{2} \rfloor}}^{(n)}\}$ (with the remaining set discarded if $|\tilde{\mathcal{C}}_n|$ is odd). Notice that, by (15), the value $s_n$ must belong to more than half of the sets $\mathcal{F}_{m_p}^{(n)}$, for $p = 1, \ldots, 2\lfloor \frac{|\tilde{\mathcal{C}}_n|}{2} \rfloor$ (even after discarding a set when $|\tilde{\mathcal{C}}_n|$ is odd). Hence, there must exist $p^* \in \{1, 2, \ldots, \lfloor \frac{|\tilde{\mathcal{C}}_n|}{2} \rfloor\}$ such that both $y_{m_{2p^*-1}}$ and $y_{m_{2p^*}}$ are noiseless. In this case, $s_n$ must belong to both sets $\mathcal{F}_{m_{2p^*-1}}^{(n)}$ and $\mathcal{F}_{m_{2p^*}}^{(n)}$ (i.e., $s_n \in \mathcal{F}_{m_{2p^*-1}}^{(n)} \cap \mathcal{F}_{m_{2p^*}}^{(n)}$). This implies that the desired value $s_n$ must belong to the set $\bigcup_{p=1}^{\lfloor \frac{|\tilde{\mathcal{C}}_n|}{2} \rfloor} \left\{\mathcal{F}_{m_{2p-1}}^{(n)} \cap \mathcal{F}_{m_{2p}}^{(n)}\right\} \triangleq \tilde{\mathcal{F}}^{(n)}$ and, thus, (13) reduces to the following:

$$\hat{s}_n = \arg \max_{s \in \tilde{\mathcal{F}}^{(n)}} \sum_{m \in \tilde{\mathcal{C}}_n} 1\left\{s \in \mathcal{F}_m^{(n)}\right\}. \tag{16}$$

Notice that, since any two circles intersect at two points or less, it follows that $|\mathcal{F}_{m_{2p-1}}^{(n)} \cap \mathcal{F}_{m_{2p}}^{(n)}| \leq 2$, for all $p$, and thus the number of candidate solutions is $|\tilde{\mathcal{F}}^{(n)}| \leq |\tilde{\mathcal{C}}_n|$, which is finite. This reduces the complexity of the search significantly. The following theorem provides a sufficient condition guaranteeing exact signal recovery using (16) under sparse noise corruption.

*Theorem 4.2:* Under the same setting as in Theorem 4.1, the proposed estimate of $s_n$ given in (16) exactly recovers the $K$-sparse signal $\mathbf{s}$ from the measurement vector given in (11).

*Proof:* See Appendix D. ∎

Theorems 4.1 and 4.2 show that, under sparse noise corruption, the true sparse vector **s** can be perfectly and efficiently recovered subject to a stricter sparsity condition, compared to the noiseless case (c.f., Theorem 3.1), and a larger detection threshold $\eta$.

*B. Case II: Non-sparse Bounded Noise*

Here, we assume that the noise vector **v** is not necessarily sparse, but bounded, i.e., there exists $\epsilon > 0$ such that

$$\|\mathbf{v}\|_\infty < \epsilon. \tag{17}$$

In this case, it is possible that $y_m \neq |b_m|^2$, for all $m \in \mathcal{C}_n$ irrespective of whether $n \in \mathcal{T}$ or not, rendering exact support identification using (5) impossible. To overcome this difficulty, we exploit the difference between $y_m$ and $|b_m|^2$ as an indication of whether the signal participates in $y_m$ or not. Indeed, if $y_m$ misses the signal, i.e., $\underline{\phi}_m \mathbf{s} = 0$, we have $y_m = |b_m|^2 + v_m$ and, thus

$$\left| y_m - |b_m|^2 \right| = |v_m| < \epsilon. \tag{18}$$

On the contrary, if $y_m$ captures the signal and $|\underline{\phi}_m \mathbf{s}|$ is reasonably large, it is expected that

$$\left| y_m - |b_m|^2 \right| > \epsilon. \tag{19}$$

Consequently, if $n \in \mathcal{T}$, (19) would imply that the majority of the measurements $\{y_m\}_{m \in \mathcal{C}_n}$ are at least $\epsilon$ away from $|b_m|^2$. By assuming that $\epsilon$ is known, the support identification rule in (5) can be modified as

$$\hat{\hat{\mathcal{T}}} = \left\{ n \in \{1, \ldots, N\} \middle| \sum_{m \in \mathcal{C}_n} \mathbf{1}\left\{ |y_m - |b_m|^2| > \epsilon \right\} > \eta \right\}. \tag{20}$$

It is easy to see that the modified support estimate in (20) is reduced to (5) in the noise-free case with $\epsilon = 0$. The following theorem establishes that exact support identification (i.e., $\hat{\hat{\mathcal{T}}} = \mathcal{T}$) can be achieved under mild conditions.

*Theorem 4.3:* Let us consider the signal model in (11) under bounded noise corruption, i.e., $\|\mathbf{v}\|_\infty < \epsilon$, and let $\phi_{\min} = \min_{n \in \{1,\ldots,N\}} \min_{m \in \mathcal{C}_n} |\phi_{m,n}|$, $\delta_{\min} = \min_{n \in \mathcal{T}} |s_n|$, and $b_{\max} = \max_{m \in \{1,\ldots,M\}} |b_m|$. For $\delta_{\min} \geq b_{\max}/\phi_{\min} + \sqrt{b_{\max}^2/\phi_{\min}^2 + 2\epsilon/\phi_{\min}^2}$ and $\frac{d+r}{2r} > K$, $\hat{\hat{\mathcal{T}}}$ in (20) with $\eta \in [Kr, d - Kr + r)$ correctly identifies the signal support $\mathcal{T}$.

*Proof:* See Appendix E. ∎

Theorem 4.3 shows that perfect support estimation $\hat{\tilde{\mathcal{T}}} = \mathcal{T}$ can be achieved if the sensing matrices and the bias vectors are chosen so that the ratio $b_{\max}/\phi_{\min}$ is sufficiently small.

By assuming perfect support estimation $\hat{\tilde{\mathcal{T}}} = \mathcal{T}$, we go on to address the issue of recovering $s_n$, for all $n \in \mathcal{T}$. Since it is possible that $v_m \neq 0$, for all $m$, in the non-sparse bounded noise case, the desired value $s_n$ may not belong to any of the sets $\{\mathcal{F}_m^{(n)}\}_{m \in \tilde{\mathcal{C}}_n}$. In this case, exact signal reconstruction using (13) is no longer possible. However, the affine sampling structure in (12) enables us to develop an efficient low-complexity signal reconstruction scheme with an analytic performance guarantee, as elaborated below. When there is no noise, recall from Theorem 3.2, that the unknown $s_n$ can be perfectly recovered by taking the intersection of the solution sets corresponding to (at least) three measurement equations given in (7). In the presence of noise, we instead adopt the least-squares (LS) approach to find $s_n$ that best fits the $|\tilde{\mathcal{C}}_n|$ measurements in (12). To do so, let us first consider the pair of measurements

$$y_{m'} = |\phi_{m',n} s_n + b_{m'}|^2 + v_{m'} \tag{21a}$$

$$y_{m''} = |\phi_{m'',n} s_n + b_{m''}|^2 + v_{m''} \tag{21b}$$

for $m' \neq m'' \in \tilde{\mathcal{C}}_n$, which can be rearranged as

$$\underbrace{\frac{y_{m'}}{|\phi_{m',n}|^2} - \left|\frac{b_{m'}}{\phi_{m',n}}\right|^2}_{\triangleq \tilde{y}_{m'}} = |s_n|^2 + s_n^* \underbrace{\frac{b_{m'}}{\phi_{m',n}}}_{\triangleq \tilde{b}_{m'}} + s_n \frac{b_{m'}^*}{\phi_{m',n}^*} + \underbrace{\frac{v_{m'}}{|\phi_{m',n}|^2}}_{\triangleq \tilde{v}_{m'}} \tag{22a}$$

$$\underbrace{\frac{y_{m''}}{|\phi_{m'',n}|^2} - \left|\frac{b_{m''}}{\phi_{m'',n}}\right|^2}_{\triangleq \tilde{y}_{m''}} = |s_n|^2 + s_n^* \underbrace{\frac{b_{m''}}{\phi_{m'',n}}}_{\triangleq \tilde{b}_{m''}} + s_n \frac{b_{m''}^*}{\phi_{m'',n}^*} + \underbrace{\frac{v_{m''}}{|\phi_{m'',n}|^2}}_{\triangleq \tilde{v}_{m''}} \tag{22b}$$

By letting $\tilde{y}_m \triangleq \frac{y_m}{|\phi_{m,n}|^2} - \left|\frac{b_m}{\phi_{m,n}}\right|^2$, $\tilde{b}_m \triangleq \frac{b_m}{\phi_{m,n}}$, and $\tilde{v}_m \triangleq \frac{v_m}{|\phi_{m,n}|^2}$ and by subtracting the two equations, we can do away with the common quadratic term $|s_n|^2$, leading to the following first-order equation

$$\tilde{y}_{m'} - \tilde{y}_{m''} = s_n^*(\tilde{b}_{m'} - \tilde{b}_{m''}) + s_n(\tilde{b}_{m'}^* - \tilde{b}_{m''}^*) + \tilde{v}_{m'} - \tilde{v}_{m''}. \tag{23}$$

By performing the procedures for all distinct pairs of measurements in $\tilde{\mathcal{C}}_n = \{m_1, m_2, \ldots, m_{|\tilde{\mathcal{C}}_n|}\}$, we can obtain a system of $|\tilde{\mathcal{C}}_n|(|\tilde{\mathcal{C}}_n|-1)/2$ linear equations which can be written in the following matrix-vector form:

$$\mathbf{G}_n \tilde{\mathbf{y}}_n = s_n^* \mathbf{G}_n \tilde{\mathbf{b}}_n + s_n \mathbf{G}_n \tilde{\mathbf{b}}_n^* + \mathbf{G}_n \tilde{\mathbf{v}}_n = \mathbf{G}_n \begin{bmatrix} \tilde{\mathbf{b}}_n & \tilde{\mathbf{b}}_n^* \end{bmatrix} \begin{bmatrix} s_n^* \\ s_n \end{bmatrix} + \mathbf{G}_n \tilde{\mathbf{v}}_n, \tag{24}$$

where $\tilde{\mathbf{y}}_n \triangleq [\tilde{y}_{m_1} \cdots \tilde{y}_{m_{\tilde{C}_n}}]^T$, $\tilde{\mathbf{b}}_n \triangleq [\tilde{b}_{m_1} \cdots \tilde{b}_{m_{\tilde{C}_n}}]^T$, $\tilde{\mathbf{v}}_n \triangleq [\tilde{v}_{m_1} \cdots \tilde{v}_{m_{\tilde{C}_n}}]^T$, and $\mathbf{G}_n \triangleq [\mathbf{g}_{1,2}\ \mathbf{g}_{1,3}\ \cdots\ \mathbf{g}_{1,|\tilde{C}_n|}\ \mathbf{g}_{2,3}\ \mathbf{g}_{2,4}\ \cdots\ \mathbf{g}_{2,|\tilde{C}_n|}\ \mathbf{g}_{3,4}\ \cdots\ \mathbf{g}_{|\tilde{C}_n|-1,|\tilde{C}_n|}]^T$ with $\mathbf{g}_{i,j}$ being a $|\tilde{C}_n|$-dimensional vector of zeros except with $1$ and $-1$ in the $i$th and $j$th entries, respectively. The value of $s_n$ that best fits the $|\tilde{C}_n|(|\tilde{C}_n|-1)/2$ equations can then be found by the LS approach where the signal estimate is given by

$$\hat{s}_n = \arg\min_{s \in \mathbb{C}} \left\| \mathbf{G}_n \tilde{\mathbf{y}}_n - \mathbf{G}_n \begin{bmatrix} \tilde{\mathbf{b}}_n & \tilde{\mathbf{b}}_n^* \end{bmatrix} \begin{bmatrix} s^* \\ s \end{bmatrix} \right\|_2. \tag{25}$$

The solution $\hat{s}_n$ can be evaluated explicitly as shown in the following theorem.

*Theorem 4.4:* Under Assumption 2, the matrix $\mathbf{G}_n[\tilde{\mathbf{b}}_n\ \tilde{\mathbf{b}}_n^*]$ is full-rank and, thus, and the resultant LS estimate of $s_n$ can be written as

$$\hat{s}_n = \frac{\tilde{\mathbf{b}}_{0,n}^T \tilde{\mathbf{b}}_{0,n} \tilde{\mathbf{b}}_{0,n}^H \tilde{\mathbf{y}}_{0,n} - \|\tilde{\mathbf{b}}_{0,n}\|_2^2 \tilde{\mathbf{b}}_{0,n}^T \tilde{\mathbf{y}}_{0,n}}{\left|\tilde{\mathbf{b}}_{0,n}^T \tilde{\mathbf{b}}_{0,n}\right|^2 - \|\tilde{\mathbf{b}}_{0,n}\|_2^4}, \quad n \in \mathcal{T}, \tag{26}$$

where $\tilde{\mathbf{b}}_{0,n} = \left(\mathbf{I} - \frac{\mathbf{1}\mathbf{1}^T}{|\tilde{C}_n|}\right)\tilde{\mathbf{b}}_n \triangleq \tilde{\mathbf{b}}_n - \bar{b}_n \mathbf{1}$ and $\tilde{\mathbf{y}}_{0,n} = \left(\mathbf{I} - \frac{\mathbf{1}\mathbf{1}^T}{|\tilde{C}_n|}\right)\tilde{\mathbf{y}}_n \triangleq \tilde{\mathbf{y}}_n - \bar{y}_n \mathbf{1}$ are the "centered" bias and measurement vectors, respectively, $\bar{b}_n \triangleq \frac{1}{|\tilde{C}_n|}\sum_{l=1}^{|\tilde{C}_n|} \tilde{b}_{m_l}$ and $\bar{y}_n \triangleq \frac{1}{|\tilde{C}_n|}\sum_{l=1}^{|\tilde{C}_n|} \tilde{y}_{m_l}$ are the sample means, and $\mathbf{1}$ is the $|\tilde{C}_n|$-dimensional all-one vector.

*Proof:* See Appendix F. ■

It is worthwhile to note that, in the absence of noise (i.e., when $\tilde{\mathbf{v}}_n = \mathbf{0}$), the LS solution in (26) reduces to the solution in (10). That is, the approach yields perfect recovery in the noiseless case. In the presence of non-sparse bounded noise, perfect signal recovery is not possible, but performance guarantees can be obtained as shown in the following subsection.

### C. Performance Guarantee for the Non-sparse Bounded Noise Case

In this subsection, we provide performance guarantees on the recovery error of the LS solution given in (26). The results are summarized in the following theorem.

*Theorem 4.5:* Under Assumption 2 and with perfect support estimation (i.e., $\hat{\mathcal{T}} = \mathcal{T}$), the LS solution given in (26) satisfies

$$\|\hat{\mathbf{s}} - \mathbf{s}\|_2 < \frac{\sqrt{K|\tilde{C}_{n^*}|}}{\|\tilde{\mathbf{b}}_{0,n^*}\|_2 \left(1 - \left|\left\langle \frac{\tilde{\mathbf{b}}_{0,n^*}}{\|\tilde{\mathbf{b}}_{0,n^*}\|_2}, \frac{\tilde{\mathbf{b}}_{0,n^*}^*}{\|\tilde{\mathbf{b}}_{0,n^*}^*\|_2}\right\rangle\right|\right)\phi_{\min}^2} \epsilon, \tag{27}$$

where

$$n^* = \arg\max_{n \in \mathcal{T}} \frac{\sqrt{|\tilde{C}_n|}}{\|\tilde{\mathbf{b}}_{0,n}\|_2 \left(1 - \left|\left\langle \frac{\tilde{\mathbf{b}}_{0,n}}{\|\tilde{\mathbf{b}}_{0,n}\|_2}, \frac{\tilde{\mathbf{b}}_{0,n}^*}{\|\tilde{\mathbf{b}}_{0,n}^*\|_2}\right\rangle\right|\right)}. \tag{28}$$

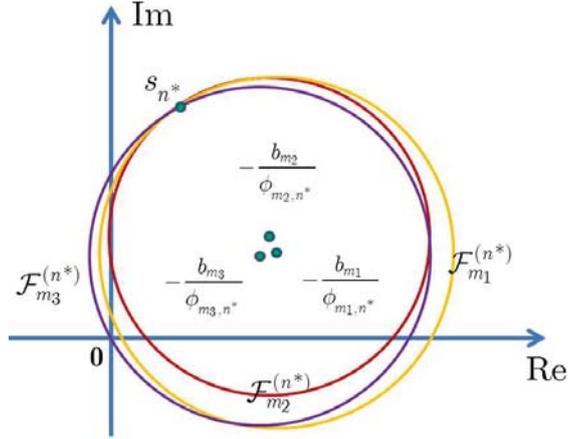

Fig. 3. Depiction of three reference points with small $\|\tilde{\mathbf{b}}_{0,n^*}\|_2$ in the noise-free case. Clearly, since the resultant distance metrics $|s_{n^*} + b_{m_l}/\phi_{m_l,n^*}|$'s tend to be homogeneous, the circles depicting the solution sets $\mathcal{F}_{m_l}^{(n^*)}$'s (9) then largely coincide.

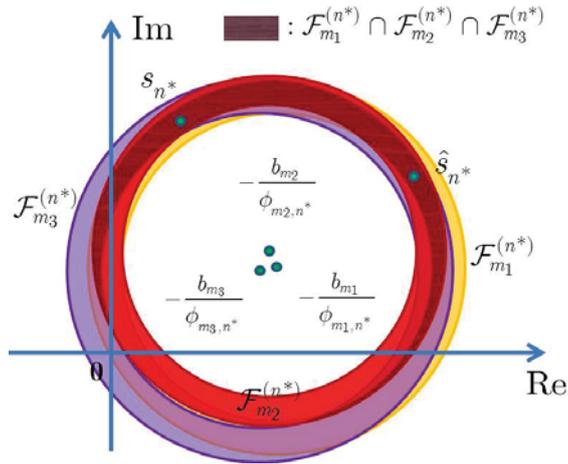

Fig. 4. Example of three reference points with small $\|\tilde{\mathbf{b}}_{0,n^*}\|_2$ in the bounded noise case. The three annular regions represent the solution sets of three equations in (12), respectively. Clearly, since these reference points $-b_{m_l}/\phi_{m_l,n^*}$'s cluster together, the intersection $\mathcal{F}_{m_1}^{(n^*)} \cap \mathcal{F}_{m_2}^{(n^*)} \cap \mathcal{F}_{m_3}^{(n^*)}$ of these three annular regions is fairly large.

*Proof:* See Appendix G. ∎

Recall that the reference points $\{-\tilde{b}_{m_l}\}_{l=1}^{|\tilde{\mathcal{C}}_{n^*}|}$ (i.e., $\{-b_{m_l}/\phi_{m_l,n^*}\}_{l=1}^{|\tilde{\mathcal{C}}_{n^*}|}$) in the proposed affine measurements help remove the global phase ambiguity, since the magnitude $|s_{n^*} + b_{m_l}/\phi_{m_l,n^*}|$ varies with the phase rotation on $s_{n^*}$. The results of Theorem 4.5 offer interesting insights into the geometry of the reference points against the bounded noise, as discussed below.

First, notice that the error bound in (27) decreases with $\|\tilde{\mathbf{b}}_{0,n^*}\|_2 = \|\tilde{\mathbf{b}}_{n^*} - \bar{b}_{n^*}\mathbf{1}\|_2$, which

is the square root of the sum of the square of the centered reference points. When $\|\tilde{\mathbf{b}}_{0,n^*}\|_2$ is small, i.e., the reference points $\{-b_{m_l}/\phi_{m_l,n^*}\}_{l=1}^{|\tilde{\mathcal{C}}_{n^*}|}$ become close to each other, and thus the circles representing the solution sets $\mathcal{F}_{m_l}^{(n^*)}$, for all $l$, almost coincide, as illustrated in Fig. 3. This is not a problem in the noise-free case since the circles intersect at a finite number of points that include the desired solution $s_n$. However, in the presence of non-sparse bounded noise, the solution instead lies in the intersection of the annular regions described by $\tilde{\mathcal{F}}_{m_l}^{(n^*)} = \left\{ s \in \mathbb{C} \,\middle|\, s = -\frac{b_{m_l}}{\phi_{m_l,n^*}} + \frac{\sqrt{y_{m_l}-v_{m_l}}}{|\phi_{m_l,n^*}|} e^{i\theta_{m_l}}, \text{ for any } \theta_{m_l} \in (0, 2\pi] \text{ and } |v_{m_l}| < \epsilon \right\}$. The annular regions overlap significantly in this case, as illustrated in Fig. 4, causing large uncertainty in the signal recovery. Hence, a larger $\|\tilde{\mathbf{b}}_{0,n^*}\|_2$ would lead to less uncertainty, and thus reduce the signal recovery error.

Second, the error bound in (27) is also seen to decrease with the absolute inner product $\left| \left\langle \frac{\tilde{\mathbf{b}}_{0,n^*}}{\|\tilde{\mathbf{b}}_{0,n^*}\|_2}, \frac{\tilde{\mathbf{b}}_{0,n^*}^*}{\|\tilde{\mathbf{b}}_{0,n^*}^*\|_2} \right\rangle \right|$, which, in brief, gauges to what extent the reference points $\{-b_{m_l}/\phi_{m_l,n^*}\}_{l=1}^{|\tilde{\mathcal{C}}_{n^*}|}$ become collinear (as will be shown at the end of this section). A large $\left| \left\langle \frac{\tilde{\mathbf{b}}_{0,n^*}}{\|\tilde{\mathbf{b}}_{0,n^*}\|_2}, \frac{\tilde{\mathbf{b}}_{0,n^*}^*}{\|\tilde{\mathbf{b}}_{0,n^*}^*\|_2} \right\rangle \right|$ implies the reference points are nearly aligned with each other. The resultant intersected annular region is split into disjoint parts (see the blue region in Fig. 5), among which those "aliasing pieces" do not contain the true $s_{n^*}$ but incur a higher reconstruction error. Hence, a small $\left| \left\langle \frac{\tilde{\mathbf{b}}_{0,n^*}}{\|\tilde{\mathbf{b}}_{0,n^*}\|_2}, \frac{\tilde{\mathbf{b}}_{0,n^*}^*}{\|\tilde{\mathbf{b}}_{0,n^*}^*\|_2} \right\rangle \right|$, i.e., the reference points are far from being collinear, is beneficial to accurate signal recovery. We end this section by providing a mathematical theorem underpinning how the absolute inner product $\left| \left\langle \frac{\tilde{\mathbf{b}}_{0,n^*}}{\|\tilde{\mathbf{b}}_{0,n^*}\|_2}, \frac{\tilde{\mathbf{b}}_{0,n^*}^*}{\|\tilde{\mathbf{b}}_{0,n^*}^*\|_2} \right\rangle \right|$ reflects the collinearity of the reference points. Recall the following perhaps the most intuitive test of collinearity: consider the LS line fit [46] of the point set $\{-b_{m_l}/\phi_{m_l,n^*}\}_{l=1}^{|\tilde{\mathcal{C}}_{n^*}|}$, compute the distance from each $-b_{m_l}/\phi_{m_l,n^*}$ to the line, and use the sum of all such distances, dubbed as the *total residual* and denoted by $r^*$, as a measure. Clearly, the smaller the value of $r^*$ is, the larger the extent that the reference points are aligned (see Fig. 6 for an illustration). The following theorem establishes an explicit connection between $\left| \left\langle \frac{\tilde{\mathbf{b}}_{0,n^*}}{\|\tilde{\mathbf{b}}_{0,n^*}\|_2}, \frac{\tilde{\mathbf{b}}_{0,n^*}^*}{\|\tilde{\mathbf{b}}_{0,n^*}^*\|_2} \right\rangle \right|$ and $r^*$.

*Theorem 4.6:* Let $n^*$ be defined as in (28), and let $r^*$ be the total residual of the LS line fit on the complex points $\{-b_{m_l}/\phi_{m_l,n^*}\}_{l=1}^{|\tilde{\mathcal{C}}_{n^*}|}$. Then we have

$$\left| \left\langle \frac{\tilde{\mathbf{b}}_{0,n^*}}{\|\tilde{\mathbf{b}}_{0,n^*}\|_2}, \frac{\tilde{\mathbf{b}}_{0,n^*}^*}{\|\tilde{\mathbf{b}}_{0,n^*}^*\|_2} \right\rangle \right| = 1 - \frac{2r^*}{\|\tilde{\mathbf{b}}_{0,n^*}\|_2^2}. \tag{29}$$

*Proof:* See Appendix H. ∎

We can clearly see from (29) that the total residual $r^*$ decreases to zero (i.e., the reference points become more and more aligned to each other) as $\left|\left\langle \frac{\tilde{\mathbf{b}}_{0,n^*}}{\|\tilde{\mathbf{b}}_{0,n^*}\|_2}, \frac{\tilde{\mathbf{b}}^*_{0,n^*}}{\|\tilde{\mathbf{b}}^*_{0,n^*}\|_2}\right\rangle\right|$ increases to one. Notably, exact alignment, i.e., $r^* = 0$, occurs if and only if $\left|\left\langle \frac{\tilde{\mathbf{b}}_{0,n^*}}{\|\tilde{\mathbf{b}}_{0,n^*}\|_2}, \frac{\tilde{\mathbf{b}}^*_{0,n^*}}{\|\tilde{\mathbf{b}}^*_{0,n^*}\|_2}\right\rangle\right| = 1$, which is precluded thanks to Assumption 2. In fact, without Assumption 2, it can be readily deduced using Cauchy-Schwarz inequality that the two vectors $\tilde{\mathbf{b}}_{0,n^*}$ and $\tilde{\mathbf{b}}^*_{0,n^*}$ are linearly dependent, which in turn implies that the rank of $\mathbf{G}_{n^*}[\tilde{\mathbf{b}}_{n^*}\ \tilde{\mathbf{b}}^*_{n^*}]$ is less than 2, contradicting with the assertion of Theorem 4.4.

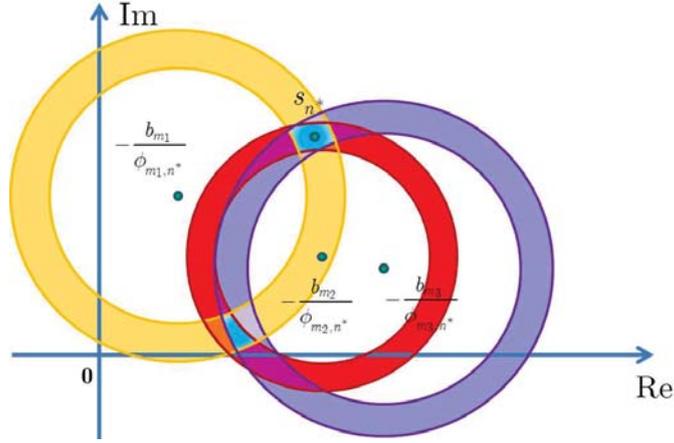

Fig. 5. Depiction of three reference points $-\frac{b_{m_1}}{\phi_{m_1,n^*}}$, $-\frac{b_{m_2}}{\phi_{m_2,n^*}}$ and $-\frac{b_{m_3}}{\phi_{m_3,n^*}}$ with getting close to alignment. Clearly, the intersection (the blue regions) of three associated annular regions is disconnected and, therefore, the aliasing piece, which does not contain $s_{n^*}$, exists, leading to signal reconstruction performance degradation.

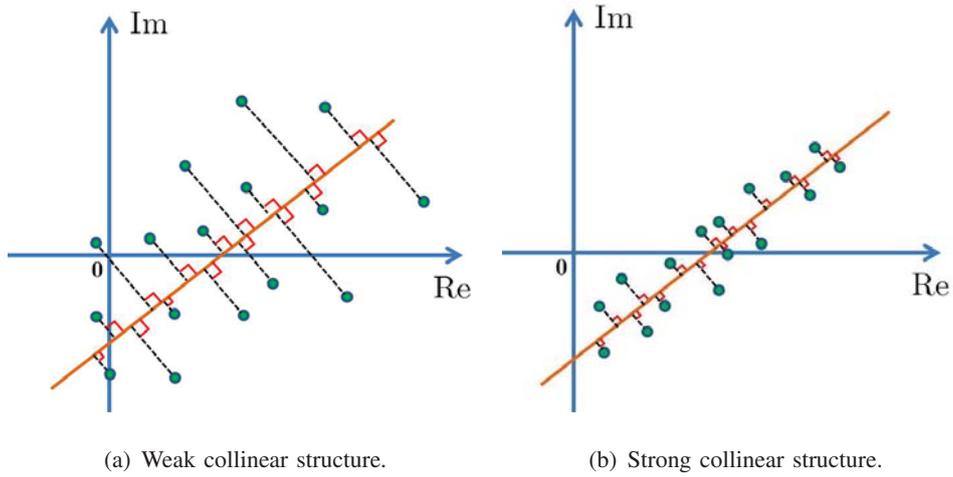

(a) Weak collinear structure.  (b) Strong collinear structure.

Fig. 6. Illustration of collinearlity level of two point sets with different geometric structures, where the orange line is the LS fitting line. Clearly, compared to the point set in Fig. 6(a), the points in 6(b) exhibit more strong collinear structure since they are more close to the best fitting line (orange line), yielding smaller value of the total residual $r^*$.

## V. Construction of $\boldsymbol{\Phi}$ and $\mathbf{b}$

The mathematical performance guarantee shown in Theorem 4.5 offers a guideline on the construction of the sensing matrix $\boldsymbol{\Phi}$ and the bias vector $\mathbf{b}$. To improve signal estimation accuracy, $\boldsymbol{\Phi}$ and $\mathbf{b}$ can be chosen to minimize the error bound given in (27), or equivalently, to minimize $\sqrt{|\tilde{\mathcal{C}}_{n^*}|}\big/\|\tilde{\mathbf{b}}_{0,n^*}\|_2\Big(1-\Big|\Big\langle\frac{\tilde{\mathbf{b}}_{0,n^*}}{\|\tilde{\mathbf{b}}_{0,n^*}\|_2},\frac{\tilde{\mathbf{b}}_{0,n^*}^*}{\|\tilde{\mathbf{b}}_{0,n^*}^*\|_2}\Big\rangle\Big|\Big)$. Toward this end, we first derive a lower bound on $\sqrt{|\tilde{\mathcal{C}}_{n^*}|}\big/\|\tilde{\mathbf{b}}_{0,n^*}\|_2\Big(1-\Big|\Big\langle\frac{\tilde{\mathbf{b}}_{0,n^*}}{\|\tilde{\mathbf{b}}_{0,n^*}\|_2},\frac{\tilde{\mathbf{b}}_{0,n^*}^*}{\|\tilde{\mathbf{b}}_{0,n^*}^*\|_2}\Big\rangle\Big|\Big)$, as presented in the following lemma.

*Lemma 5.1:* Let $\phi_{\min}$ and $b_{\max}$ be defined as in Theorem 4.3, and let $n^*$ be defined as in (28). Then, it follows that

$$\frac{\sqrt{|\tilde{\mathcal{C}}_{n^*}|}}{\|\tilde{\mathbf{b}}_{0,n^*}\|_2\Big(1-\Big|\Big\langle\frac{\tilde{\mathbf{b}}_{0,n^*}}{\|\tilde{\mathbf{b}}_{0,n^*}\|_2},\frac{\tilde{\mathbf{b}}_{0,n^*}^*}{\|\tilde{\mathbf{b}}_{0,n^*}^*\|_2}\Big\rangle\Big|\Big)} \geq \frac{\phi_{\min}}{b_{\max}}, \tag{30}$$

where the equality holds if (i) $\big\langle \tilde{\mathbf{b}}_{n^*}, \tilde{\mathbf{b}}_{n^*}^* \big\rangle = 0$ and $\big\langle \tilde{\mathbf{b}}_{n^*}, \mathbf{1} \big\rangle = 0$ and (ii) $\phi_{m,n^*}$ and $b_m$ are respectively equal-amplitude with $|\phi_{m,n^*}| = \phi_{\min}$ and $|b_m| = b_{\max}$, for all $m \in \tilde{\mathcal{C}}_{n^*}$.

*Proof:* See Appendix I. ■

The theorem shows that, under conditions (i) and (ii), the bound in (30) is tight and, therefore, the error bound in (27) reduces to

$$\|\hat{\mathbf{s}} - \mathbf{s}\|_2 < \frac{\sqrt{K|\tilde{\mathcal{C}}_{n^*}|}}{\|\tilde{\mathbf{b}}_{0,n^*}\|_2\Big(1-\Big|\Big\langle\frac{\tilde{\mathbf{b}}_{0,n^*}}{\|\tilde{\mathbf{b}}_{0,n^*}\|_2},\frac{\tilde{\mathbf{b}}_{0,n^*}^*}{\|\tilde{\mathbf{b}}_{0,n^*}^*\|_2}\Big\rangle\Big|\Big)\phi_{\min}^2}\epsilon = \frac{\sqrt{K}}{b_{\max}\phi_{\min}}\epsilon. \tag{31}$$

Since the signal support $\mathcal{T}$ is not known in advance, it is by no means possible to determine the index $n^*$ in (28) so as to fulfill the optimality conditions, in particular, the pair-wise orthogonality requirement (i), asserted in Lemma 5.1. Despite this, a near optimal performance can be achieved with high probability by resorting to probabilistic reconstruction. To see this, let us commence with the equal-magnitude requirement (ii) in Lemma 5.1 that is easy to realize, and propose the following random generation scheme:

---

Random construction of $\boldsymbol{\Phi}$ and $\mathbf{b}$:

1) The nonzero entries of $\boldsymbol{\Phi}$ are i.i.d. and uniformly drawn from the circle of radius $\phi_c$, i.e., $\phi_{m,n} = \phi_c e^{i\varphi_{m,n}}$, where $\varphi_{m,n} \sim \mathcal{U}(0, 2\pi)$, for all $1 \leq n \leq N$ and $m \in \mathcal{C}_n$.
2) The entries of $\mathbf{b}$, say, $b_m$ for $1 \leq m \leq M$, are i.i.d. and uniformly drawn from the circle of radius $b_c$ such that $b_m = b_c e^{i\theta_m}$, where $\theta_m \sim \mathcal{U}(0, 2\pi)$.

We note that the equal-amplitude solution is reasonable since the sampling process considered here is non-adaptive, and no prior knowledge about signal and noise (other than signal support size and noise level) is available beforehand. In accordance with the proposed random construction scheme, we have the following theorem.

*Theorem 5.2:* Let $n' \triangleq \arg\min_{n \in \mathcal{T}} |\tilde{\mathcal{C}}_n|$. Under the same setting as in Theorem 4.5, if the cardinality of $\tilde{\mathcal{C}}_{n'}$ satisfies $|\tilde{\mathcal{C}}_{n'}| \geq Ct^{-2}\log(K)$, for some positive constant $C$ and for every $0 \leq t \leq \frac{-1+\sqrt{1+\rho^2}}{2}$, where $\rho \triangleq b_c/\phi_c$, the estimated signal using (26) obeys

$$\|\hat{\mathbf{s}} - \mathbf{s}\|_2 < (1 + \delta(t))\frac{\sqrt{K}}{b_c \phi_c}\epsilon, \tag{32}$$

where $\delta(t) = \frac{4(t+1)/\rho^2}{1-4(t^2+t)/\rho^2}t$, with probability exceeding $1 - \exp(-c_1|\tilde{\mathcal{C}}_{n'}|t^2)$, for some positive constant $c_1$.

*Proof:* See Appendix J. ∎

It is easy to verify that $\delta(t) < 1$ if $t < \frac{-1+\sqrt{1+\rho^2/2}}{2}$. Hence, Theorem 5.2 shows that when $|\tilde{\mathcal{C}}_{n'}|$ is sufficiently large so that $|\tilde{\mathcal{C}}_{n'}| \geq Ct^{-2}\log(K)$ holds for a fairly small $t$, then it occurs with overwhelming probability that the error bound in (32) is at most a small constant multiple of the minimal error bound in (31). Also recall that, as long as $\hat{\mathcal{T}} = \mathcal{T}$, the nonzero entry $s_n$ is estimated by solving the $|\tilde{\mathcal{C}}_n|$ equations in (12). Theorem 5.2 asserts that, under the proposed random construction, $|\tilde{\mathcal{C}}_n| = \mathcal{O}(\log K)$ equations is required for every $n \in \mathcal{T}$ to achieve near optimal performance in terms of (32) with an overwhelming probability. This implies that $M = \mathcal{O}(K|\tilde{\mathcal{C}}_n|) = \mathcal{O}(K \log K)$ measurements are needed in the second stage (i.e., the signal recovery stage) of the proposed signal reconstruction scheme. Notably, by following similar statements as in Theorem 3.1, $M = \mathcal{O}(K^2 \log N)$ is required to guarantee $\hat{\mathcal{T}} = \mathcal{T}$ in the first stage (i.e., the support identification stage). Therefore, the number of measurements that is required for the overall proposed algorithm to succeed must scale as $M = \mathcal{O}(K^2 \log N)$.

## VI. EXPERIMENTAL RESULTS

In this section, numerical simulations are used to illustrate the effectiveness of the proposed scheme. The ambient signal dimension is set as $N = 7500$, the signal support $\mathcal{T}$ is selected uniformly at random, and the nonzero signal entries $s_n$, for $n \in \mathcal{T}$, are generated according to an i.i.d. circularly-symmetric complex Gaussian distribution with variance $\sigma_s^2 = 2$. The bias terms $b_m$, for $m = 1, \ldots, M$, are randomly and uniformly drawn from a circle with radius $\sqrt{2}$ centered at origin. The sparse sensing matrix $\mathbf{\Phi}$ is constructed following DeVore's recipe [45], where the number of measurements satisfies $M = |\mathcal{C}_n|^2 = d^2$, with $d$ being some prime number; as such and with $N = 7500$, $d = \sqrt{M} < \sqrt{N} \approx 86.6$. Throughout all experiments, the size of the overlapping support between any two columns of $\mathbf{\Phi}$ is at most $r = 2$. The nonzero entries of $\mathbf{\Phi}$ are randomly and uniformly drawn from the circle centered at origin with radius $\sqrt{2}$ in complex plane. We compare the proposed scheme with SPARTA [26] and CRAF [27], which employ dense sensing matrices with entries that are i.i.d. standard complex Gaussian variables (i.e., zero mean and unit variance). In order to evaluate the quality of the signal recovery, we consider the following two metrics: the *relative error* (RE) [47]

$$\text{RE} \triangleq \frac{\|\hat{\mathbf{s}} - \mathbf{s}\|_2}{\|\mathbf{s}\|_2}, \tag{33}$$

and the *ambiguity-removed relative error* (AR-RE) [27]

$$\text{AR-RE} \triangleq \min_{\omega \in [0, 2\pi)} \left( \frac{\|\hat{\mathbf{s}} e^{i\omega} - \mathbf{s}\|_2}{\|\mathbf{s}\|_2} \right). \tag{34}$$

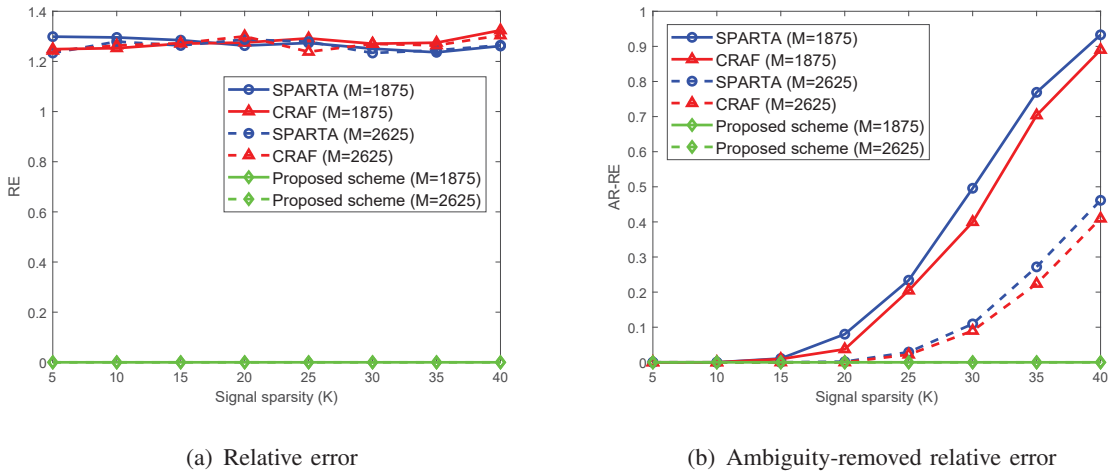

(a) Relative error  (b) Ambiguity-removed relative error

Fig. 7. Performance comparisons of SPARTA, CRAF, and the proposed scheme for different values of the signal sparsity $K$.

All results are averaged over 250 trials, and all experiments are conducted using MATLAB R2019b on a desktop with an Intel Core i9 CPU at 2.9GHz and 16GB RAM.

## A. Synthetic Data: Noise-free Case

We first compare the performance of all methods in the noise-free case. The decision threshold in (5) is set as $\eta = d - 1$. For measurement size $M = 1875$ and $2625$,[1] Figure 7(a) plots the

---

[1] For our method, we first construct a matrix $\mathbf{\Phi}'$ of size $43^2 \times 7500$ ($47^2 \times 7500$) following DeVore's recipe. Then the sensing matrix $\mathbf{\Phi}$ of size $1875 \times 7500$ ($2625 \times 7500$) is generated by appending an $(1875 - 43^2) \times 7500$ (($2625 - 47^2) \times 7500$) all-zero matrix to the bottom of $\mathbf{\Phi}'$.

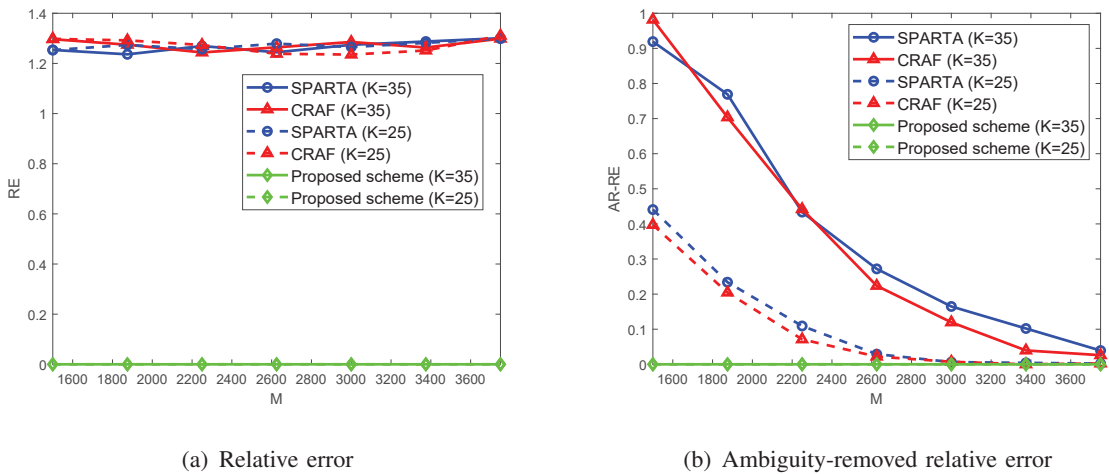

(a) Relative error
(b) Ambiguity-removed relative error

Fig. 8. Performance comparisons of SPARTA, CRAF, and the proposed scheme for different values of $M$.

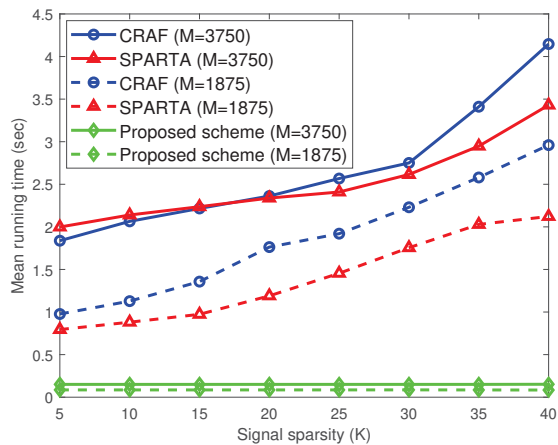

Fig. 9. Mean running time of all methods for different signal sparsity levels.

RE of all methods with respect to different sparsity levels. The result shows that the proposed scheme is able to achieve perfect signal recovery whereas SPARTA and CRAF can result in RE as high as 1.273. This is expected since both SPARTA and CRAF can only recover the signal up to a global phase ambiguity, i.e., $\hat{\mathbf{s}} = \mathbf{s}e^{i\theta}$ for some $\theta \in [0, 2\pi)$. In this case, the expected value of RE is $E\{\|\hat{\mathbf{s}} - \mathbf{s}\|_2/\|\mathbf{s}\|_2\} = \int_0^{2\pi} |e^{i\theta} - 1|/2\pi d\theta = 4/\pi \approx 1.273$. Figure 7(b) depicts AR-RE versus the sparsity level $K$. We can see that, for both values of $M$, the AR-RE of both SPARTA and CRAF increases with the signal sparsity $K$ whereas the proposed scheme again yields perfect signal recovery in all cases. For two sparsity levels $K = 25$ and $35$, Figures 8(a) and 8(b) respectively show the RE and AR-RE curves of all methods with respect to different values of $M$. The result in Figure 8(a) again confirms that the proposed scheme achieves perfect signal recovery, while SPARTA and CRAF can only recover the signal up to a global phase ambiguity. Moreover, it can be observed from Figure 8(b) that the AR-RE of SPARTA and CRAF deviates from $0$ significantly when $M < 2200$. This is because the two methods recover signal based on the restricted isometry property of the sensing matrix which is difficult to attain when $M$ is not sufficiently large. For $M = 1875$ and $3750$, Figure 9 further plots the average running times of all methods as a function of the sparsity level $K$. The result shows that our proposed scheme is at least $8$ times faster than SPARTA and CRAF. Moreover, when the number of measurements $M$ doubles (i.e., $M = 1875$ increases to $M = 3750$), the average running time of the proposed scheme increases only slightly, as opposed to the other algorithms. This is expected since our proposed scheme involves only a simple counting rule for support identification followed by a linear LS estimation of the nonzero signal elements. Note that we can also evaluate the performance of different methods in terms of the success rate, which is defined as the ratio of the number of successful trials (if $AR - RE < 10^{-5}$) to a total of 250 independent runs. Figure 10 plots the success rates of all methods with respect to different sparsity level $K$ and measurement size $M$. As the figure shows, the proposed scheme outperforms SPARTA and CRAF in all cases. Besides, it can be observed from Figures 10(b) and 10(c) that the performance of both CRAF and SPARTA deteriorates when $K$ increases or $M$ decreases, and CRAF has higher success rate than SPARTA since a more stable initialization is adopted.

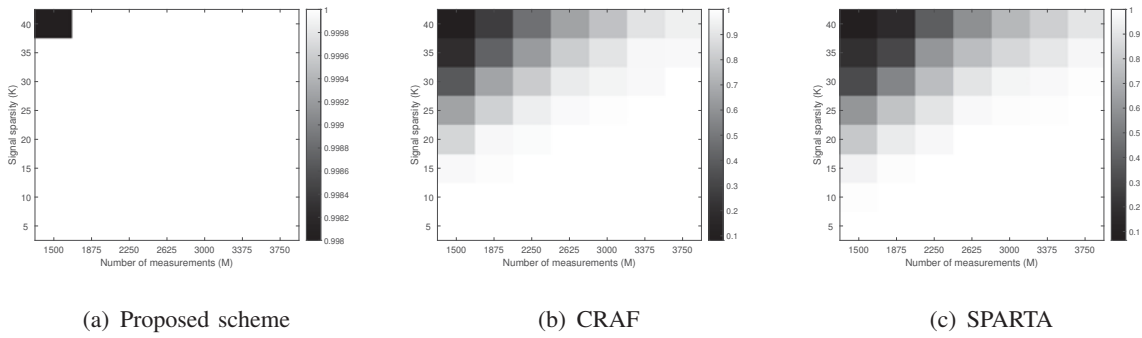

(a) Proposed scheme  (b) CRAF  (c) SPARTA

Fig. 10. Phase transition plots for all algorithms with signal length $N = 7500$.

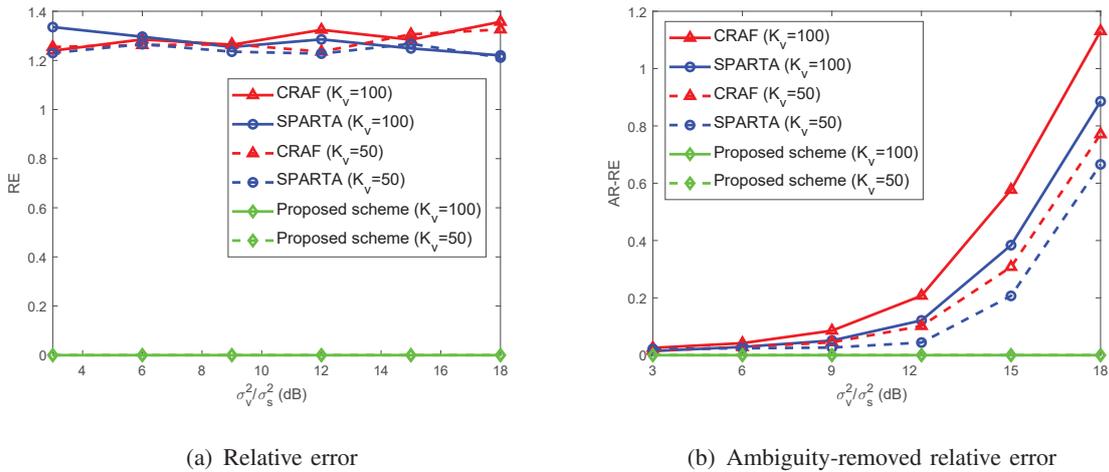

(a) Relative error

(b) Ambiguity-removed relative error

Fig. 11. Performance comparisons of all methods for different values of $\sigma_v^2/\sigma_s^2$ under sparse noise corruption when $K_v = 50$ and 100.

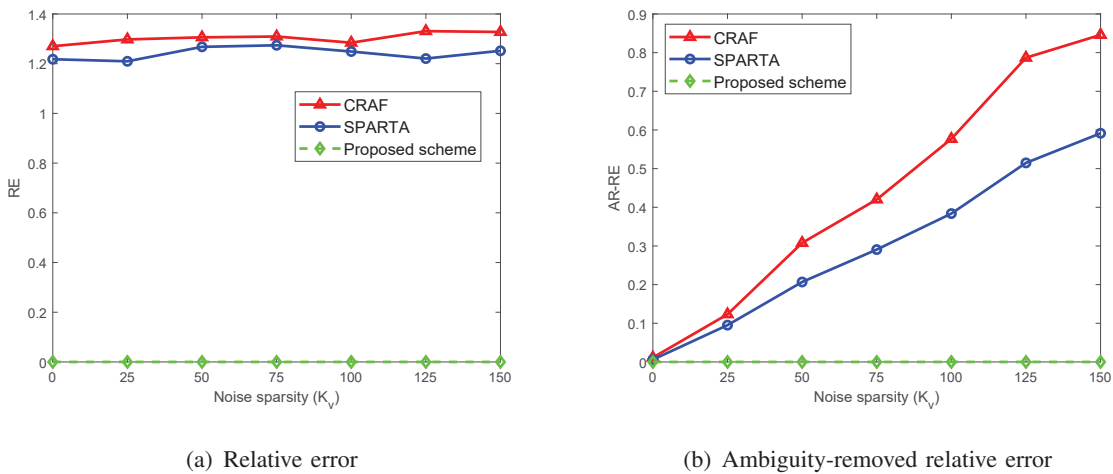

(a) Relative error

(b) Ambiguity-removed relative error

Fig. 12. Performance comparisons of all methods for different values of $K_v$ under sparse noise corruption ($\sigma_v^2/\sigma_s^2 = 15$ dB).

## B. Synthetic Data: Sparse Noise Case

In the second experiment, we examine the performance of all methods under sparse noise corruption. The noise support $\mathcal{V}$ is selected uniformly at random and the nonzero entries of $\mathbf{v}$ are generated according to i.i.d. zero-mean Gaussian distribution with variance $\sigma_v^2$. The decision threshold $\eta$ in (5) is set to be $\eta = d - 1$, as before. For $K = 15$ and $M = 1875$, Figure 11(a) plots the RE curve of all methods for different ratios $\sigma_v^2/\sigma_s^2$ (equal to $\sigma_v^2/2$) when $K_v = 50$ and 100. As the figure shows, with compression ratio $M/N = 0.25$, the proposed scheme can achieve perfect signal recovery in all cases under sparse noise corruption, whereas a global phase ambiguity is inevitable for SPARTA and CRAF. The resultant AR-RE of all methods is then exhibited in Figure 11(b); clearly, it can be seen that the reconstruction error of the two comparisons increases with the ratio $\sigma_v^2/\sigma_s^2$. For $\sigma_v^2/\sigma_s^2 = 15$ dB and $M = 1875$, Figures 12(a) and 12(b) further compare the RE and AR-RE, respectively, of the three methods versus the noise support size $K_v$. It can be observed from Figure 12(a) that the proposed scheme can exactly recover signal against sparse noise (outliers), but SPARTA and CRAF cannot remove global phase ambiguity. In addition, Figure 12(b) clearly shows that when $K_v$ increases, i.e., more measurements are corrupted by outliers, the signal reconstruction error of SPARTA and CRAF increases. This is expected since less reliable measurements can be exploited to reconstruct signal, leading to worse AR-RE. For $K_v = 25$ and $\sigma_v^2/\sigma_s^2 = 6$ dB, Figures 13(a) and 13(b) plot the RE and AR-RE curves, respectively, versus different signal sparsity $K$ when $M = 1875$ and $2625$. As expected, the proposed scheme is observed to achieve perfect signal recovery, while the AR-RE of the other two methods increases with $K$ but decreases with $M$.

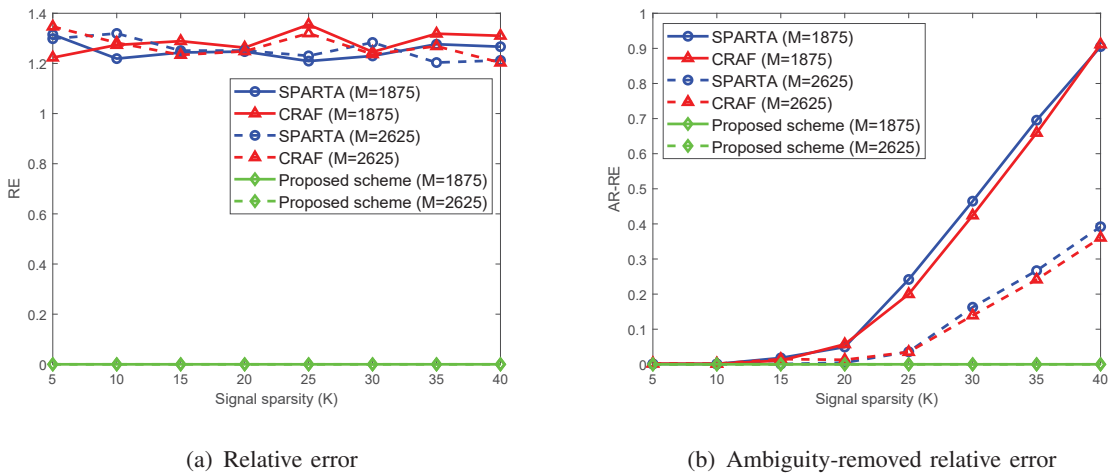

(a) Relative error

(b) Ambiguity-removed relative error

Fig. 13. Performance comparisons of all methods for different values of the signal sparsity $K$ under sparse noise corruption when $M = 1875$ and $2625$.

## C. Synthetic Data: Non-sparse Bounded Noise Case

In the third experiment, we study the performance of all methods under non-sparse bounded noise corruption. In this case, the nonzero entries of s are randomly and uniformly drawn from a circle of radius 5 centered at origin. The bounded noise $v_m$'s are i.i.d. uniform random variables in the interval $(-\epsilon, \epsilon)$, i.e., $v_m \sim \mathcal{U}(-\epsilon, \epsilon)$ for all $1 \leq m \leq M$. For $M = 2825$ ($M/N \approx 0.377$) and $K = 11$, Figure 14(a) plots the RE curves of all methods with respect to different signal-to-

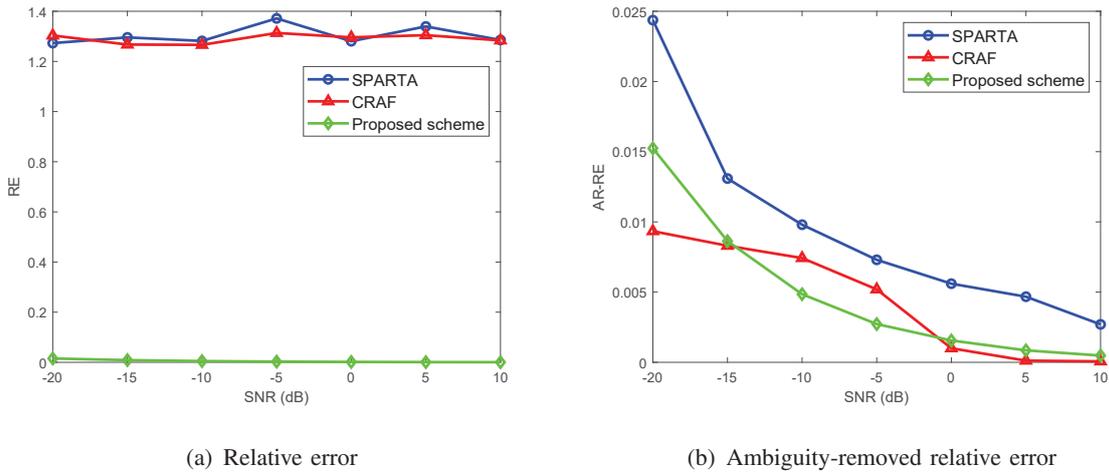

(a) Relative error

(b) Ambiguity-removed relative error

Fig. 14. Performance comparisons of SPARTA, CRAF, and the proposed scheme for different values of SNR under bounded noise corruption.

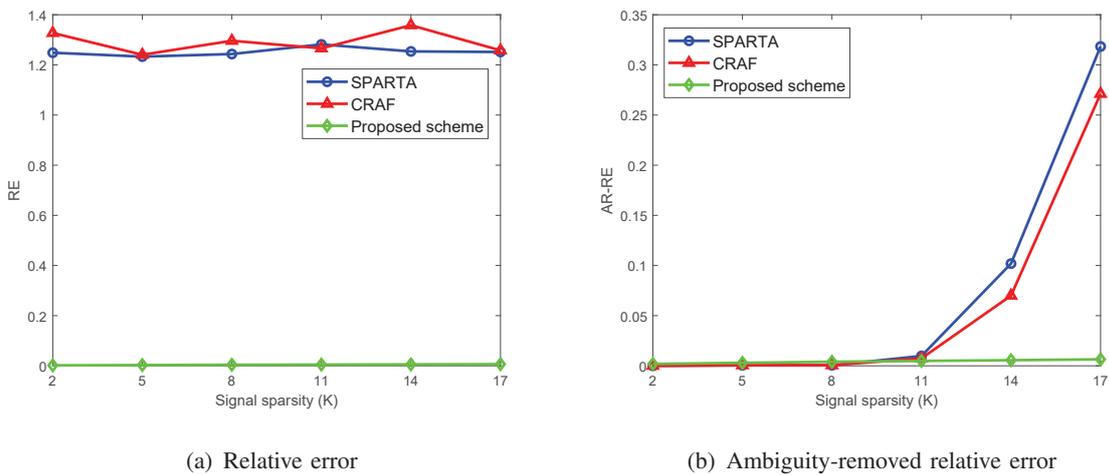

(a) Relative error

(b) Ambiguity-removed relative error

Fig. 15. Performance comparisons of SPARTA, CRAF, and the proposed scheme for different values of the signal sparsity $K$ under bounded noise corruption (SNR $= -10$ dB).

noise ratio (SNR), defined to be $E\{\|\mathbf{s}\|_2^2\}/E\{\|\mathbf{v}\|_2^2\}$ (equal to $75K/M\epsilon^2$ in our setting). As the figure shows, the value of RE for SPARTA and CRAF is around $1.273$, whereas the proposed scheme is free from the phase ambiguity and yields a much lower RE. Figure 14(b) then plots the resultant AR-RE curves. While the performances of all methods degrade as SNR decreases, the proposed scheme exhibits greater robustness to noise as compared to SPARTA. When SNR is low ($< -15$ dB), CRAF yields a lower AR-RE; this is because, in this case, support identification via the proposed simple counting rule (20) is not guaranteed to be perfect, thereby degrading the signal reconstruction performance. When SNR is above $-15$ dB, it can be observed from

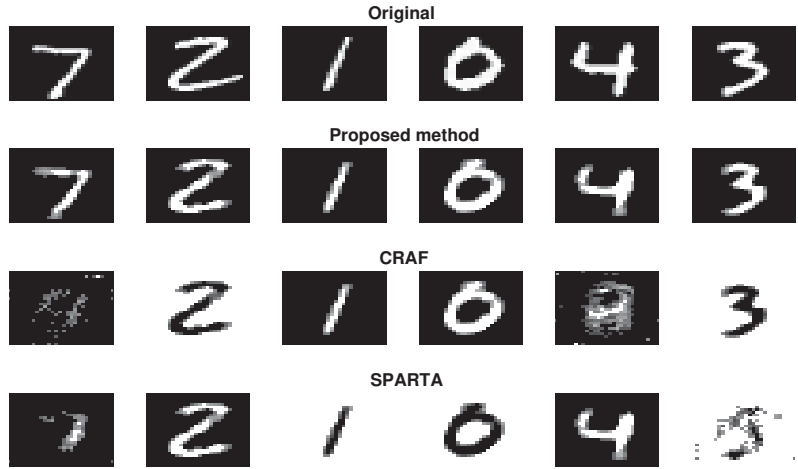

Fig. 16. Six recovered test images (from the MNIST handwritten dataset) using the three methods.

TABLE I

PERFORMANCE COMPARISONS OF ALL METHODS FOR MNIST DATASET.

| Dataset | Algorithm | RE (per pixel) | AR-RE (per pixel) | Mean running time (sec) |
|---|---|---|---|---|
| MNIST | Proposed method | $5.68 \times 10^{-5}$ | $5.68 \times 10^{-5}$ | $6.2 \times 10^{-3}$ |
|  | CRAF | $1.5 \times 10^{-3}$ | $6.55 \times 10^{-4}$ | $8.96 \times 10^{-2}$ |
|  | SPARTA | $1.5 \times 10^{-3}$ | $7.27 \times 10^{-4}$ | $4.17 \times 10^{-2}$ |

figure that the AR-RE curves of the proposed scheme and CRAF are close, meaning that, in the medium-to-high SNR region, our proposed method and CRAF with ambiguity removed perform comparably. We should note that the proposed method adopts a sparse sensing matrix whose Frobenius norm is at least one fifth smaller than the dense complex Gaussian random matrix used in SPARTA and CRAF. This result together Figures 14(a) and 14(b) confirm that the proposed scheme can achieve good performance with significantly fewer data storage cost and sensing power consumption. Figures 15(a) and 15(b) further plot the resultant RE and AR-RE for different values of the signal sparsity $K$ when SNR $= -10$ dB. The results again confirm that the proposed scheme can remove the global phase ambiguity and exhibits greater robustness to noise as compared to SPARTA and CRAF.

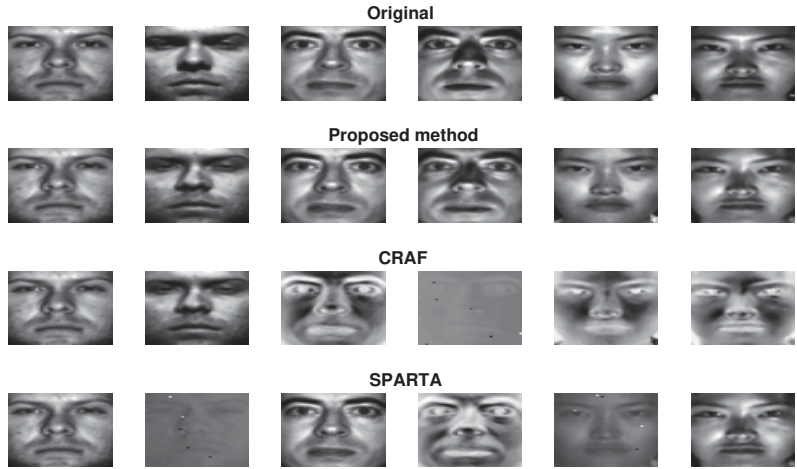

Fig. 17. Six recovered test images (from the Extended Yale B human face dataset) using the three methods.

## D. MNIST and Real Human Face Data

Finally, we examine the performance of the proposed scheme for the image recovery task using the MNIST [48] and Extended Yale B human face datasets [49]. We first consider the MNIST dataset consisting of 60000 training images and 10000 testing images (the size of each image is $28 \times 28 = 784$). Based on the training dataset, the standard PCA technique is used to obtain a proper basis for the testing data, yielding a sparse approximation with $K = 15$. Fig. 16 shows the recovery result of six testing images with $M = 289$ measurements. As can be seen, the proposed scheme, while being free from the sign ambiguity, achieves the highest reconstruction quality. The resultant RE, AR-RE and mean running time of all methods over all trials for MNIST are listed in Table I. As the table shows, our proposed scheme is at least 6 times faster than SPARTA and CRAF, and results in the lowest reconstruction error. We then consider the Extended Yale B dataset consisting of 2452 human face images (each of size $48 \times 42 = 2016$). According to [49], these human face images have a well sparse approximation with $K = 10$ over the PCA-based basis. Under this setup, with $M = 169$ measurements, Figure 17 shows the reconstructed 6 images chosen from the Extended Yale B's test set. As can be seen, our proposed scheme still outperforms SPARTA and CRAF and without global phase ambiguity. Table II lists the RE, AR-RE and mean running time of all the three methods, confirming again that the proposed scheme is faster and achieves higher image reconstruction accuracy.

TABLE II

Performance comparisons of all methods for Extended Yale B human face dataset.

| Dataset | Algorithm | RE (per pixel) | AR-RE (per pixel) | Mean running time (sec) |
|---|---|---|---|---|
| EYB-fc | Proposed method | $1.02 \times 10^{-4}$ | $1.02 \times 10^{-4}$ | $1.3 \times 10^{-2}$ |
| | CRAF | $5.47 \times 10^{-4}$ | $2.1 \times 10^{-4}$ | $2.46 \times 10^{-1}$ |
| | SPARTA | $5.36 \times 10^{-4}$ | $2.23 \times 10^{-4}$ | $5.28 \times 10^{-2}$ |

## VII. Conclusion

Sparse phase retrieval problems focus on the signal reconstruction from an incomplete set of magnitude-squared measurements. In this paper, we proposed the novel concept of sparse affine sensing, and developed a two-step sparse signal reconstruction scheme which first identifies the signal support and subsequently conducts ambiguity-free unknown signal retrieval. To the best of our knowledge, the proposed recovery scheme is the first one in the literature that can achieve unambiguous signal reconstruction in phase retrieval problems. Our analytic results first confirmed that under mild conditions, perfect support identification can be achieved via a simple counting-based rule not only in the noise-free case, but also in the sparse noise and the non-sparse bounded noise cases. After obtaining the support knowledge, ambiguity-free/robust recovery of the unknown signal entries was achieved by leveraging the affine sampling structure. Specifically, with the aid of non-collinear reference points, we first showed that in the noise-free case, unknown signal entries can be exactly recovered element-wise in closed-form. When the measurements are corrupted by sparse noise (i.e., outlier), we showed that exact entry-wise ambiguity-free signal recovery continues to hold through a simple majority rule, provided that the noise support set is sufficiently small. An extension of our study to the non-sparse bounded noise case was also investigated. By leveraging the affine sampling structure, we proposed an efficient element-wise LS estimator, and provided analytic bounds on the signal recovery errors to justify its stability. We then derived the optimal sensing matrices and bias vectors that minimize the obtained theoretical error bound. While deterministic construction of the optimal solutions was hard to accomplish, our analytic results confirmed that near-optimal solutions can be obtained with an overwhelming probability by resorting to probabilistic construction, say, using random entries uniformly distributed on the circle. Finally, simulation results using both synthetic and real-world datasets demonstrated the effectiveness of the proposed scheme.

# APPENDIX A
## PROOF OF THEOREM 3.1

Note that, since $\frac{d+r-2}{2r} > K$, the set $[Kr, d-Kr+r-2)$ must be non-empty. The following lemma, which we prove later in this appendix, is essential to the proof of Theorem 3.1.

*Lemma A.1:* In the noise-free case (i.e., (4)) and under Assumptions 1 and 2, we have

1) $\sum_{m \in \mathcal{C}_n} 1\{y_m \neq |b_m|^2\} \leq Kr$, if $n \notin \mathcal{T}$.
2) $\sum_{m \in \mathcal{C}_n} 1\{y_m \neq |b_m|^2\} \geq d - Kr + r - 2$, if $n \in \mathcal{T}$.

Suppose that $n \in \hat{\mathcal{T}}$. Then, by the definition of $\hat{\mathcal{T}}$ in (5) and the fact that $\eta \in [Kr, d - Kr + r - 2)$, we have $\sum_{m \in \mathcal{C}_n} 1\{y_m \neq |b_m|^2\} > \eta \geq Kr$, which together with part 1) of Lemma A.1 immediately implies that $n \in \mathcal{T}$. Hence, $\hat{\mathcal{T}} \subseteq \mathcal{T}$. Conversely, for $n \in \mathcal{T}$, it follows from part 2) of Lemma A.1 that $\sum_{m \in \mathcal{C}_n} 1\{y_m \neq |b_m|^2\} \geq d - Kr + r - 2$. Since $\eta < d - Kr + r - 2$, we can immediately deduce that $n \in \hat{\mathcal{T}}$. Therefore, $\mathcal{T} \subseteq \hat{\mathcal{T}}$ and, thus, $\hat{\mathcal{T}} = \mathcal{T}$, which proves Theorem 3.1. The proof of Lemma A.1 is then given as follows.

*[Proof of Lemma A.1]:* 1) In the noise-free case, it can be directly seen from (2) that

$$y_m \neq |b_m|^2 \text{ only if } \underline{\phi}_m \mathbf{s} \neq 0, \quad 1 \leq m \leq M. \tag{35}$$

Let $\mathcal{A}_m \subset \{1, \ldots, N\}$ be the support of $\underline{\phi}_m$. Then, with some manipulation, we can obtain

$$\sum_{m \in \mathcal{C}_n} 1\{y_m \neq |b_m|^2\} \overset{(a)}{\leq} \sum_{m \in \mathcal{C}_n} 1\{\underline{\phi}_m \mathbf{s} \neq 0\} \overset{(b)}{\leq} \sum_{m \in \mathcal{C}_n} 1\{\mathcal{T} \cap \mathcal{A}_m \neq \emptyset\} = \left|\mathcal{C}_n \cap \left(\bigcup_{j \in \mathcal{T}} \mathcal{C}_j\right)\right|$$

$$= \left|\bigcup_{j \in \mathcal{T}} (\mathcal{C}_n \cap \mathcal{C}_j)\right| \leq \sum_{j \in \mathcal{T}} |\mathcal{C}_n \cap \mathcal{C}_j| \overset{(c)}{\leq} Kr, \tag{36}$$

where (a) follows from (35), (b) holds since $\mathcal{T} \cap \mathcal{A}_m \neq \emptyset$ is necessary for $\underline{\phi}_m \mathbf{s} \neq 0$, and (c) holds by Assumption 1. Thus, the proof of part 1) is completed.

2) For $n \in \mathcal{T}$, it can be observed from (7) that the measurements $\{y_m\}_{m \in \tilde{\mathcal{C}}_n}$, where $\tilde{\mathcal{C}}_n$ is defined in (6), satisfy $\phi_{m,n} s_n = \underline{\phi}_m \mathbf{s} \neq 0$. We claim that at most two measurements among $\{y_m\}_{m \in \tilde{\mathcal{C}}_n}$ satisfies $y_m = |\phi_{m,n} s_n + b_m|^2 = |\underline{\phi}_m \mathbf{s} + b_m|^2 = |b_m|^2$, i.e.,

$$\sum_{m \in \tilde{\mathcal{C}}_n} 1\{y_m = |b_m|^2\} \leq 2. \tag{37}$$

Then, with (37), it follows that

$$\sum_{m \in \mathcal{C}_n} 1\{y_m \neq |b_m|^2\} \overset{(a)}{\geq} \sum_{m \in \tilde{\mathcal{C}}_n} 1\{y_m \neq |b_m|^2\} \geq |\tilde{\mathcal{C}}_n| - 2, \tag{38}$$

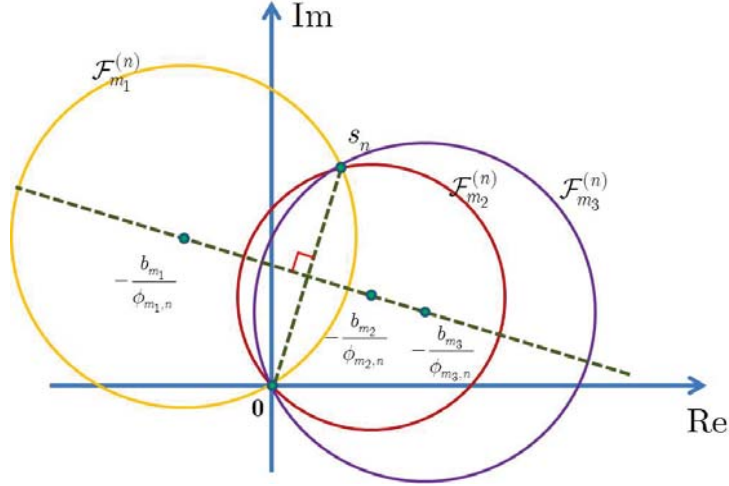

Fig. 18. Depiction of the intersection of three solution sets $\mathcal{F}_{m_1}^{(n)}$, $\mathcal{F}_{m_2}^{(n)}$ and $\mathcal{F}_{m_3}^{(n)}$ obtained from (40), where the associated centers $-\frac{b_{m_1}}{\phi_{m_1,n}}$, $-\frac{b_{m_2}}{\phi_{m_2,n}}$ and $-\frac{b_{m_3}}{\phi_{m_3,n}}$ are collinear.

where (a) holds due to $\tilde{\mathcal{C}}_n \subseteq \mathcal{C}_n$. Hence, the assertion of part 2) directly follows from (8) and (38).

The claim in (37) can be shown by contradiction. Suppose otherwise that there exist three distinct indices $m_1, m_2, m_3 \in \tilde{\mathcal{C}}_n$ such that

$$y_{m_j} = \left|\phi_{m_j,n} s_n + b_{m_j}\right|^2 = \left|b_{m_j}\right|^2, \tag{39}$$

for $j = 1, 2, 3$. Then, with (39), direct manipulation shows that

$$\left|s_n + \frac{b_{m_j}}{\phi_{m_j,n}}\right| = \frac{\left|b_{m_j}\right|}{\left|\phi_{m_j,n}\right|}, \quad 1 \leq j \leq 3. \tag{40}$$

Clearly, the solution set $\mathcal{F}_{m_j}^{(n)}$ of the $j$th equation in (40) can be depicted as a circle centered at $-\frac{b_{m_j}}{\phi_{m_j,n}}$ with radius $\left|\frac{b_{m_j}}{\phi_{m_j,n}}\right|$ passing through the origin $0$. Since, according to (40), $s_n \neq 0$ also belongs to $\mathcal{F}_{m_j}^{(n)}$, thereby lying on the circle, it can be shown that the center $-\frac{b_{m_j}}{\phi_{m_j,n}}$ is on the perpendicular bisector of the line segment connecting points $s_n$ and $0$. This statement is true for all $1 \leq j \leq 3$ and, consequently, it can be readily deduced that the three centers $-\frac{b_{m_1}}{\phi_{m_1,n}}$, $-\frac{b_{m_2}}{\phi_{m_2,n}}$ and $-\frac{b_{m_3}}{\phi_{m_3,n}}$ lie on the same line (i.e., they are collinear, see Fig. 18 for an illustration), contradicting with Assumption 2. As a result, (37) holds. □

# APPENDIX B

## PROOF OF THEOREM 3.2

Recall that $\mathcal{F}_{m_1}^{(n)}$, $\mathcal{F}_{m_2}^{(n)}$, and $\mathcal{F}_{m_3}^{(n)}$ are solutions to (7), for $m = m_1, m_2$, and $m_3$. By reorganizing the equations and by subtracting that of $m_2$ from $m_1$ and that of $m_3$ from $m_1$, respectively, we have

$$\frac{y_{m_1} - |b_{m_1}|^2}{|\phi_{m_1,n}|^2} - \frac{y_{m_2} - |b_{m_2}|^2}{|\phi_{m_2,n}|^2} = s_n^* \left( \frac{b_{m_1}}{\phi_{m_1,n}} - \frac{b_{m_2}}{\phi_{m_2,n}} \right) + s_n \left( \frac{b_{m_1}}{\phi_{m_1,n}} - \frac{b_{m_2}}{\phi_{m_2,n}} \right)^*; \quad (41)$$

$$\frac{y_{m_1} - |b_{m_1}|^2}{|\phi_{m_1,n}|^2} - \frac{y_{m_3} - |b_{m_3}|^2}{|\phi_{m_3,n}|^2} = s_n^* \left( \frac{b_{m_1}}{\phi_{m_1,n}} - \frac{b_{m_3}}{\phi_{m_3,n}} \right) + s_n \left( \frac{b_{m_1}}{\phi_{m_1,n}} - \frac{b_{m_3}}{\phi_{m_3,n}} \right)^*. \quad (42)$$

Using (41) and (42) to eliminate $s_n^*$ yields

$$\left( \frac{b_{m_1}}{\phi_{m_1,n}} - \frac{b_{m_2}}{\phi_{m_2,n}} \right) \left( \frac{y_{m_1} - |b_{m_1}|^2}{|\phi_{m_1,n}|^2} - \frac{y_{m_3} - |b_{m_3}|^2}{|\phi_{m_3,n}|^2} \right) - \left( \frac{b_{m_1}}{\phi_{m_1,n}} - \frac{b_{m_3}}{\phi_{m_3,n}} \right) \left( \frac{y_{m_1} - |b_{m_1}|^2}{|\phi_{m_1,n}|^2} - \frac{y_{m_2} - |b_{m_2}|^2}{|\phi_{m_2,n}|^2} \right)$$

$$= i2\mathrm{Im}\left( \left( \frac{b_{m_1}}{\phi_{m_1,n}} - \frac{b_{m_2}}{\phi_{m_2,n}} \right) \left( \frac{b_{m_1}}{\phi_{m_1,n}} - \frac{b_{m_3}}{\phi_{m_3,n}} \right)^* \right) s_n. \quad (43)$$

It follows from Assumption 2 that

$$2\mathrm{Im}\left( \left( \frac{b_{m_1}}{\phi_{m_1,n}} - \frac{b_{m_2}}{\phi_{m_2,n}} \right) \left( \frac{b_{m_1}}{\phi_{m_1,n}} - \frac{b_{m_3}}{\phi_{m_3,n}} \right)^* \right) \neq 0. \quad (44)$$

Then, the assertion in (10) immediately follows from (43) and (44).

Notice that, (44), can be shown by contradiction. In fact, suppose that

$$2\mathrm{Im}\left( \left( \frac{b_{m_1}}{\phi_{m_1,n}} - \frac{b_{m_2}}{\phi_{m_2,n}} \right) \left( \frac{b_{m_1}}{\phi_{m_1,n}} - \frac{b_{m_3}}{\phi_{m_3,n}} \right)^* \right)$$

$$= \left( \frac{b_{m_1}}{\phi_{m_1,n}} - \frac{b_{m_2}}{\phi_{m_2,n}} \right) \left( \frac{b_{m_1}}{\phi_{m_1,n}} - \frac{b_{m_3}}{\phi_{m_3,n}} \right)^* - \left( \frac{b_{m_1}}{\phi_{m_1,n}} - \frac{b_{m_2}}{\phi_{m_2,n}} \right)^* \left( \frac{b_{m_1}}{\phi_{m_1,n}} - \frac{b_{m_3}}{\phi_{m_3,n}} \right) = 0. \quad (45)$$

Let $\theta_{m_1,m_2}^{(n)}, \theta_{m_1,m_3}^{(n)} \in (0, 2\pi]$ be the polar angles of $\frac{b_{m_1}}{\phi_{m_1,n}} - \frac{b_{m_2}}{\phi_{m_2,n}}$ and $\frac{b_{m_1}}{\phi_{m_1,n}} - \frac{b_{m_3}}{\phi_{m_3,n}}$, respectively. Then, with some manipulation, we can obtain from (45) that

$$2\left( \theta_{m_1,m_2}^{(n)} - \theta_{m_1,m_3}^{(n)} \right) = 2\pi q, \quad (46)$$

where $q$ is some integer. Hence, we have

$$\frac{b_{m_1}}{\phi_{m_1,n}} - \frac{b_{m_2}}{\phi_{m_2,n}} = \left| \frac{b_{m_1}}{\phi_{m_1,n}} - \frac{b_{m_2}}{\phi_{m_2,n}} \right| e^{i\theta_{m_1,m_2}^{(n)}} \stackrel{(a)}{=} (-1)^q \left| \frac{b_{m_1}}{\phi_{m_1,n}} - \frac{b_{m_2}}{\phi_{m_2,n}} \right| e^{i\theta_{m_1,m_3}^{(n)}}$$

$$= (-1)^q \frac{\left| \frac{b_{m_1}}{\phi_{m_1,n}} - \frac{b_{m_2}}{\phi_{m_2,n}} \right|}{\left| \frac{b_{m_1}}{\phi_{m_1,n}} - \frac{b_{m_3}}{\phi_{m_3,n}} \right|} \left( \frac{b_{m_1}}{\phi_{m_1,n}} - \frac{b_{m_3}}{\phi_{m_3,n}} \right), \quad (47)$$

where (a) follows from (46). According to (47), it can be deduced that the three points $\frac{b_{m_1}}{\phi_{m_1,n}}$, $\frac{b_{m_2}}{\phi_{m_2,n}}$ and $\frac{b_{m_3}}{\phi_{m_3,n}}$ are collinear on the complex plane, which contradicts with Assumption 2. The proof is thus completed. $\square$

# APPENDIX C
## PROOF OF THEOREM 4.1

The idea is basically the same as the proof of Theorem 3.1. Notably, since $\frac{d+r-2}{2} > Kr + K_v$, the set $[Kr + K_v, d - Kr - K_v + r - 2)$ must be non-empty. The following lemma (whose proof is placed at the end of this appendix) generalizes the result in Lemma A.1 to deal with $K_v$-sparse noise and is essential to the proof of Theorem 4.1.

*Lemma C.1:* Suppose that $\mathbf{v}$ is $K_v$-sparse. Under Assumptions 1 and 2, we have

1) $\sum_{m \in \mathcal{C}_n} 1\{y_m \neq |b_m|^2\} \leq Kr + K_v$, if $n \notin \mathcal{T}$.
2) $\sum_{m \in \mathcal{C}_n} 1\{y_m \neq |b_m|^2\} \geq d - Kr - K_v + r - 2$, if $n \in \mathcal{T}$.

Let $n \in \hat{\mathcal{T}}$. Then by definition of $\hat{\mathcal{T}}$ in (5) and the fact that $\eta \in [Kr + K_v, d - Kr - K_v + r - 2)$, we obtain $\sum_{m \in \mathcal{C}_n} 1\{y_m \neq |b_m|^2\} > \eta \geq Kr + K_v$, which together with part 1) of Lemma C.1 asserts that $n \in \mathcal{T}$. Therefore, $\hat{\mathcal{T}} \subseteq \mathcal{T}$. Next, we will prove $\mathcal{T} \subseteq \hat{\mathcal{T}}$. Let $n \in \mathcal{T}$ so that $\sum_{m \in \mathcal{C}_n} 1\{y_m \neq |b_m|^2\} \geq d - Kr - K_v + r - 2$, due to part 2) of Lemma C.1. Since $\eta < d - Kr - K_v + r - 2$ (thus, $\sum_{m \in \mathcal{C}_n} 1\{y_m \neq |b_m|^2\} > \eta$), it immediately follows that $n \in \hat{\mathcal{T}}$. This establishes $\mathcal{T} \subseteq \hat{\mathcal{T}}$ and, thus, $\hat{\mathcal{T}} = \mathcal{T}$. □

*[Proof of Lemma C.1]:* 1) The proof is similar to that of Lemma A.1. By considering the subset $\mathcal{C}_n \setminus \mathcal{V}$ and by following derivations similar to that in (36), it can be shown that

$$\sum_{m \in \mathcal{C}_n \setminus \mathcal{V}} 1\{y_m \neq |b_m|^2\} \leq Kr. \tag{48}$$

Using (48) together with some manipulation, we obtain

$$\sum_{m \in \mathcal{C}_n} 1\{y_m \neq |b_m|^2\} = \sum_{m \in \mathcal{C}_n \setminus \mathcal{V}} 1\{y_m \neq |b_m|^2\} + \sum_{m \in \mathcal{C}_n \cap \mathcal{V}} 1\{y_m \neq |b_m|^2\}$$

$$\leq Kr + \sum_{m \in \mathcal{C}_n \cap \mathcal{V}} 1\{y_m \neq |b_m|^2\} \leq Kr + |\mathcal{C}_n \cap \mathcal{V}|$$

$$\leq Kr + |\mathcal{V}| = Kr + K_v. \tag{49}$$

The proof of part 1) is thus completed.

2) Again using the similar arguments for deriving (38) (together with some manipulation), it can be shown that

$$\sum_{m \in \tilde{\mathcal{C}}_n \setminus \mathcal{V}} 1\{y_m \neq |b_m|^2\} \geq |\tilde{\mathcal{C}}_n \setminus \mathcal{V}| - 2 \geq |\tilde{\mathcal{C}}_n| - K_v - 2 \overset{(a)}{\geq} d - (K-1)r - K_v - 2, \tag{50}$$

where (a) holds by (8). Therefore, the assertion of part 2) directly follows from (50) since $\tilde{\mathcal{C}}_n \setminus \mathcal{V} \subseteq \mathcal{C}_n$. □

## APPENDIX D
## PROOF OF THEOREM 4.2

Based on the assumption $\frac{d+r-2}{2} > Kr + K_v$, as it has been shown in Section IV-A that $s_n \in \tilde{\mathcal{F}}^{(n)}$ holds for every $n \in \mathcal{T}$. Moreover, it follows from (14) that, since $|\tilde{\mathcal{C}}_n \setminus \mathcal{V}| > Kr + K_v + 2$, there exist at least $Kr + K_v + 3$ indices, say, $m_1, \ldots, m_{Kr+K_v+3} \in \tilde{\mathcal{C}}_n$ such that

$$y_{m_j} = \left|\phi_{m_j,n} s_n + b_{m_j}\right|^2, \tag{51}$$

for $j = 1, \ldots, Kr + K_v + 3$. Accordingly, the signal component $s_n$ belongs to $\mathcal{F}_{m_j}^{(n)}$ for all $1 \leq j \leq Kr + K_v + 3$, leading to

$$\sum_{j=1}^{|\tilde{\mathcal{C}}_n|} 1\{s_n \in \mathcal{F}_{m_j}^{(n)}\} > Kr + K_v + 2. \tag{52}$$

Let $s'_n \in \tilde{\mathcal{F}}^{(n)}$ and $s'_n \neq s_n$. We claim that

$$\sum_{j=1}^{|\tilde{\mathcal{C}}_n|} 1\{s'_n \in \mathcal{F}_{m_j}^{(n)}\} \leq K_v + 2. \tag{53}$$

It then directly follows from (52) and (53) that

$$\sum_{j=1}^{|\tilde{\mathcal{C}}_n|} 1\{s_n \in \mathcal{F}_{m_j}^{(n)}\} > \sum_{j=1}^{|\tilde{\mathcal{C}}_n|} 1\{s'_n \in \mathcal{F}_{m_j}^{(n)}\}, \tag{54}$$

for all $s'_n \in \tilde{\mathcal{F}}^{(n)}$ not equal to $s_n$. Hence, we have $\hat{s}_n = s_n$. $\square$

*[Proof of (53)]:* The proof is done by contradiction. Suppose otherwise $\sum_{j=1}^{|\tilde{\mathcal{C}}_n|} 1\{s'_n \in \mathcal{F}_{m_j}^{(n)}\} > K_v + 2$. Since $\tilde{\mathcal{C}}_n$ can be decomposed into $\tilde{\mathcal{C}}_n = (\tilde{\mathcal{C}}_n \setminus \mathcal{V}) \bigcup (\tilde{\mathcal{C}}_n \cap \mathcal{V})$, assume without loss of generality that $m_j \in \tilde{\mathcal{C}}_n \setminus \mathcal{V}$ if $1 \leq j \leq |\tilde{\mathcal{C}}_n \setminus \mathcal{V}|$ and $m_j \in \tilde{\mathcal{C}}_n \cap \mathcal{V}$ when $|\tilde{\mathcal{C}}_n \setminus \mathcal{V}| + 1 \leq j \leq |\tilde{\mathcal{C}}_n|$. Then we can obtain

$$\sum_{j=1}^{|\tilde{\mathcal{C}}_n \setminus \mathcal{V}|} 1\{s'_n \in \mathcal{F}_{m_j}^{(n)}\} \geq K_v + 3 - \sum_{j=|\tilde{\mathcal{C}}_n \setminus \mathcal{V}|+1}^{|\tilde{\mathcal{C}}_n|} 1\{s'_n \in \mathcal{F}_{m_j}^{(n)}\}. \tag{55}$$

Since $|\tilde{\mathcal{C}}_n \cap \mathcal{V}| \leq |\mathcal{V}| = K_v$, it follows from (55) that

$$\sum_{j=1}^{|\tilde{\mathcal{C}}_n \setminus \mathcal{V}|} 1\{s'_n \in \mathcal{F}_{m_j}^{(n)}\} \geq K_v + 3 - |\tilde{\mathcal{C}}_n \cap \mathcal{V}| \geq 3, \tag{56}$$

meaning that at least three elements $m_j$, $m_k$ and $m_l$, where $1 \leq j \neq k \neq l \leq |\tilde{\mathcal{C}}_n \setminus \mathcal{V}|$, belong to the set $\tilde{\mathcal{C}}_n \setminus \mathcal{V}$ so that $s'_n \in \mathcal{F}_{m_j}^{(n)} \cap \mathcal{F}_{m_k}^{(n)} \cap \mathcal{F}_{m_l}^{(n)}$. Notably, $s_n$ also belongs to $\mathcal{F}_{m_j}^{(n)}$, $\mathcal{F}_{m_k}^{(n)}$ and $\mathcal{F}_{m_l}^{(n)}$, and recall from (9) that $\mathcal{F}_{m_j}^{(n)}$, $\mathcal{F}_{m_k}^{(n)}$ and $\mathcal{F}_{m_l}^{(n)}$ can be depicted as circles. As a result, using similar

arguments for deriving (37), it can be shown that $-\frac{b_{m_j}}{\phi_{m_j,n}}$, $-\frac{b_{m_k}}{\phi_{m_k,n}}$ and $-\frac{b_{m_l}}{\phi_{m_l,n}}$ (the centers of the circles $\mathcal{F}_{m_j}^{(n)}$, $\mathcal{F}_{m_k}^{(n)}$ and $\mathcal{F}_{m_l}^{(n)}$, respectively) lie on the same line (the perpendicular bisector of the line segment connecting points $s_n$ and $s'_n$, see Fig. 18 for a similar illustration), contradicting with Assumption 2. The assertion of the claim is thus proved. $\square$

## APPENDIX E
## PROOF OF THEOREM 4.3

The following lemma (whose proof is placed at the end of this appendix) is needed to prove Theorem 4.3.

*Lemma E.1:* Consider the noisy model (11) with $\|\mathbf{v}\|_\infty < \epsilon$. Under Assumptions 1 and 2, we have

1) $\sum_{m \in \mathcal{C}_n} 1\left\{|y_m - |b_m|^2| > \epsilon\right\} \leq Kr$, if $n \notin \mathcal{T}$.
2) $\sum_{m \in \mathcal{C}_n} 1\left\{|y_m - |b_m|^2| > \epsilon\right\} \geq d - Kr + r$, if $n \in \mathcal{T}$.

With the aid of Lemma E.1, we start to prove Theorem 4.3. Let $n \in \hat{\hat{\mathcal{T}}}$. By definition of $\hat{\hat{\mathcal{T}}}$ in (20) and the fact that $\eta \in [Kr, d-Kr+r)$, we can obtain $\sum_{m \in \mathcal{C}_n} 1\left\{|y_m - |b_m|^2| > \epsilon\right\} > \eta \geq Kr$, which together with part 1) of Lemma E.1 asserts that $n \in \mathcal{T}$. Therefore, $\hat{\hat{\mathcal{T}}} \subseteq \mathcal{T}$. Now, it remains to prove $\mathcal{T} \subseteq \hat{\hat{\mathcal{T}}}$. Let $n \in \mathcal{T}$ so that $\sum_{m \in \mathcal{C}_n} 1\left\{|y_m - |b_m|^2| > \epsilon\right\} \geq d - Kr + r$, due to part 2) of Lemma E.1. Since $\eta \in [Kr, d-Kr+r)$, it immediately follows $\sum_{m \in \mathcal{C}_n} 1\left\{|y_m - |b_m|^2| > \epsilon\right\} > \eta$, which implies that $n \in \hat{\hat{\mathcal{T}}}$. This establishes $\mathcal{T} \subseteq \hat{\hat{\mathcal{T}}}$, and, thereby, $\hat{\hat{\mathcal{T}}} = \mathcal{T}$. We end this appendix by proving Lemma E.1.

*[Proof of Lemma E.1]:* 1) We first note from (18) that $\underline{\phi}_m \mathbf{s} \neq 0$ is a necessary condition for guaranteeing $|y_m - |b_m|^2| > \epsilon$ for all $m$. As a result, we have

$$\sum_{m \in \mathcal{C}_n} 1\left\{|y_m - |b_m|^2| > \epsilon\right\} \leq \sum_{m \in \mathcal{C}_n} 1\{\underline{\phi}_m \mathbf{s} \neq 0\}. \tag{57}$$

Moreover, since $\sum_{m \in \mathcal{C}_n} 1\{\underline{\phi}_m \mathbf{s} \neq 0\} \leq Kr$ for all $n \notin \mathcal{T}$, as shown in (36), the assertion $\sum_{m \in \mathcal{C}_n} 1\left\{|y_m - |b_m|^2| > \epsilon\right\} \leq Kr$ immediately follows.

2) For $n \in \mathcal{T}$, (8) asserts that there exist at least $d - Kr + r$ indices, say, $m_1, \ldots, m_{d-Kr+r} \in \tilde{\mathcal{C}}_n$ such that

$$y_{m_j} = \left|\phi_{m_j,n} s_n + b_{m_j}\right|^2 + v_{m_j}, \quad 1 \leq j \leq d - Kr + r. \tag{58}$$

Since

$$\left|\phi_{m_j,n} s_n\right| \geq \phi_{\min} \delta_{\min} \geq b_{\max} + \sqrt{b_{\max}^2 + 2\epsilon}, \quad 1 \leq j \leq d - Kr + r, \tag{59}$$

we have

$$y_{m_j} \geq \left(|\phi_{m_j,n} s_n| - |b_{m_j}|\right)^2 + v_{m_j}$$
$$\geq b_{\max}^2 + 2\epsilon + v_{m_j}$$
$$> |b_{m_j}|^2 + \epsilon, \quad 1 \leq j \leq d - Kr + r, \tag{60}$$

which then implies $\sum_{m \in \mathcal{C}_n} \mathbb{1}\{|y_m - |b_m|^2| > \epsilon\} \geq d - Kr + r$. The proof is thus completed. $\square$

## APPENDIX F
## PROOF OF THEOREM 4.4

Since $\mathbf{G}_n[\tilde{\mathbf{b}}_n \ \tilde{\mathbf{b}}_n^*]$ is full column rank based on Assumption 2 (the proof is placed at the end of this appendix), the unknown signal entry $s_n$ can be estimated as the LS solutions to (24), i.e.,

$$\hat{s}_n = \frac{\tilde{\mathbf{b}}_n^T \mathbf{G}_n^T \mathbf{G}_n \tilde{\mathbf{b}}_n \tilde{\mathbf{b}}_n^H \mathbf{G}_n^T \mathbf{G}_n \tilde{\mathbf{y}}_n - \|\mathbf{G}_n \tilde{\mathbf{b}}_n\|_2^2 \tilde{\mathbf{b}}_n^T \mathbf{G}_n^T \mathbf{G}_n \tilde{\mathbf{y}}_n}{\left|\tilde{\mathbf{b}}_n^T \mathbf{G}_n^T \mathbf{G}_n \tilde{\mathbf{b}}_n\right|^2 - \|\mathbf{G}_n \tilde{\mathbf{b}}_n\|_2^4}, \tag{61}$$

for all $n \in \mathcal{T}$. Hence, with some manipulation, the assertion in (26) immediately follows from (61) and the fact that $\mathbf{G}_n^T \mathbf{G}_n = |\tilde{\mathcal{C}}_n|(\mathbf{I} - \mathbf{1}\mathbf{1}^T/|\tilde{\mathcal{C}}_n|)$, where $(\mathbf{I} - \mathbf{1}\mathbf{1}^T/|\tilde{\mathcal{C}}_n|)$ is an orthogonal projection matrix satisfying $(\mathbf{I} - \mathbf{1}\mathbf{1}^T/|\tilde{\mathcal{C}}_n|)^2 = (\mathbf{I} - \mathbf{1}\mathbf{1}^T/|\tilde{\mathcal{C}}_n|)$. We end this appendix by proving $\mathbf{G}_n[\tilde{\mathbf{b}}_n \ \tilde{\mathbf{b}}_n^*]$ is a full column rank matrix.

It is equivalent to show that $(\mathbf{G}_n[\tilde{\mathbf{b}}_n \ \tilde{\mathbf{b}}_n^*])^H(\mathbf{G}_n[\tilde{\mathbf{b}}_n \ \tilde{\mathbf{b}}_n^*])$ is nonsingular, or equivalently, its determinant is not zero, i.e.,

$$\det\left((\mathbf{G}_n[\tilde{\mathbf{b}}_n \ \tilde{\mathbf{b}}_n^*])^H(\mathbf{G}_n[\tilde{\mathbf{b}}_n \ \tilde{\mathbf{b}}_n^*])\right) = \det\left([\tilde{\mathbf{b}}_n \ \tilde{\mathbf{b}}_n^*]^H |\tilde{\mathcal{C}}_n|(\mathbf{I} - \mathbf{1}\mathbf{1}^T/|\tilde{\mathcal{C}}_n|)[\tilde{\mathbf{b}}_n \ \tilde{\mathbf{b}}_n^*]\right)$$
$$\stackrel{(a)}{=} |\tilde{\mathcal{C}}_n|^2 \left|[\tilde{\mathbf{b}}_{0,n} \ \tilde{\mathbf{b}}_{0,n}^*]^H[\tilde{\mathbf{b}}_{0,n} \ \tilde{\mathbf{b}}_{0,n}^*]\right|$$
$$= \|\tilde{\mathbf{b}}_{0,n}\|_2^2 \|\tilde{\mathbf{b}}_{0,n}^*\|_2^2 - \left|\tilde{\mathbf{b}}_{0,n}^H \tilde{\mathbf{b}}_{0,n}^*\right|^2 \neq 0, \tag{62}$$

where (a) holds since $\mathbf{I} - \mathbf{1}\mathbf{1}^T/|\tilde{\mathcal{C}}_n|$ is an orthogonal projection matrix. Towards this end, we first note that Cauchy-Schwarz inequality implies $\left|\tilde{\mathbf{b}}_{0,n}^H \tilde{\mathbf{b}}_{0,n}^*\right| \leq \|\tilde{\mathbf{b}}_{0,n}\|_2 \|\tilde{\mathbf{b}}_{0,n}^*\|_2$. Moreover, we claim that, under Assumption 2, $\left|\tilde{\mathbf{b}}_{0,n}^H \tilde{\mathbf{b}}_{0,n}^*\right| \neq \|\tilde{\mathbf{b}}_{0,n}\|_2 \|\tilde{\mathbf{b}}_{0,n}^*\|_2$ holds; hence, we have $\left|\tilde{\mathbf{b}}_{0,n}^H \tilde{\mathbf{b}}_{0,n}^*\right| < \|\tilde{\mathbf{b}}_{0,n}\|_2 \|\tilde{\mathbf{b}}_{0,n}^*\|_2$, which immediately guarantees (62). Now, we prove the claim by contradiction. Assume otherwise $\left|\tilde{\mathbf{b}}_{0,n}^H \tilde{\mathbf{b}}_{0,n}^*\right| = \|\tilde{\mathbf{b}}_{0,n}\|_2 \|\tilde{\mathbf{b}}_{0,n}^*\|_2$. Then, based on Cauchy-Schwarz inequality, it can be deduced that there exists $\underline{c} \in \mathbb{C}$ such that

$$\tilde{\mathbf{b}}_{0,n} = \underline{c} \tilde{\mathbf{b}}_{0,n}^*. \tag{63}$$

Let $\tilde{b}_j^{(0,n)} = |\tilde{b}_j^{(0,n)}|e^{i\theta_j^{(0,n)}}$ be the $j$th entry of $\tilde{\mathbf{b}}_{0,n}$, where $1 \leq j \leq |\tilde{\mathcal{C}}_n|$ and $\theta_j^{(0,n)} \in (0, 2\pi]$. Then we can obtain from (63) that $\underline{c} = e^{i2\theta_j^{(0,n)}}$ for all $1 \leq j \leq |\tilde{\mathcal{C}}_n|$, which in turn implies

$$e^{i2\theta_j^{(0,n)}} = e^{i2\theta_k^{(0,n)}} \text{ for all } 1 \leq j \neq k \leq |\tilde{\mathcal{C}}_n|. \tag{64}$$

With the aid of (64), it can be directly deduced that the centered reference points $\{\tilde{b}_j^{(0,n)}\}_{j=1}^{|\tilde{\mathcal{C}}_n|}$ are alinged with each other (i.e., they are collinear on the complex plane), which together with the translational invariance property implies that the reference points $\{-b_{m_j}/\phi_{m_j,n}\}_{j=1}^{|\tilde{\mathcal{C}}_n|}$ are also aligned each other, contradicting with Assumption 2. $\square$

## APPENDIX G
## PROOF OF THEOREM 4.5

Let $\tilde{\mathbf{v}}_{0,n} = (\mathbf{I} - \mathbf{1}\mathbf{1}^T/|\tilde{\mathcal{C}}_n|)\tilde{\mathbf{v}}_n$ be the "centered" vector of $\tilde{\mathbf{v}}_n$. Then we have

$$|\hat{s}_n - s_n| = \left| \frac{\tilde{\mathbf{b}}_{0,n}^T \tilde{\mathbf{b}}_{0,n} \tilde{\mathbf{b}}_{0,n}^H \tilde{\mathbf{y}}_{0,n} - \|\tilde{\mathbf{b}}_{0,n}\|_2^2 \tilde{\mathbf{b}}_{0,n}^T \tilde{\mathbf{y}}_{0,n}}{|\tilde{\mathbf{b}}_{0,n}^H \tilde{\mathbf{b}}_{0,n}^*|^2 - \|\tilde{\mathbf{b}}_{0,n}\|_2^4} - s_n \right|$$

$$\stackrel{(a)}{=} \left| \frac{\tilde{\mathbf{b}}_{0,n}^T \tilde{\mathbf{b}}_{0,n} \tilde{\mathbf{b}}_{0,n}^H \tilde{\mathbf{v}}_{0,n} - \|\tilde{\mathbf{b}}_{0,n}\|_2^2 \tilde{\mathbf{b}}_{0,n}^T \tilde{\mathbf{v}}_{0,n}}{|\tilde{\mathbf{b}}_{0,n}^H \tilde{\mathbf{b}}_{0,n}^*|^2 - \|\tilde{\mathbf{b}}_{0,n}\|_2^4} \right|$$

$$\leq \frac{|\langle \tilde{\mathbf{b}}_{0,n}, \tilde{\mathbf{v}}_{0,n}\rangle| \left( |\langle \tilde{\mathbf{b}}_{0,n}, \tilde{\mathbf{b}}_{0,n}^*\rangle| + \|\tilde{\mathbf{b}}_{0,n}\|_2^2 \right)}{\left( |\langle \tilde{\mathbf{b}}_{0,n}, \tilde{\mathbf{b}}_{0,n}^*\rangle| + \|\tilde{\mathbf{b}}_{0,n}\|_2^2 \right)\left( \|\tilde{\mathbf{b}}_{0,n}\|_2^2 - |\langle \tilde{\mathbf{b}}_{0,n}, \tilde{\mathbf{b}}_{0,n}^*\rangle| \right)}$$

$$= \frac{\left|\langle \tilde{\mathbf{b}}_{0,n}, \tilde{\mathbf{v}}_{0,n}\rangle\right|}{\|\tilde{\mathbf{b}}_{0,n}\|_2^2 \left(1 - \left|\left\langle \frac{\tilde{\mathbf{b}}_{0,n}}{\|\tilde{\mathbf{b}}_{0,n}\|_2}, \frac{\tilde{\mathbf{b}}_{0,n}^*}{\|\tilde{\mathbf{b}}_{0,n}^*\|_2}\right\rangle\right|\right)}$$

$$\stackrel{(b)}{\leq} \frac{\|\tilde{\mathbf{v}}_{0,n}\|_2}{\|\tilde{\mathbf{b}}_{0,n}\|_2 \left(1 - \left|\left\langle \frac{\tilde{\mathbf{b}}_{0,n}}{\|\tilde{\mathbf{b}}_{0,n}\|_2}, \frac{\tilde{\mathbf{b}}_{0,n}^*}{\|\tilde{\mathbf{b}}_{0,n}^*\|_2}\right\rangle\right|\right)}$$

$$\stackrel{(c)}{\leq} \frac{\|\tilde{\mathbf{v}}_n\|_2}{\|\tilde{\mathbf{b}}_{0,n}\|_2 \left(1 - \left|\left\langle \frac{\tilde{\mathbf{b}}_{0,n}}{\|\tilde{\mathbf{b}}_{0,n}\|_2}, \frac{\tilde{\mathbf{b}}_{0,n}^*}{\|\tilde{\mathbf{b}}_{0,n}^*\|_2}\right\rangle\right|\right)}$$

$$\stackrel{(d)}{<} \frac{\sqrt{|\tilde{\mathcal{C}}_n|}}{\|\tilde{\mathbf{b}}_{0,n}\|_2 \left(1 - \left|\left\langle \frac{\tilde{\mathbf{b}}_{0,n}}{\|\tilde{\mathbf{b}}_{0,n}\|_2}, \frac{\tilde{\mathbf{b}}_{0,n}^*}{\|\tilde{\mathbf{b}}_{0,n}^*\|_2}\right\rangle\right|\right)\phi_{\min}^2}\epsilon,$$

where (a) holds since $s_n = \frac{\tilde{\mathbf{b}}_{0,n}^T \tilde{\mathbf{b}}_{0,n} \tilde{\mathbf{b}}_{0,n}^H (\tilde{\mathbf{y}}_{0,n} - \tilde{\mathbf{v}}_{0,n}) - \|\tilde{\mathbf{b}}_{0,n}\|_2^2 \tilde{\mathbf{b}}_{0,n}^T (\tilde{\mathbf{y}}_{0,n} - \tilde{\mathbf{v}}_{0,n})}{|\tilde{\mathbf{b}}_{0,n}^H \tilde{\mathbf{b}}_{0,n}^*|^2 - \|\tilde{\mathbf{b}}_{0,n}\|_2^4}$, (b) holds by Cauchy-Schwarz inequality, (c) holds since $\|\tilde{\mathbf{v}}_{0,n}\|_2 = \|(\mathbf{I} - \mathbf{1}\mathbf{1}^T/|\tilde{\mathcal{C}}_n|)\tilde{\mathbf{v}}_n\|_2 \leq \|\tilde{\mathbf{v}}_n\|_2$ (by property of

orthogonal projection), and (d) follows from the definition of $\tilde{\mathbf{v}}_n$. Note that $\hat{s}_n = s_n = 0$ holds for all $n \notin \mathcal{T}$, since $\hat{\tilde{\mathcal{T}}} = \mathcal{T}$. Thereby, we have

$$\|\hat{\mathbf{s}} - \mathbf{s}\|_\infty = \max_{n \in \mathcal{T}} |\hat{s}_n - s_n| < \frac{\sqrt{|\tilde{\mathcal{C}}_{n^*}|}}{\|\tilde{\mathbf{b}}_{0,n^*}\|_2 \left(1 - \left|\left\langle \frac{\tilde{\mathbf{b}}_{0,n^*}}{\|\tilde{\mathbf{b}}_{0,n^*}\|_2}, \frac{\tilde{\mathbf{b}}_{0,n^*}^*}{\|\tilde{\mathbf{b}}_{0,n^*}^*\|_2} \right\rangle \right|\right) \phi_{\min}^2} \epsilon, \tag{65}$$

where $n^*$ is defined in (28). Hence, the assertion (27) immediately follows from (65) and the inequality $\|\hat{\mathbf{s}} - \mathbf{s}\|_2 \leq \sqrt{K} \|\hat{\mathbf{s}} - \mathbf{s}\|_\infty$. □

## APPENDIX H
### PROOF OF THEOREM 4.6

The following two lemmas are needed to prove Theorem 4.6; to ease reading, their proofs are relegated to the end of this appendix.

*Lemma H.1:* Let $\tilde{\mathbf{b}}_n = \tilde{\mathbf{b}}_n^{(\text{Re})} + i\tilde{\mathbf{b}}_n^{(\text{Im})}$, $\tilde{\mathbf{b}}_{0,n} = \tilde{\mathbf{b}}_{0,n}^{(\text{Re})} + i\tilde{\mathbf{b}}_{0,n}^{(\text{Im})}$ and $\bar{b}_n = \bar{b}_n^{(\text{Re})} + i\bar{b}_n^{(\text{Im})}$ be defined as in (24) and Theorem 4.4, respectively. Then we have

1) $\|\tilde{\mathbf{b}}_{0,n}\|_2^2 = \|\tilde{\mathbf{b}}_n\|_2^2 - |\tilde{\mathcal{C}}_n| |\bar{b}_n|^2$;
2) $\left|\langle \tilde{\mathbf{b}}_{0,n}, \tilde{\mathbf{b}}_{0,n}^* \rangle\right|^2 = 4\left(\langle \tilde{\mathbf{b}}_n^{(\text{Re})}, \tilde{\mathbf{b}}_n^{(\text{Im})} \rangle - |\tilde{\mathcal{C}}_n| \bar{b}_n^{(\text{Re})} \bar{b}_n^{(\text{Im})}\right)^2$
$$+ \left(\|\tilde{\mathbf{b}}_n^{(\text{Re})}\|_2^2 - \|\tilde{\mathbf{b}}_n^{(\text{Im})}\|_2^2 - |\tilde{\mathcal{C}}_n| \bar{b}_n^{(\text{Re})2} + |\tilde{\mathcal{C}}_n| \bar{b}_n^{(\text{Im})2}\right)^2.$$
□

*Lemma H.2:* With $n^*$ defined in (28), the total residual of the LS line fit of the point set $\{-b_{m_l}/\phi_{m_l,n^*}\}_{l=1}^{|\tilde{\mathcal{C}}_{n^*}|}$ can be expressed as

$$r^* = \frac{\|\tilde{\mathbf{b}}_{n^*}\|_2^2 - |\tilde{\mathcal{C}}_{n^*}| |\bar{b}_{n^*}|^2}{2} - $$
$$\frac{1}{2}\sqrt{\left(\|\tilde{\mathbf{b}}_{n^*}^{(\text{Re})}\|_2^2 - \|\tilde{\mathbf{b}}_{n^*}^{(\text{Im})}\|_2^2 - |\tilde{\mathcal{C}}_{n^*}| \bar{b}_{n^*}^{(\text{Re})2} + |\tilde{\mathcal{C}}_{n^*}| \bar{b}_{n^*}^{(\text{Im})2}\right)^2 + 4\left(\langle \tilde{\mathbf{b}}_{n^*}^{(\text{Re})}, \tilde{\mathbf{b}}_{n^*}^{(\text{Im})} \rangle - |\tilde{\mathcal{C}}_{n^*}| \bar{b}_{n^*}^{(\text{Re})} \bar{b}_{n^*}^{(\text{Im})}\right)^2}, \tag{66}$$

where $\tilde{\mathbf{b}}_{n^*} = \tilde{\mathbf{b}}_{n^*}^{(\text{Re})} + i\tilde{\mathbf{b}}_{n^*}^{(\text{Im})}$ and $\bar{b}_{n^*} = \bar{b}_{n^*}^{(\text{Re})} + i\bar{b}_{n^*}^{(\text{Im})}$ are defined in (24) and Theorem 4.4, respectively. □

Based on Lemmas H.1 and H.2, we start to prove Theorem 4.6. With the aid of Lemma H.1, we have

$$\|\tilde{\mathbf{b}}_{0,n^*}\|_2^2 - \left|\langle \tilde{\mathbf{b}}_{0,n^*}, \tilde{\mathbf{b}}_{0,n^*}^* \rangle\right| = \|\tilde{\mathbf{b}}_{n^*}\|_2^2 - |\tilde{\mathcal{C}}_{n^*}| |\bar{b}_{n^*}|^2$$
$$- \sqrt{4\left(\langle \tilde{\mathbf{b}}_{n^*}^{(\text{Re})}, \tilde{\mathbf{b}}_{n^*}^{(\text{Im})} \rangle - |\tilde{\mathcal{C}}_{n^*}| \bar{b}_{n^*}^{(\text{Re})} \bar{b}_{n^*}^{(\text{Im})}\right)^2 + \left(\|\tilde{\mathbf{b}}_{n^*}^{(\text{Re})}\|_2^2 - \|\tilde{\mathbf{b}}_{n^*}^{(\text{Im})}\|_2^2 - |\tilde{\mathcal{C}}_{n^*}| \bar{b}_{n^*}^{(\text{Re})2} + |\tilde{\mathcal{C}}_{n^*}| \bar{b}_{n^*}^{(\text{Im})2}\right)^2}, \tag{67}$$

which along with (66) yields

$$\left\|\tilde{\mathbf{b}}_{0,n^*}\right\|_2^2 - \left|\langle\tilde{\mathbf{b}}_{0,n^*}, \tilde{\mathbf{b}}_{0,n^*}^*\rangle\right| = 2r^*. \tag{68}$$

The assertion (29) follows immediately from (68), and the proof of Theorem 4.6 is thus completed. $\square$

*[Proof of Lemma H.1]:* By definition of $\tilde{\mathbf{b}}_{0,n}$, we have

$$\left\|\tilde{\mathbf{b}}_{0,n}\right\|_2^2 = \left\|\tilde{\mathbf{b}}_n - \bar{b}_n\mathbf{1}\right\|_2^2 = \left\|\tilde{\mathbf{b}}_n\right\|_2^2 + |\tilde{\mathcal{C}}_n||\bar{b}_n|^2 - 2\bar{b}_n\mathbf{1}^T\tilde{\mathbf{b}}_n^* = \left\|\tilde{\mathbf{b}}_n\right\|_2^2 - |\tilde{\mathcal{C}}_n||\bar{b}_n|^2, \tag{69}$$

which proves the assertion of part 1).

Next, we go on to prove part 2). By definition the inner product $\langle\tilde{\mathbf{b}}_{0,n}, \tilde{\mathbf{b}}_{0,n}^*\rangle$ reads

$$\langle\tilde{\mathbf{b}}_{0,n}, \tilde{\mathbf{b}}_{0,n}^*\rangle = \langle\tilde{\mathbf{b}}_n - \bar{b}_n\mathbf{1}, \tilde{\mathbf{b}}_n^* - \bar{b}_n^*\mathbf{1}\rangle = \langle\tilde{\mathbf{b}}_n, \tilde{\mathbf{b}}_n^*\rangle - |\tilde{\mathcal{C}}_n|\left(\bar{b}_n^*\right)^2, \tag{70}$$

which can be further rearranged into

$$\langle\tilde{\mathbf{b}}_{0,n}, \tilde{\mathbf{b}}_{0,n}^*\rangle = \left\|\tilde{\mathbf{b}}_n^{(\text{Re})}\right\|_2^2 - \left\|\tilde{\mathbf{b}}_n^{(\text{Im})}\right\|_2^2 - i2\langle\tilde{\mathbf{b}}_n^{(\text{Re})}, \tilde{\mathbf{b}}_n^{(\text{Im})}\rangle - |\tilde{\mathcal{C}}_n|\left(\bar{b}_n^{(\text{Re})} - i\bar{b}_n^{(\text{Im})}\right)^2$$
$$= \left\|\tilde{\mathbf{b}}_n^{(\text{Re})}\right\|_2^2 - \left\|\tilde{\mathbf{b}}_n^{(\text{Im})}\right\|_2^2 - |\tilde{\mathcal{C}}_n|\bar{b}_n^{(\text{Re})^2} + |\tilde{\mathcal{C}}_n|\bar{b}_n^{(\text{Im})^2} - i2\left(\langle\tilde{\mathbf{b}}_n^{(\text{Re})}, \tilde{\mathbf{b}}_n^{(\text{Im})}\rangle - |\tilde{\mathcal{C}}_n|\bar{b}_n^{(\text{Re})}\bar{b}_n^{(\text{Im})}\right). \tag{71}$$

Hence, the assertion of part 2) then follows directly from (71). The proof of Lemma H.1 is thus completed. $\square$

*[Proof of Lemma H.2]:* Let $\boldsymbol{\alpha} = \mathbf{c} + t\mathbf{a}$ be the parametric equation of a line, where $t \in \mathbb{R}$ and $\boldsymbol{\alpha}, \mathbf{c}, \mathbf{a} \in \mathbb{R}^2$. Without loss of generality, we assume that $\mathbf{a}$ is unit-norm. As such, for any $1 \leq j \leq |\tilde{\mathcal{C}}_{n^*}|$, the squared distance $r_j^2 \geq 0$ between the line and the reference point $-\frac{b_{m_j}}{\phi_{m_j,n^*}}$ is expressed as

$$r_j^2 = \left\|\left(\mathbf{I} - \mathbf{a}\mathbf{a}^T\right)(\mathbf{p}_j - \mathbf{c})\right\|_2^2, \ 1 \leq j \leq |\tilde{\mathcal{C}}_{n^*}|, \tag{72}$$

where

$$\mathbf{p}_j = \left[\text{Re}\left(-\frac{b_{m_j}}{\phi_{m_j,n^*}}\right) \ \ \text{Im}\left(-\frac{b_{m_j}}{\phi_{m_j,n^*}}\right)\right]^T \in \mathbb{R}^2 \tag{73}$$

is the position vector of $-\frac{b_{m_j}}{\phi_{m_j,n^*}}$. Hence, with the aid of (72), the line of best fit can be obtained by solving the following optimization problem

$$(P1) \quad \min_{\mathbf{a},\mathbf{c}\in\mathbb{R}^2} \sum_{j=1}^{|\tilde{\mathcal{C}}_{n^*}|} r_j^2 = \sum_{j=1}^{|\tilde{\mathcal{C}}_{n^*}|} \left\|\left(\mathbf{I} - \mathbf{a}\mathbf{a}^T\right)(\mathbf{p}_j - \mathbf{c})\right\|_2^2 \quad \text{subject to } \|\mathbf{a}\|_2 = 1.$$

The objective function of ($P1$) can be rearranged into

$$\sum_{j=1}^{|\tilde{\mathcal{C}}_{n^*}|} \left\| \left(\mathbf{I} - \mathbf{a}\mathbf{a}^T\right)(\mathbf{p}_j - \mathbf{c}) \right\|_2^2 \stackrel{(a)}{=} \sum_{j=1}^{|\tilde{\mathcal{C}}_{n^*}|} (\mathbf{p}_j - \mathbf{c})^T \left(\mathbf{I} - \mathbf{a}\mathbf{a}^T\right)(\mathbf{p}_j - \mathbf{c})$$

$$= \sum_{j=1}^{|\tilde{\mathcal{C}}_{n^*}|} (\mathbf{p}_j - \mathbf{c})^T (\mathbf{p}_j - \mathbf{c}) - (\mathbf{p}_j - \mathbf{c})^T \mathbf{a}\mathbf{a}^T (\mathbf{p}_j - \mathbf{c}), \quad (74)$$

where (a) holds since $\mathbf{I} - \mathbf{a}\mathbf{a}^T$ is an orthogonal projection matrix. With (74), the associated Lagrangian then reads

$$L(\mathbf{a}, \mathbf{c}, \tilde{\lambda}) = \sum_{j=1}^{|\tilde{\mathcal{C}}_{n^*}|} (\mathbf{p}_j - \mathbf{c})^T (\mathbf{p}_j - \mathbf{c}) - (\mathbf{p}_j - \mathbf{c})^T \mathbf{a}\mathbf{a}^T (\mathbf{p}_j - \mathbf{c}) + \tilde{\lambda}\left(\mathbf{a}^T\mathbf{a} - 1\right), \quad (75)$$

where $\tilde{\lambda} \in \mathbb{R}$ is the Lagrange multiplier. Based on (75) and with some manipulation, the first-order necessary condition $\nabla_\mathbf{c} L(\mathbf{a}, \mathbf{c}, \tilde{\lambda}) = \mathbf{0}$ and $\nabla_\mathbf{a} L(\mathbf{a}, \mathbf{c}, \tilde{\lambda}) = \mathbf{0}$ are given by, respectively,

$$\left(\mathbf{I} - \mathbf{a}\mathbf{a}^T\right)\left(\frac{1}{|\tilde{\mathcal{C}}_{n^*}|} \sum_{j=1}^{|\tilde{\mathcal{C}}_{n^*}|} \mathbf{p}_j - \mathbf{c}\right) = \mathbf{0}, \quad (76)$$

and

$$\sum_{j=1}^{|\tilde{\mathcal{C}}_{n^*}|} (\mathbf{p}_j - \mathbf{c})(\mathbf{p}_j - \mathbf{c})^T \mathbf{a} = \tilde{\lambda}\mathbf{a}. \quad (77)$$

From the condition (76), we can obtain

$$\mathbf{c} = \underbrace{\frac{1}{|\tilde{\mathcal{C}}_{n^*}|} \sum_{j=1}^{|\tilde{\mathcal{C}}_{n^*}|} \mathbf{p}_j}_{\triangleq \bar{\mathbf{p}}} + \bar{t}\mathbf{a}, \ \bar{t} \in \mathbb{R}. \quad (78)$$

For simplicity yet without lose of generality, we assume $\bar{t} = 0$ in (78) (hence, $\mathbf{c} = \bar{\mathbf{p}}$). Then using (78), the condition (77) becomes

$$\sum_{j=1}^{|\tilde{\mathcal{C}}_{n^*}|} (\mathbf{p}_j - \bar{\mathbf{p}})(\mathbf{p}_j - \bar{\mathbf{p}})^T \mathbf{a} = \tilde{\lambda}\mathbf{a}. \quad (79)$$

After some manipulation, the matrix $\sum_{j=1}^{|\tilde{\mathcal{C}}_{n^*}|} (\mathbf{p}_j - \bar{\mathbf{p}}) (\mathbf{p}_j - \bar{\mathbf{p}})^T$ in (79) can be expressed as

$$\sum_{j=1}^{|\tilde{\mathcal{C}}_{n^*}|} (\mathbf{p}_j - \bar{\mathbf{p}}) (\mathbf{p}_j - \bar{\mathbf{p}})^T = \sum_{j=1}^{|\tilde{\mathcal{C}}_{n^*}|} \mathbf{p}_j \mathbf{p}_j^T - \sum_{j=1}^{|\tilde{\mathcal{C}}_{n^*}|} \mathbf{p}_j \bar{\mathbf{p}}^T - \bar{\mathbf{p}} \sum_{j=1}^{|\tilde{\mathcal{C}}_{n^*}|} \mathbf{p}_j^T + |\tilde{\mathcal{C}}_{n^*}| \bar{\mathbf{p}} \bar{\mathbf{p}}^T$$

$$= \sum_{j=1}^{|\tilde{\mathcal{C}}_{n^*}|} \mathbf{p}_j \mathbf{p}_j^T - |\tilde{\mathcal{C}}_{n^*}| \bar{\mathbf{p}} \bar{\mathbf{p}}^T - |\tilde{\mathcal{C}}_{n^*}| \bar{\mathbf{p}} \bar{\mathbf{p}}^T + |\tilde{\mathcal{C}}_{n^*}| \bar{\mathbf{p}} \bar{\mathbf{p}}^T = \sum_{j=1}^{|\tilde{\mathcal{C}}_{n^*}|} \mathbf{p}_j \mathbf{p}_j^T - |\tilde{\mathcal{C}}_{n^*}| \bar{\mathbf{p}} \bar{\mathbf{p}}^T$$

$$\stackrel{(a)}{=} \begin{bmatrix} \sum_{j=1}^{|\tilde{\mathcal{C}}_{n^*}|} \operatorname{Re}\left(-\frac{b_{m_j}}{\phi_{m_j,n^*}}\right)^2 & \sum_{j=1}^{|\tilde{\mathcal{C}}_{n^*}|} \operatorname{Re}\left(-\frac{b_{m_j}}{\phi_{m_j,n^*}}\right) \operatorname{Im}\left(-\frac{b_{m_j}}{\phi_{m_j,n^*}}\right) \\ \sum_{j=1}^{|\tilde{\mathcal{C}}_{n^*}|} \operatorname{Re}\left(-\frac{b_{m_j}}{\phi_{m_j,n^*}}\right) \operatorname{Im}\left(-\frac{b_{m_j}}{\phi_{m_j,n^*}}\right) & \sum_{j=1}^{|\tilde{\mathcal{C}}_{n^*}|} \operatorname{Im}\left(-\frac{b_{m_j}}{\phi_{m_j,n^*}}\right)^2 \end{bmatrix} - \begin{bmatrix} |\tilde{\mathcal{C}}_{n^*}| \bar{b}_{n^*}^{(\operatorname{Re})2} & |\tilde{\mathcal{C}}_{n^*}| \bar{b}_{n^*}^{(\operatorname{Re})} \bar{b}_{n^*}^{(\operatorname{Im})} \\ |\tilde{\mathcal{C}}_{n^*}| \bar{b}_{n^*}^{(\operatorname{Re})} \bar{b}_{n^*}^{(\operatorname{Im})} & |\tilde{\mathcal{C}}_{n^*}| \bar{b}_{n^*}^{(\operatorname{Im})2} \end{bmatrix}$$

$$\stackrel{(b)}{=} \underbrace{\begin{bmatrix} \|\tilde{\mathbf{b}}_{n^*}^{(\operatorname{Re})}\|_2^2 - |\tilde{\mathcal{C}}_{n^*}| \bar{b}_{n^*}^{(\operatorname{Re})2} & \langle \tilde{\mathbf{b}}_{n^*}^{(\operatorname{Re})}, \tilde{\mathbf{b}}_{n^*}^{(\operatorname{Im})} \rangle - |\tilde{\mathcal{C}}_{n^*}| \bar{b}_{n^*}^{(\operatorname{Re})} \bar{b}_{n^*}^{(\operatorname{Im})} \\ \langle \tilde{\mathbf{b}}_{n^*}^{(\operatorname{Re})}, \tilde{\mathbf{b}}_{n^*}^{(\operatorname{Im})} \rangle - |\tilde{\mathcal{C}}_{n^*}| \bar{b}_{n^*}^{(\operatorname{Re})} \bar{b}_{n^*}^{(\operatorname{Im})} & \|\tilde{\mathbf{b}}_{n^*}^{(\operatorname{Im})}\|_2^2 - |\tilde{\mathcal{C}}_{n^*}| \bar{b}_{n^*}^{(\operatorname{Im})2} \end{bmatrix}}_{\triangleq \boldsymbol{\Psi} \in \mathbb{R}^{2 \times 2}}, \quad (80)$$

where (a) holds by the definition of $\mathbf{p}_j$ in (73) and using the fact that $\bar{\mathbf{p}} = [\bar{b}_{n^*}^{(\operatorname{Re})} \ \bar{b}_{n^*}^{(\operatorname{Im})}]^T$, and (b) follows from the definition of $\tilde{\mathbf{b}}_{n^*}$ in (24). Therefore, based on (80), (79) admits the following expression:

$$\boldsymbol{\Psi} \mathbf{a} = \tilde{\lambda} \mathbf{a}, \quad (81)$$

which implies that $\mathbf{a}$ and $\tilde{\lambda}$ are the eigenvector and eigenvalue of $\boldsymbol{\Psi}$, respectively. With (78), (80) and (81), the objective function of $(P1)$ in (74) then becomes

$$\sum_{j=1}^{|\tilde{\mathcal{C}}_{n^*}|} (\mathbf{p}_j - \mathbf{c})^T (\mathbf{p}_j - \mathbf{c}) - (\mathbf{p}_j - \mathbf{c})^T \mathbf{a} \mathbf{a}^T (\mathbf{p}_j - \mathbf{c})$$

$$\stackrel{(a)}{=} \sum_{j=1}^{|\tilde{\mathcal{C}}_{n^*}|} (\mathbf{p}_j - \bar{\mathbf{p}})^T (\mathbf{p}_j - \bar{\mathbf{p}}) - \sum_{j=1}^{|\tilde{\mathcal{C}}_{n^*}|} (\mathbf{p}_j - \bar{\mathbf{p}})^T \mathbf{a} \mathbf{a}^T (\mathbf{p}_j - \bar{\mathbf{p}})$$

$$= \sum_{j=1}^{|\tilde{\mathcal{C}}_{n^*}|} \|\mathbf{p}_j\|_2^2 - 2\bar{\mathbf{p}}^T \sum_{j=1}^{|\tilde{\mathcal{C}}_{n^*}|} \mathbf{p}_j + |\tilde{\mathcal{C}}_{n^*}| \|\bar{\mathbf{p}}\|_2^2 - \sum_{j=1}^{|\tilde{\mathcal{C}}_{n^*}|} (\mathbf{p}_j - \bar{\mathbf{p}})^T \mathbf{a} \mathbf{a}^T (\mathbf{p}_j - \bar{\mathbf{p}})$$

$$\stackrel{(b)}{=} \sum_{j=1}^{|\tilde{\mathcal{C}}_{n^*}|} \|\mathbf{p}_j\|_2^2 - |\tilde{\mathcal{C}}_{n^*}| \|\bar{\mathbf{p}}\|_2^2 - \sum_{j=1}^{|\tilde{\mathcal{C}}_{n^*}|} (\mathbf{p}_j - \bar{\mathbf{p}})^T \mathbf{a} \mathbf{a}^T (\mathbf{p}_j - \bar{\mathbf{p}})$$

$$\stackrel{(c)}{=} \sum_{j=1}^{|\tilde{\mathcal{C}}_{n^*}|} \|\mathbf{p}_j\|_2^2 - |\tilde{\mathcal{C}}_{n^*}| \|\bar{\mathbf{p}}\|_2^2 - \mathbf{a}^T \boldsymbol{\Psi} \mathbf{a}$$

$$\stackrel{(d)}{=} \sum_{j=1}^{|\tilde{\mathcal{C}}_{n^*}|} \|\mathbf{p}_j\|_2^2 - |\tilde{\mathcal{C}}_{n^*}| \|\bar{\mathbf{p}}\|_2^2 - \tilde{\lambda}, \quad (82)$$

where (a) follows from (78) (with $\bar{t}=0$), (b) holds since $\sum_{j=1}^{|\tilde{\mathcal{C}}_{n^*}|}\mathbf{p}_j = |\tilde{\mathcal{C}}_{n^*}|\bar{\mathbf{p}}$, (c) holds by (80), and (d) is due to (81). Since $\tilde{\lambda}$ in (82) is an eigenvalue of $\boldsymbol{\Psi}$ (see (81)), it can be observed from (82) that the minimum value of $\sum_{j=1}^{|\tilde{\mathcal{C}}_{n^*}|} r_j^2$, the objective function of (P1), is

$$\sum_{j=1}^{|\tilde{\mathcal{C}}_{n^*}|} \|\mathbf{p}_j\|_2^2 - |\tilde{\mathcal{C}}_{n^*}|\,\|\bar{\mathbf{p}}\|_2^2 - \tilde{\lambda}_{\max}, \tag{83}$$

where

$$\tilde{\lambda}_{\max} = \frac{\|\tilde{\mathbf{b}}_{n^*}\|_2^2 - |\tilde{\mathcal{C}}_{n^*}||\bar{b}_{n^*}|^2}{2} + \frac{1}{2}\sqrt{\left(\|\tilde{\mathbf{b}}_{n^*}^{(\text{Re})}\|_2^2 - \|\tilde{\mathbf{b}}_{n^*}^{(\text{Im})}\|_2^2 - |\tilde{\mathcal{C}}_{n^*}|\bar{b}_{n^*}^{(\text{Re})2} + |\tilde{\mathcal{C}}_{n^*}|\bar{b}_{n^*}^{(\text{Im})2}\right)^2 + 4\left(\langle\tilde{\mathbf{b}}_{n^*}^{(\text{Re})},\tilde{\mathbf{b}}_{n^*}^{(\text{Im})}\rangle - |\tilde{\mathcal{C}}_{n^*}|\bar{b}_{n^*}^{(\text{Re})}\bar{b}_{n^*}^{(\text{Im})}\right)^2}. \tag{84}$$

is the largest eigenvalue of $\boldsymbol{\Psi}$. Therefore, based on (83) and (84), the expression for $r^*$ in (66) is then obtained by using the facts that $\sum_{j=1}^{|\tilde{\mathcal{C}}_{n^*}|}\|\mathbf{p}_j\|_2^2 = \|\tilde{\mathbf{b}}_{n^*}\|_2^2$ and $\|\bar{\mathbf{p}}\|_2^2 = |\bar{b}_{n^*}|^2$. □

## APPENDIX I

### PROOF OF LEMMA 5.1

We first note that according to part 1) of Lemma H.1, $\|\tilde{\mathbf{b}}_{0,n^*}\|_2^2$ is bounded above by

$$\|\tilde{\mathbf{b}}_{0,n^*}\|_2^2 \stackrel{(a)}{\leq} \|\tilde{\mathbf{b}}_{n^*}\|_2^2 = \sum_{j=1}^{|\tilde{\mathcal{C}}_{n^*}|}\left|\frac{b_{m_j}}{\phi_{m_j,n^*}}\right|^2 \stackrel{(b)}{\leq} \sum_{j=1}^{|\tilde{\mathcal{C}}_{n^*}|}\frac{b_{\max}^2}{\phi_{\min}^2} = |\tilde{\mathcal{C}}_{n^*}|\frac{b_{\max}^2}{\phi_{\min}^2}, \tag{85}$$

where the equality in (a) holds if $\langle\tilde{\mathbf{b}}_{n^*},\mathbf{1}\rangle = 0$ (thus, $\bar{b}_{n^*}=0$), and the equality in (b) holds provided that $|\phi_{m_j,n^*}| = \phi_{\min}$ and $|b_{m_j}| = b_{\max}$ are true for all $1 \leq j \leq |\tilde{\mathcal{C}}_{n^*}|$. Therefore, with the aid of (85), we can obtain the following inequality

$$\frac{\sqrt{|\tilde{\mathcal{C}}_{n^*}|}}{\|\tilde{\mathbf{b}}_{0,n^*}\|_2\left(1-\left|\left\langle\frac{\tilde{\mathbf{b}}_{0,n^*}}{\|\tilde{\mathbf{b}}_{0,n^*}\|_2},\frac{\tilde{\mathbf{b}}_{0,n^*}^*}{\|\tilde{\mathbf{b}}_{0,n^*}^*\|_2}\right\rangle\right|\right)} \geq \frac{\phi_{\min}}{b_{\max}\left(1-\left|\left\langle\frac{\tilde{\mathbf{b}}_{0,n^*}}{\|\tilde{\mathbf{b}}_{0,n^*}\|_2},\frac{\tilde{\mathbf{b}}_{0,n^*}^*}{\|\tilde{\mathbf{b}}_{0,n^*}^*\|_2}\right\rangle\right|\right)}. \tag{86}$$

Since it can be observed from (70) that

$$\langle\tilde{\mathbf{b}}_{0,n},\tilde{\mathbf{b}}_{0,n}^*\rangle = 0 \text{ if } \langle\tilde{\mathbf{b}}_{n^*},\tilde{\mathbf{b}}_{n^*}^*\rangle = 0 \text{ and } \bar{b}_{n^*}^* = 0, \tag{87}$$

which in turn implies

$$\left|\left\langle\frac{\tilde{\mathbf{b}}_{0,n^*}}{\|\tilde{\mathbf{b}}_{0,n^*}\|_2},\frac{\tilde{\mathbf{b}}_{0,n^*}^*}{\|\tilde{\mathbf{b}}_{0,n^*}^*\|_2}\right\rangle\right| = 0 \text{ if } \langle\tilde{\mathbf{b}}_{n^*},\tilde{\mathbf{b}}_{n^*}^*\rangle = 0 \text{ and } \langle\tilde{\mathbf{b}}_{n^*},\mathbf{1}\rangle = 0. \tag{88}$$

The assertion (30) immediately follows from (86) and (88). The proof of Lemma 5.1 is thus completed. □

# APPENDIX J
## PROOF OF THEOREM 5.2

The proof basically consists of two parts. We will first derive a probability lower bound for the event that $\sqrt{|\tilde{\mathcal{C}}_{n^*}|}\big/\|\tilde{\mathbf{b}}_{0,n^*}\|_2 \left(1 - \left|\left\langle \frac{\tilde{\mathbf{b}}_{0,n^*}}{\|\tilde{\mathbf{b}}_{0,n^*}\|_2}, \frac{\tilde{\mathbf{b}}_{0,n^*}^*}{\|\tilde{\mathbf{b}}_{0,n^*}^*\|_2}\right\rangle\right|\right)$ is smaller than certain threshold. Then, the proof is completed with the aid of (27) and some manipulation. The following lemma (whose proof is placed at the end of this appendix) is needed to prove the first part.

*Lemma J.1:* Suppose that the nonzero entries of $\boldsymbol{\Phi}$ are i.i.d. and uniformly drawn from the circle of radius $\phi_c$, and that the entries of $\mathbf{b}$ are i.i.d. and uniformly drawn from the circle of radius $b_c$. Then for every $n \in \mathcal{T}$ and every $0 \leq t \leq \frac{-1+\sqrt{1+\rho^2}}{2}$, where $\rho \triangleq b_c/\phi_c$, the inequality

$$\frac{\sqrt{|\tilde{\mathcal{C}}_n|}}{\|\tilde{\mathbf{b}}_{0,n}\|_2\left(1 - \left|\left\langle \frac{\tilde{\mathbf{b}}_{0,n}}{\|\tilde{\mathbf{b}}_{0,n}\|_2}, \frac{\tilde{\mathbf{b}}_{0,n}^*}{\|\tilde{\mathbf{b}}_{0,n}^*\|_2}\right\rangle\right|\right)} \leq \frac{1}{\rho}\left(1 + \frac{4(t+1)/\rho^2}{1 - (4t^2+4t)/\rho^2}t\right) \tag{89}$$

holds with probability at least $1 - 5e\exp(-c_2|\tilde{\mathcal{C}}_n|t^2)$, in which $c_2 > 0$ is a constant. $\square$

We first note that the inequality

$$\frac{\sqrt{|\tilde{\mathcal{C}}_{n^*}|}}{\|\tilde{\mathbf{b}}_{0,n^*}\|_2\left(1 - \left|\left\langle \frac{\tilde{\mathbf{b}}_{0,n^*}}{\|\tilde{\mathbf{b}}_{0,n^*}\|_2}, \frac{\tilde{\mathbf{b}}_{0,n^*}^*}{\|\tilde{\mathbf{b}}_{0,n^*}^*\|_2}\right\rangle\right|\right)} \leq \frac{1}{\rho}\left(1 + \frac{4(t+1)/\rho^2}{1 - (4t^2+4t)/\rho^2}t\right) \tag{90}$$

holds if and only if (89) holds for all $n \in \mathcal{T}$. Hence, based on Lemma J.1 and by invoking the union bound, we conclude that the inequality (90) holds at least

$$1 - 5eK\exp\left(-c_2'|\tilde{\mathcal{C}}_{n'}|t^2\right) = 1 - \exp\left(\log(5e) + \log(K) - c_2'|\tilde{\mathcal{C}}_{n'}|t^2\right)$$
$$\stackrel{(a)}{\geq} 1 - \exp\left(-\frac{c_2'}{2}|\tilde{\mathcal{C}}_{n'}|t^2\right), \tag{91}$$

where (a) follows by the assumption $|\tilde{\mathcal{C}}_{n'}| \geq Ct^{-2}\log(K)$ with sufficiently large constant $C$. Let $c_1 = c_2'/2$. Then by means of (27), (90) and (91), the assertion (32) immediately follows. $\square$

*[Proof of Lemma J.1]:* The following lemma (proof is given at the end of the appendix) is needed to prove Lemma J.1.

*Lemma J.2:* Let $\tilde{\mathbf{b}}_n = \tilde{\mathbf{b}}_n^{(\text{Re})} + i\tilde{\mathbf{b}}_n^{(\text{Im})}$ and $\bar{\mathbf{b}}_n = \bar{b}_n^{(\text{Re})} + i\bar{b}_n^{(\text{Im})}$ be defined as in (24) and Theorem 4.4, respectively. Under the same assumptions on $\boldsymbol{\Phi}$ and $\mathbf{b}$ as in Lemma J.1, for every $n \in \mathcal{T}$ and every $t \geq 0$, we have the following results.

1) $\Pr\left(\left|\frac{1}{|\tilde{\mathcal{C}}_n|}\langle \tilde{\mathbf{b}}_n^{(\text{Re})}, \tilde{\mathbf{b}}_n^{(\text{Im})}\rangle\right| \geq t\right) \leq e\exp\left(-\frac{c_3|\tilde{\mathcal{C}}_n|t^2}{\rho^4}\right)$;
2) $\Pr\left(\left|\bar{b}_n^{(\text{Re})}\right| \geq t\right) \leq e\exp\left(-\frac{c_4|\tilde{\mathcal{C}}_n|t^2}{\rho^2}\right)$;

3) $\Pr\left(\left|\bar{b}_n^{(\text{Im})}\right| \geq t\right) \leq e \exp\left(-\frac{c_5|\tilde{\mathcal{C}}_n|t^2}{\rho^2}\right);$

4) $\Pr\left(\left|\left\|\frac{\tilde{\mathbf{b}}_n^{(\text{Re})}}{\sqrt{|\tilde{\mathcal{C}}_n|}}\right\|_2^2 - \frac{1}{2}\rho^2\right| \geq t\right) \leq e \exp\left(-\frac{c_6|\tilde{\mathcal{C}}_n|t^2}{\rho^4}\right);$

5) $\Pr\left(\left|\left\|\frac{\tilde{\mathbf{b}}_n^{(\text{Im})}}{\sqrt{|\tilde{\mathcal{C}}_n|}}\right\|_2^2 - \frac{1}{2}\rho^2\right| \geq t\right) \leq e \exp\left(-\frac{c_7|\tilde{\mathcal{C}}_n|t^2}{\rho^4}\right),$

where $\rho \triangleq b_c/\phi_c$ and $c_3, c_4, c_5, c_6$ and $c_7$ are positive constants. $\square$

Let us rewrite

$$\frac{\sqrt{|\tilde{\mathcal{C}}_n|}}{\left\|\tilde{\mathbf{b}}_{0,n}\right\|_2\left(1 - \left|\left\langle \frac{\tilde{\mathbf{b}}_{0,n}}{\|\tilde{\mathbf{b}}_{0,n}\|_2}, \frac{\tilde{\mathbf{b}}_{0,n}^*}{\|\tilde{\mathbf{b}}_{0,n}^*\|_2}\right\rangle\right|\right)} = \frac{\sqrt{|\tilde{\mathcal{C}}_n|}\left\|\tilde{\mathbf{b}}_{0,n}\right\|_2}{\left\|\tilde{\mathbf{b}}_{0,n}\right\|_2^2 - \left|\langle \tilde{\mathbf{b}}_{0,n}, \tilde{\mathbf{b}}_{0,n}^*\rangle\right|}. \quad (92)$$

The denominator on the right-hand-side of (92) is bounded below according to

$$\left\|\tilde{\mathbf{b}}_{0,n}\right\|_2^2 - \left|\langle \tilde{\mathbf{b}}_{0,n}, \tilde{\mathbf{b}}_{0,n}^*\rangle\right|$$
$$\stackrel{(a)}{=} \left\|\tilde{\mathbf{b}}_n\right\|_2^2 - |\tilde{\mathcal{C}}_n||\bar{b}_n|^2 -$$
$$\sqrt{4\left(\langle \tilde{\mathbf{b}}_n^{(\text{Re})}, \tilde{\mathbf{b}}_n^{(\text{Im})}\rangle - |\tilde{\mathcal{C}}_n|\bar{b}_n^{(\text{Re})}\bar{b}_n^{(\text{Im})}\right)^2 + \left(\left\|\tilde{\mathbf{b}}_n^{(\text{Re})}\right\|_2^2 - \left\|\tilde{\mathbf{b}}_n^{(\text{Im})}\right\|_2^2 - |\tilde{\mathcal{C}}_n|\bar{b}_n^{(\text{Re})2} + |\tilde{\mathcal{C}}_n|\bar{b}_n^{(\text{Im})2}\right)^2}$$
$$\geq \left\|\tilde{\mathbf{b}}_n\right\|_2^2 - |\tilde{\mathcal{C}}_n||\bar{b}_n|^2 -$$
$$\sqrt{\left[2\left|\langle \tilde{\mathbf{b}}_n^{(\text{Re})}, \tilde{\mathbf{b}}_n^{(\text{Im})}\rangle - |\tilde{\mathcal{C}}_n|\bar{b}_n^{(\text{Re})}\bar{b}_n^{(\text{Im})}\right| + \left|\left\|\tilde{\mathbf{b}}_n^{(\text{Re})}\right\|_2^2 - \left\|\tilde{\mathbf{b}}_n^{(\text{Im})}\right\|_2^2 - |\tilde{\mathcal{C}}_n|\bar{b}_n^{(\text{Re})2} + |\tilde{\mathcal{C}}_n|\bar{b}_n^{(\text{Im})2}\right|\right]^2}$$
$$= \left\|\tilde{\mathbf{b}}_n^{(\text{Re})}\right\|_2^2 - |\tilde{\mathcal{C}}_n|\bar{b}_n^{(\text{Re})2} + \left\|\tilde{\mathbf{b}}_n^{(\text{Im})}\right\|_2^2 - |\tilde{\mathcal{C}}_n|\bar{b}_n^{(\text{Im})2} - \left|\left\|\tilde{\mathbf{b}}_n^{(\text{Re})}\right\|_2^2 - |\tilde{\mathcal{C}}_n|\bar{b}_n^{(\text{Re})2} - \left\|\tilde{\mathbf{b}}_n^{(\text{Im})}\right\|_2^2 + |\tilde{\mathcal{C}}_n|\bar{b}_n^{(\text{Im})2}\right|$$
$$- 2\left|\langle \tilde{\mathbf{b}}_n^{(\text{Re})}, \tilde{\mathbf{b}}_n^{(\text{Im})}\rangle - |\tilde{\mathcal{C}}_n|\bar{b}_n^{(\text{Re})}\bar{b}_n^{(\text{Im})}\right|$$
$$= 2|\tilde{\mathcal{C}}_n|\left[\min\left(\left\|\frac{\tilde{\mathbf{b}}_n^{(\text{Re})}}{\sqrt{|\tilde{\mathcal{C}}_n|}}\right\|_2^2 - \bar{b}_n^{(\text{Re})2}, \left\|\frac{\tilde{\mathbf{b}}_n^{(\text{Im})}}{\sqrt{|\tilde{\mathcal{C}}_n|}}\right\|_2^2 - \bar{b}_n^{(\text{Im})2}\right) - \left|\frac{1}{|\tilde{\mathcal{C}}_n|}\langle \tilde{\mathbf{b}}_n^{(\text{Re})}, \tilde{\mathbf{b}}_n^{(\text{Im})}\rangle - \bar{b}_n^{(\text{Re})}\bar{b}_n^{(\text{Im})}\right|\right]$$
$$\geq 2|\tilde{\mathcal{C}}_n|\left[\min\left(\left\|\frac{\tilde{\mathbf{b}}_n^{(\text{Re})}}{\sqrt{|\tilde{\mathcal{C}}_n|}}\right\|_2^2 - \bar{b}_n^{(\text{Re})2}, \left\|\frac{\tilde{\mathbf{b}}_n^{(\text{Im})}}{\sqrt{|\tilde{\mathcal{C}}_n|}}\right\|_2^2 - \bar{b}_n^{(\text{Im})2}\right) - \frac{1}{|\tilde{\mathcal{C}}_n|}\left|\langle \tilde{\mathbf{b}}_n^{(\text{Re})}, \tilde{\mathbf{b}}_n^{(\text{Im})}\rangle\right| - \left|\bar{b}_n^{(\text{Re})}\bar{b}_n^{(\text{Im})}\right|\right],$$
$$(93)$$

where (a) follows from Lemma H.1. In addition, using part 1) of Lemma H.1 we can obtain

$$\left\|\tilde{\mathbf{b}}_{0,n}\right\|_2^2 \leq \left\|\tilde{\mathbf{b}}_n\right\|_2^2 = |\tilde{\mathcal{C}}_n|\rho^2. \quad (94)$$

Combining (92), (93) and (94), we have

$$\frac{\sqrt{|\tilde{\mathcal{C}}_n|}}{\|\tilde{\mathbf{b}}_{0,n}\|_2 \left(1 - \left|\left\langle \frac{\tilde{\mathbf{b}}_{0,n}}{\|\tilde{\mathbf{b}}_{0,n}\|_2}, \frac{\tilde{\mathbf{b}}_{0,n}^*}{\|\tilde{\mathbf{b}}_{0,n}^*\|_2}\right\rangle\right|\right)} = \frac{\sqrt{|\tilde{\mathcal{C}}_n|}\|\tilde{\mathbf{b}}_{0,n}\|_2}{\|\tilde{\mathbf{b}}_{0,n}\|_2^2 - |\langle \tilde{\mathbf{b}}_{0,n}, \tilde{\mathbf{b}}_{0,n}^*\rangle|} \leq \frac{|\tilde{\mathcal{C}}_n|\rho}{\|\tilde{\mathbf{b}}_{0,n}\|_2^2 - |\langle \tilde{\mathbf{b}}_{0,n}, \tilde{\mathbf{b}}_{0,n}^*\rangle|}$$

$$\leq \frac{\rho}{2\left[\min\left(\left\|\frac{\tilde{\mathbf{b}}_n^{(\text{Re})}}{\sqrt{|\tilde{\mathcal{C}}_n|}}\right\|_2^2 - \bar{b}_n^{(\text{Re})2}, \left\|\frac{\tilde{\mathbf{b}}_n^{(\text{Im})}}{\sqrt{|\tilde{\mathcal{C}}_n|}}\right\|_2^2 - \bar{b}_n^{(\text{Im})2}\right) - \frac{1}{|\tilde{\mathcal{C}}_n|}\left|\langle \tilde{\mathbf{b}}_n^{(\text{Re})}, \tilde{\mathbf{b}}_n^{(\text{Im})}\rangle\right| - |\bar{b}_n^{(\text{Re})}||\bar{b}_n^{(\text{Im})}|\right]}. \quad (95)$$

Using (95) and Lemma J.2, we conclude that, for every $0 \leq t \leq \frac{-1+\sqrt{1+\rho^2}}{2}$, the inequality

$$\frac{\sqrt{|\tilde{\mathcal{C}}_n|}}{\|\tilde{\mathbf{b}}_{0,n}\|_2 \left(1 - \left|\left\langle \frac{\tilde{\mathbf{b}}_{0,n}}{\|\tilde{\mathbf{b}}_{0,n}\|_2}, \frac{\tilde{\mathbf{b}}_{0,n}^*}{\|\tilde{\mathbf{b}}_{0,n}^*\|_2}\right\rangle\right|\right)} \leq \frac{\rho}{\rho^2 - 4t^2 - 4t} = \frac{1}{\rho}\left(1 + \frac{4t(t+1)/\rho^2}{1-(4t^2+4t)/\rho^2}\right) \quad (96)$$

holds with probability exceeding

$$1 - e\left(\exp\left(-\frac{c_3|\tilde{\mathcal{C}}_n|t^2}{\rho^4}\right) + \exp\left(-\frac{c_4|\tilde{\mathcal{C}}_n|t^2}{\rho^2}\right) + \exp\left(-\frac{c_5|\tilde{\mathcal{C}}_n|t^2}{\rho^2}\right) + \exp\left(-\frac{c_6|\tilde{\mathcal{C}}_n|t^2}{\rho^4}\right) + \exp\left(-\frac{c_7|\tilde{\mathcal{C}}_n|t^2}{\rho^4}\right)\right)$$
$$\geq 1 - 5e\exp\left(-c_2|\tilde{\mathcal{C}}_n|t^2\right), \quad (97)$$

where the inequality follows for $c_2 > 0$ sufficiently small (e.g., may choose $c_2 = \min\left(\frac{c_3}{\rho^4}, \frac{c_4}{\rho^2}, \frac{c_5}{\rho^2}, \frac{c_6}{\rho^4}, \frac{c_7}{\rho^4}\right)$). Hence the proof of Lemma J.1 is completed. We end this appendix by proving Lemma J.2.

*[Proof of Lemma J.2]:* We first note that, according to the proposed random construction scheme of $\boldsymbol{\Phi}$ and $\mathbf{b}$, the real part and imaginary part of $\frac{b_{m_j}}{\phi_{m_j,n}}$, $j = 1, \ldots, |\tilde{\mathcal{C}}_n|$, can be respectively expressed as

$$\text{Re}\left(\frac{b_{m_j}}{\phi_{m_j,n}}\right) = \frac{b_c}{\phi_c}\cos\left(\theta_{m_j} - \varphi_{m_j,n}\right), \quad (98)$$

and

$$\text{Im}\left(\frac{b_{m_j}}{\phi_{m_j,n}}\right) = \frac{b_c}{\phi_c}\sin\left(\theta_{m_j} - \varphi_{m_j,n}\right). \quad (99)$$

Let $X_j = \text{Re}\left(\frac{b_{m_j}}{\phi_{m_j,n}}\right)\text{Im}\left(\frac{b_{m_j}}{\phi_{m_j,n}}\right)$, where $j = 1, \ldots, |\tilde{\mathcal{C}}_n|$. Then using (98) and (99), and based on the proposed random construction scheme, it can be shown that $X_j = \frac{\rho^2}{2}\sin(2\theta_{m_j} - 2\varphi_{m_j,n})$ is a bounded random variable with zero mean. Moreover, Hoeffding's lemma [50] asserts that bounded random variables are also sub-Gaussian. Hence, $X_1, \ldots, X_{|\tilde{\mathcal{C}}_n|}$ are independent mean-zero sub-Gaussian random variables with the sub-Gaussian norm $\|X_j\|_{\psi_2} \leq \frac{\rho^2}{2}$. Accordingly, we can obtain

$$\Pr\left(\left|\frac{1}{|\tilde{\mathcal{C}}_n|}\langle \tilde{\mathbf{b}}_n^{(\text{Re})}, \tilde{\mathbf{b}}_n^{(\text{Im})}\rangle\right| \geq t\right) = \Pr\left(\left|\sum_{j=1}^{|\tilde{\mathcal{C}}_n|} \frac{1}{|\tilde{\mathcal{C}}_n|} X_j\right| \geq t\right) \overset{(a)}{\leq} e\exp\left(-\frac{c'_3 t^2}{\left(\frac{\rho^2}{2}\right)^2 \left\|\frac{\mathbf{1}}{|\tilde{\mathcal{C}}_n|}\right\|_2^2}\right)$$

$$= e\exp\left(-\frac{c_3 |\tilde{\mathcal{C}}_n| t^2}{\rho^4}\right),$$

where (a) holds by using Hoeffding-type inequality [50] and $c'_3 > 0$ is an absolute constant. Therefore, the proof of part 1) is completed.

The proof of part 2) is similar. Firstly, according to (98), we can show that $\text{Re}\left(\frac{b_{m_j}}{\phi_{m_j,n}}\right)$'s are independent bounded random variables with zero mean; this in turn implies $\text{Re}\left(\frac{b_{m_j}}{\phi_{m_j,n}}\right)$'s are independent mean-zero sub-Gaussian with $\left\|\text{Re}\left(\frac{b_{m_j}}{\phi_{m_j,n}}\right)\right\|_{\psi_2} \leq \rho$. Then we have

$$\Pr\left(\left|\bar{b}_n^{(\text{Re})}\right| \geq t\right) = \Pr\left(\left|\frac{1}{|\tilde{\mathcal{C}}_n|}\langle \mathbf{1}, \tilde{\mathbf{b}}_n^{(\text{Re})}\rangle\right| \geq t\right)$$

$$= \Pr\left(\left|\sum_{j=1}^{|\tilde{\mathcal{C}}_n|} \frac{1}{|\tilde{\mathcal{C}}_n|}\text{Re}\left(\frac{b_{m_j}}{\phi_{m_j,n}}\right)\right| \geq t\right)$$

$$\overset{(a)}{\leq} e\exp\left(-\frac{c_4 t^2}{\rho^2 \left\|\frac{\mathbf{1}}{|\tilde{\mathcal{C}}_n|}\right\|_2^2}\right)$$

$$= e\exp\left(-\frac{c_4 |\tilde{\mathcal{C}}_n| t^2}{\rho^2}\right),$$

where (a) holds by applying Hoeffding-type inequality [50]. The proof of part 2) is thus completed. The proof of part 3) can also be done by following essentially the same procedures as in the proof of part 2).

Next, we go on to prove part 4). Let $\tilde{X}_j = \frac{\rho^2}{2}\cos\left(2\theta_{m_j} - 2\varphi_{m_j,n}\right)$, where $j = 1, \ldots, |\tilde{\mathcal{C}}_n|$. Then based on Hoeffding's lemma, it can be shown that $\tilde{X}_j$'s are independent mean-zero sub-Gaussian random variables with $\|\tilde{X}_j\|_{\psi_2} \leq \frac{\rho^2}{2}$. Hence, we obtain

$$\Pr\left(\left|\left\|\frac{\tilde{\mathbf{b}}_n^{(\text{Re})}}{\sqrt{|\tilde{\mathcal{C}}_n|}}\right\|_2^2 - \frac{1}{2}\rho^2\right| \geq t\right) = \Pr\left(\left|\sum_{j=1}^{|\tilde{\mathcal{C}}_n|} \frac{1}{|\tilde{\mathcal{C}}_n|}\rho^2 \cos^2\left(\theta_{m_j} - \varphi_{m_j,n}\right) - \frac{1}{2}\rho^2\right| \geq t\right)$$

$$\overset{(a)}{=} \Pr\left(\left|\sum_{j=1}^{|\tilde{\mathcal{C}}_n|} \frac{1}{|\tilde{\mathcal{C}}_n|}\tilde{X}_j\right| \geq t\right)$$

$$\overset{(b)}{\leq} e \exp\left(-\frac{c_6' t^2}{\left(\frac{\rho^2}{2}\right)^2 \left\|\frac{1}{|\tilde{\mathcal{C}}_n|}\right\|_2^2}\right)$$

$$= e \exp\left(-\frac{c_6 |\tilde{\mathcal{C}}_n| t^2}{\rho^4}\right),$$

where (a) follows from the trigonometric identity $\cos^2\left(\theta_{m_j} - \varphi_{m_j,n}\right) = \frac{1}{2}\left[1 + \cos\left(2\theta_{m_j} - 2\varphi_{m_j,n}\right)\right]$, (b) holds by using Hoeffding-type inequality and $c_6' > 0$ is an absolute constant. Hence, the proof of part 4) is completed. The result given in part 5) can also be proved by following the similar procedures as in the above proof. Therefore, the proof of Lemma J.2 is completed. □

## REFERENCES

[1] R. P. Millane, "Phase retrieval in crystallography and optics," *J. Opt. Soc. Amer. A*, vol. 7, no. 3, pp. 394–411, March 1990.

[2] H. A. Hauptman, "The phase problem of X-ray crystallography," *Rep. Prog. Phys.*, vol. 54, no. 11, pp. 1427–1454, Nov. 1991.

[3] J. C. Dainty and J. R. Fienup, "Phase retrieval and image reconstruction for astronomy," in *Image Recovery: Theory and Application*, San Diego, CA, USA: Academic, 1987, pp. 231–275.

[4] J. M. Rodenburg, "Ptychography and related diffractive imaging methods," *Adv. Imag. Electron Phys.*, vol. 150, pp. 87–184, Dec. 2008.

[5] X. Li, J. Fang, H. Duan, Z. Chen and H. Li, "Fast beam alignment for millimeter wave communications: A sparse encoding and phaseless decoding approach," *IEEE Trans. Signal Processing*, vol. 67, no. 17, pp. 4402–4417, Sept. 2019.

[6] C. Hu, X. Wang, L. Dai and J. Ma, "Partially coherent compressive phase retrieval for millimeter-wave massive MIMO channel estimation," *IEEE Trans. Signal Processing*, vol. 68, pp. 1673–1687, 2020.

[7] M. Hayes, "The reconstruction of a multidimensional sequence from the phase or magnitude of its Fourier transform," *IEEE Trans. Acoust., Speech, and Signal Processing*, vol. 30, no. 2, pp. 140–154, April 1982.

[8] R. Balan, P. Casazza and D. Edidin, "On signal reconstruction without phase," *Appl. Comput. Harmon. Anal.*, vol. 20, no. 3, pp. 345–356, May 2006.

[9] Y. Shechtman, Y. C. Eldar, O. Cohen, H. N. Chapman, J. Miao and M. Segev, "Phase retrieval with application to optical imaging: A contemporary overview," in *IEEE Signal Processing Magazine*, vol. 32, no. 3, pp. 87–109, May 2015.

[10] P. Netrapalli, P. Jain and S. Sanghavi, "Phase retrieval using alternating minimization," *IEEE Trans. Signal Processing*, vol. 63, no. 18, pp. 4814–4826, Sep. 2015.

[11] P. Chen, A. Fannjiang and G.-R. Liu, "Phase retrieval with one or two diffraction patterns by alternating projection with null initialization," *J. Fourier Anal. Appl.*, pp. 1–40, March 2017.

[12] E. J. Candès, T. Strohmer and V. Voroninski, "PhaseLift: Exact and stable signal recovery from magnitude measurements via convex programming," *Appl. Comput. Harmon. Anal.*, vol. 66, no. 8, pp. 1241–1274, Nov. 2013.

[13] E. J. Candès, Y. C. Eldar, T. Strohmer and V. Voroninski, "Phase retrieval via matrix completion," *SIAM J. Imag. Sci.*, vol. 6, no. 1, pp. 199–225, 2013.

[14] I. Waldspurger, A. d'Aspremont and S. Mallat, "Phase recovery, maxcut and complex semidefinite programming," *Math. Program.*, vol. 149, no. 1, pp. 47–81, 2015.

[15] H. Ohlsson and Y. C. Eldar, "On conditions for uniqueness in sparse phase retrieval," *2014 IEEE International Conference on Acoustics, Speech and Signal Processing (ICASSP)*, pp. 1841–1845, Aug. 2014.

[16] Y. C. Eldar and S. Mendelson, "Phase retrieval: Stability and recovery guarantees," *Appl. Comput. Harmon. Anal.*, vol. 36, no. 3, pp. 473–494, May 2014.

[17] V. Voroninski and Z. Xu, "A strong restricted isometry property, with an application to phaseless compressed sensing," *Appl. Comput. Harmon. Anal.*, vol. 40, no. 2, pp. 386–395, March 2016.

[18] X. Li and V. Voroninski, "Sparse signal recovery from quadratic measurements via convex programming," *SIAM J. Math. Anal.*, vol. 45, no. 5, pp. 3019–3033, 2013.

[19] E. J. R. Pauwels, A. Beck, Y. C. Eldar and S. Sabach, "On fienup methods for sparse phase retrieval," *IEEE Trans. Signal Processing*, vol. 66, no. 4, pp. 982–991, Feb. 2018.

[20] G. Jagatap and C. Hegde, "Fast, sample-efficient algorithms for structured phase retrieval," in *Proc. Adv. Neural Inf. Process. Syst.*, pp. 4917–4927, 2017.

[21] G. Jagatap and C. Hegde, "Sample-Efficient algorithms for recovering structured signals from magnitude-only measurements," *IEEE Trans. Information Theory*, vol. 65, no. 7, pp. 4434–4456, July 2019.

[22] T. Goldstein and C. Studer, "PhaseMax: Convex phase retrieval via basis pursuit," *IEEE Trans. Information Theory*, vol. 64, no. 4, pp. 2675–2689, April 2018.

[23] T. Qiu and D. P. Palomar, "Undersampled sparse phase retrieval via majorization–minimization," *IEEE Trans. Signal Processing*, vol. 65, no. 22, pp. 5957–5969, Nov. 2017.

[24] H. Ohlsson, A. Y. Yang, R. Dong and S. S. Sastry, "CPRL–An extension of compressive sensing to the phase retrieval problem," in *Proc. Neural Inf. Process. Syst. (NIPS)*, 2012.

[25] Y. Shechtman, A. Beck, and Y. C. Eldar, "GESPAR: Efficient phase retrieval of sparse signals," *IEEE Trans. Signal Processing*, vol. 62, no. 4, pp. 928–938, Feb. 2014.

[26] G. Wang, L. Zhang, G. B. Giannakis, M. Akçakaya and J. Chen, "Sparse phase retrieval via truncated amplitude flow," *IEEE Trans. Signal Processing*, vol. 66, no. 2, pp. 479–491, Jan. 2018.

[27] L. Zhang, G. Wang, G. B. Giannakis and J. Chen, "Compressive phase retrieval via reweighted amplitude flow," *IEEE Trans. Signal Processing*, vol. 66, no. 19, pp. 5029–5040, Oct. 2018.

[28] C. Papadimitriou and K. Steiglitz, *Combinatorial Optimization: Algorithms and Complexity*. Dover Publications, 1998.

[29] Z. Yuan and H. Wang, "Phase retrieval with background information," *Inverse Problems*, vol. 35, no. 5, 2019.

[30] V. Elser, T.-Y. Lan and T. Bendory, "Benchmark problems for phase retrieval," *SIAM J. Imag. Sci.*, vol. 11, no. 4, pp. 2429–2455, Oct. 2018.

[31] G. Baechler, M. Kreković, J. Ranieri, A. Chebira, Y. M. Lu and M. Vetterli, "Super resolution phase retrieval for sparse signals," *IEEE Trans. Signal Processing*, vol. 67, no. 18, pp. 4839–4854, Sept. 2019.

[32] T. Cai, X. Li and Z. Ma, "Optimal rates of convergence for noisy sparse phase retrieval via thresholded Wirtinger flow," *Ann. Statist.*, vol. 44, no. 5, pp. 2221–2251, 2016.

[33] B. Gao, Q. Sun, Y. Wang and Z. Xu, "Phase retrieval from the magnitudes of affine linear measurements," *Adv. Appl. Math.*, vol. 93, pp. 121–141, Feb. 2018.

[34] R. Pedarsani, D. Yin, K. Lee and K. Ramchandran, "PhaseCode: Fast and efficient compressive phase retrieval based on sparse-graph codes," *IEEE Trans. Information Theory*, vol. 63, no. 6, pp. 3663–3691, June 2017.

[35] A. Gilbert and P. Indyk, "Sparse recovery using sparse matrices," *Proc. IEEE*, vol. 98, no. 6, pp. 937–947, June 2010.

[36] M. F. Duarte, M. A. Davenport, D. Takhar, J. N. Laska, T. Sun, K. F. Kelly, and R. G. Baraniuk, "Single-pixel imaging via compressive sampling," *IEEE Signal Processing Mag.*, vol. 25, pp. 83–91, March 2008.

[37] S. Gopi, P. Netrapalli, P. Jain and A. Nori, "One-bit compressed sensing: Provable support and vector recovery," in *Proc. 30th Int. Conf. Mach. Learn. (ICML)*, pp. 154–162, 2013.

[38] M. Lotfi and M. Vidyasagar, "A fast noniterative algorithm for compressive sensing using binary measurement matrices," *IEEE Trans. Signal Processing*, vol. 66, pp. 4079–4089, Aug. 2018.

[39] J. Acharya, A. Bhattacharyya and P. Kamath, "Improved bounds for universal one-bit compressive sensing," *2017 IEEE International Symposium on Information Theory (ISIT)*, pp. 2353–2357, 2017.

[40] L. Flodin, V. Gandikota and A. Mazumdar, "Superset technique for approximate recovery in one-bit compressed sensing," in *Conf. Neur. Inf. Proc. Sys. (NeurIPS)*, 2019.

[41] S. Jukna, *Extremal Combinatorics*. New York: Springer-Verlag, 2001.

[42] D.-Z. Du and F. K. Hwang, *Combinatorial Group Testing and its Applications*. Singapore: World Scientific, 1993.

[43] G. Atia and V. Saligrama, "Boolean compressed sensing and noisy group testing," *IEEE Trans. Information Theory*, vol. 58, no. 3, pp. 1880–1901, March 2012.

[44] A. Mazumdar, "Nonadaptive group testing with random set of defectives," *IEEE Trans. Information Theory*, vol. 62, no. 12, pp. 7522–7531, Dec. 2016.

[45] R. DeVore, "Deterministic construction of compressed sensing matrices," *J. Complexity*, vol. 23, pp. 918–925, 2007.

[46] N. Kiryati and A. M. Bruckstein, "What's in a set of points? (straight line fitting)," *IEEE Trans. Pattern Anal. Mach. Intell.*, vol. 14, no. 4, pp. 496–500, April 1992.

[47] M. A. Herman and T. Strohmer, "General deviants: An analysis of perturbations in compressed sensing," *IEEE J. Sel. Topics Signal Processing*, vol. 4, no. 2, pp. 342–349, April 2010.

[48] Y. Lecun, L. Bottou, Y. Bengio and P. Haffner, "Gradient-based learning applied to document recognition," *Proc. IEEE*, vol. 86, no. 11, pp. 2278–2324, 1998.

[49] E. Elhamifar and R. Vidal, "Sparse subspace clustering: Algorithms, theory, and applications," *IEEE Trans. Pattern Anal. Mach. Intell.*, vol. 35, no. 11, pp. 2765–2781, Nov. 2013.

[50] Y. C. Eldar and G. Kutyniok, *Compressed Sensing: Theory and Applications*. Cambridge University Press, 2011.